\documentclass{aa} 

\usepackage{graphicx}
\usepackage{txfonts}
\usepackage[citecolor=blue, linkcolor=blue, urlcolor = black, colorlinks = true]{hyperref}

\usepackage{graphicx}	
\usepackage{amsmath}	
\usepackage{amssymb}	
\usepackage{xspace}

\usepackage{siunitx}
\usepackage{comment}
\usepackage{textcomp}
\usepackage{bm}
\usepackage{lscape}
\usepackage{xcolor}

\newcommand{\gDor}{$\gamma$~Dor\xspace}
\newcommand{\gmodes}{g~modes\xspace}
\newcommand{\Msun}{\,M$_{\odot}$\xspace}

\newcommand{\xc}{$X_{\rm c}$\xspace}
\newcommand{\zini}{$Z$\xspace}
\newcommand{\fov}{$f_{\rm ov}$\xspace}
\newcommand{\frot}{$f_{\rm rot}$\xspace}

\newcommand{\teff}{$T_{\rm eff}$\xspace}
\newcommand{\lteff}{$\log T_{\rm eff}$\xspace}
\newcommand{\logg}{$\log g$\xspace}
\newcommand{\logL}{$\log (L/{\rm L_\odot})$\xspace}

\newcommand{\pin}{$\Pi_0$\xspace}

\newcommand{\kn}{KIC\,9751996\xspace}
\newcommand{\edit}[1]{#1}

\newcommand{\mesa}{\texttt{MESA}\xspace}
\newcommand{\gyre}{\texttt{GYRE}\xspace}
\newcommand{\ctpo}{\texttt{C-3PO}\xspace}

\begin{document}

   \title{Constraining stellar evolution theory with asteroseismology of $\gamma$\,Doradus stars using deep learning }
   \subtitle{Stellar masses, ages, and core-boundary mixing}

   \author{J.S.G. Mombarg
          \inst{1}
          \and
          T. Van Reeth\inst{1}
          \and
          C. Aerts\inst{1,2,3}      
          }

   \institute{Institute of Astronomy, KU Leuven, Celestijnenlaan 200D, B-3001 Leuven, Belgium\\
              \email{joey.mombarg@kuleuven.be}
    \and
             Department of Astrophysics, IMAPP, Radboud University Nijmegen, PO Box 9010, 6500 GL Nijmegen, The Netherlands
    \and
        Max Planck Institute for Astronomy, Koenigstuhl 17, 69117 Heidelberg, Germany
        }

   \date{Received September, 28 2020; accepted March, 19 2021}
\titlerunning{modelling  gravito-inertial modes}
\authorrunning{Mombarg et al.}
 
  \abstract
   {The efficiency of the transport of angular momentum and chemical elements inside intermediate-mass stars lacks proper calibration, thereby introducing uncertainties on a star's evolutionary pathway. Improvements require better estimation of stellar masses, evolutionary stages, and internal mixing properties.
   }
   {We aim to develop a neural network approach for asteroseismic modelling  and test its capacity to provide stellar masses, ages, and \edit{overshooting parameter} 
for a sample of 37 $\gamma$ Doradus stars for which these parameters were previously determined from their 
effective temperature, surface gravity, near-core rotation frequency, and buoyancy travel time, $\Pi_0$. Here, our goal is to perform the parameter estimation from modelling of individual
  periods measured for dipole modes with consecutive radial order rather than from $\Pi_0$. We assess if fitting of these individual mode periods increases the capacity of the parameter estimation.}
   {We have trained neural networks to predict theoretical pulsation periods of high-order gravity modes 
   ($n \in [15, 91]$), as well as the luminosity, effective temperature, and surface gravity for a given mass, age, \edit{overshooting parameter}, \edit{diffusive envelope mixing}, metallicity, and near-core rotation frequency. We have applied our neural networks for Computing Pulsation Periods and Photospheric Observables, \ctpo, to our sample and compute grids of stellar pulsation models for the estimated parameters.}
   {We present the near-core rotation rates (from literature) as a function of 
   the inferred stellar age and critical rotation rate. We assess the  rotation rates of the sample near the start of the main sequence assuming rigid rotation. Furthermore, we measure the extent of the core overshoot region and find no correlation with mass, age, or rotation. Finally, for one star in our sample, KIC\,12066947, we find indications of mode coupling in the period spacing pattern which we cannot reproduce with mode trapping.}
   {The neural network approach developed in this study allows for the derivation of stellar properties dominant for stellar evolution -- such as mass, age, and extent of core-boundary mixing. It also opens a path for future estimation of mixing profiles throughout the radiative envelope, with the aim to infer those profiles for large samples of $\gamma$ Doradus stars. }


   \maketitle
%

\section{Introduction}

{Accurate predictions of a star's evolutionary path depend on the accuracy of the description of transport of angular momentum \citep[AM, e.g.][]{maeder2009, Aerts2019-ARAA} and chemical elements \citep[e.g.][]{Salaris2017}. The} transport mechanisms are still poorly understood, thereby introducing uncertainties in state-of-the-art stellar structure and evolution models. These uncertainties already occur during the core-hydrogen burning stage, and hence propagate into models of more evolved stars as well. Asteroseismology, the study of stellar oscillations, of low-to-intermediate mass $M_\star \lesssim 3.3\,{\rm M_\odot}$ stars constitutes a powerful tool to measure surface- and near-core rotation rates across various evolutionary phases: main sequence (MS); \citealt{Kurtz2014, Saio2015, VanReeth2016, VanReeth2018, christophe2018, ouazzani2017}, subgiants and red giants; \citealt{beck2012, mosser2012, gehan2018}, and white dwarfs; \citealt{hermes2017}.
The empirically derived rotation rates require the transport of AM to be one to two order(s) of magnitude more efficient than what is currently predicted by theory \citep[e.g.][]{cantiello2014, fuller2014, eggenberger2017, tayar2018, Eggenberger2019I, Eggenberger2019II, fuller2019, DenHartogh2019, DenHartogh2020}. It has been suggested by \cite{eggenberger2017} that the efficiency of AM transport increases with increasing mass. 

\cite{Aerts2019-ARAA} (their Figure~4) present an overview of measured core rotation rates versus the surface gravity (\logg) in the literature for stars with $M_\star \in [0.72, 7.9]\,{\rm M_\odot}$. While the different evolutionary stages can be distinguished based on \logg, the typical uncertainty is too large to infer any correlations between the rotation rate and the stellar age \citep{aerts2017}. Instead of using \logg as an age proxy, \cite{ouazzani2018} used the reduced asymptotic period spacing $\Pi_0$, 
{which represents the buoyancy travel time throughout the star, as}
derived for a sample of $\gamma$ Doradus (\gDor) stars. Such stars are of spectral type late-A to early-F ($1.4\,{\rm M_\odot} \lesssim M_\star \lesssim 1.9\,{\rm M_\odot}$) and show gravity (g) modes excited via a {convective flux blocking mechanism \citep[e.g.][]{guzik2000, dupret2005}, 
{although the $\kappa\,$mechanism also plays a role for the hotter members of the class \citep{xiong2016}.}}

In a chemically homogeneous, non-rotating, non-magnetic star, the periods of \gmodes\ -- with the same spherical degree $\ell$ and azimuthal order $m$, but consecutive radial order $n$ -- are equally spaced in period in the asymptotic regime ($n \gg \ell$), namely by $\Pi_0/\sqrt{\ell (\ell + 1)}$. 
It is therefore customary to present the oscillations in a period spacing diagram where the spacing $\Delta P = P_{n+1} - P_{n}$ 
{of each individual mode with radial order $n$
is plotted as a function of its period $P_{n}$.} The advent of the \textit{Kepler} \citep{borucki2010} and TESS \citep{Ricker2015} missions have led to numerous detections of such period spacing patterns in \gDor stars \citep{Kurtz2014, Saio2015, VanReeth2015, Keen2015, Saio2018, Li2018, Li2019, antoci2019}. A first attempt at estimating stellar masses and ages of a sample of 37 \gDor stars was done by \cite{Mombarg2019}, using $\Pi_0$, \teff, and \logg assembled in \cite{VanReeth2015b, VanReeth2016} as input for the modelling. Their results show that faster rotating stars tend to be in the early phases of the MS and there was no correlation between age and the detection of Rossby modes. In this paper we aim to refine their work {by fitting the measured periods for each of the individual dipole modes with consecutive radial order instead of just $\Pi_0$ as asteroseismic observable.} 

Besides probing stellar rotation rates, asteroseismology also allows for the scrutiny of a star's internal chemical mixing profile \citep[e.g.][]{Pedersen2018, Michielsen2019, Pedersen2021}. During the core-hydrogen burning phase, a chemical gradient is introduced in the near-core region 
as the mean molecular weight inside the core increases.

The presence of a chemical gradient causes mode trapping, which translates into characteristic dips in the period spacing pattern \citep{miglio2008}. Yet, mixing occurs 
{in the core boundary layers, due to effects such as core overshooting and a variety of phenomena occurring at the bottom of the envelope.} 
In addition, mixing throughout the radiative envelope occurs, for example as a result of shear instabilities \citep{maeder2009} or internal gravity waves \citep[e.g.][]{RogersMcElwaine2017}. Such forms of mixing alter the chemical gradient, making it possible to probe the efficiency 
{of the mixing throughout the stellar envelope and infer properties of 
the mixing profile provided that modes with suitable probing power can be detected and identified \citep{Aerts2021}. This has been achieved for a sample of B-type g-mode pulsators, revealing a large range of mixing levels ranging 
from $\sim\! 10$ to 
$\sim\! 10^6$\,cm$^2$\,s$^{-1}$ \citep{Pedersen2021}. For \gDor stars, our understanding of mixing in the envelope is less advanced because their levels of mixing at the deep bottom of the envelope, at the interface with the core overshoot zone, were found to be much lower \citep{VanReeth2016,Mombarg2019}.}

In this paper, 
{we take the first steps to develop a new modelling approach based on deep learning, with the future aim to estimate the mixing profile throughout the radiative envelope of \gDor stars. In this initial study, we only treat the mixing in the near-core boundary layer, while fixing the one in the outer envelope. We do this because we first aim to assess the precision estimation of the global stellar parameters, such as the mass, metallicity, age, and  convective-core overshooting \edit{for an exponential overshooting prescription}, 
by relying on the individual mode period spacings rather than on just \pin. In order to test the capacity of our new deep learning method,  
we re-model the measured period spacing patterns of the 37 \gDor 
stars from the sample of \cite{VanReeth2015b} and derive masses, ages, and 
near-core mixing efficiencies. These are then compared to those 
obtained earlier by \citet{Mombarg2019}. If we achieve a better modelling strategy based on our initial deep learning approach, then future applications in a much higher-dimensional parameter space become possible, which would allow for the additional estimation of the envelope mixing profiles responsible for the observed morphologies of the period spacing patterns in terms of isolated or recurring dips as observed by \citep{VanReeth2015b,Li2018,Li2019}.}

\section{Deep Learning}

\begin{figure}
    \centering
    \includegraphics[width = 0.48\textwidth]{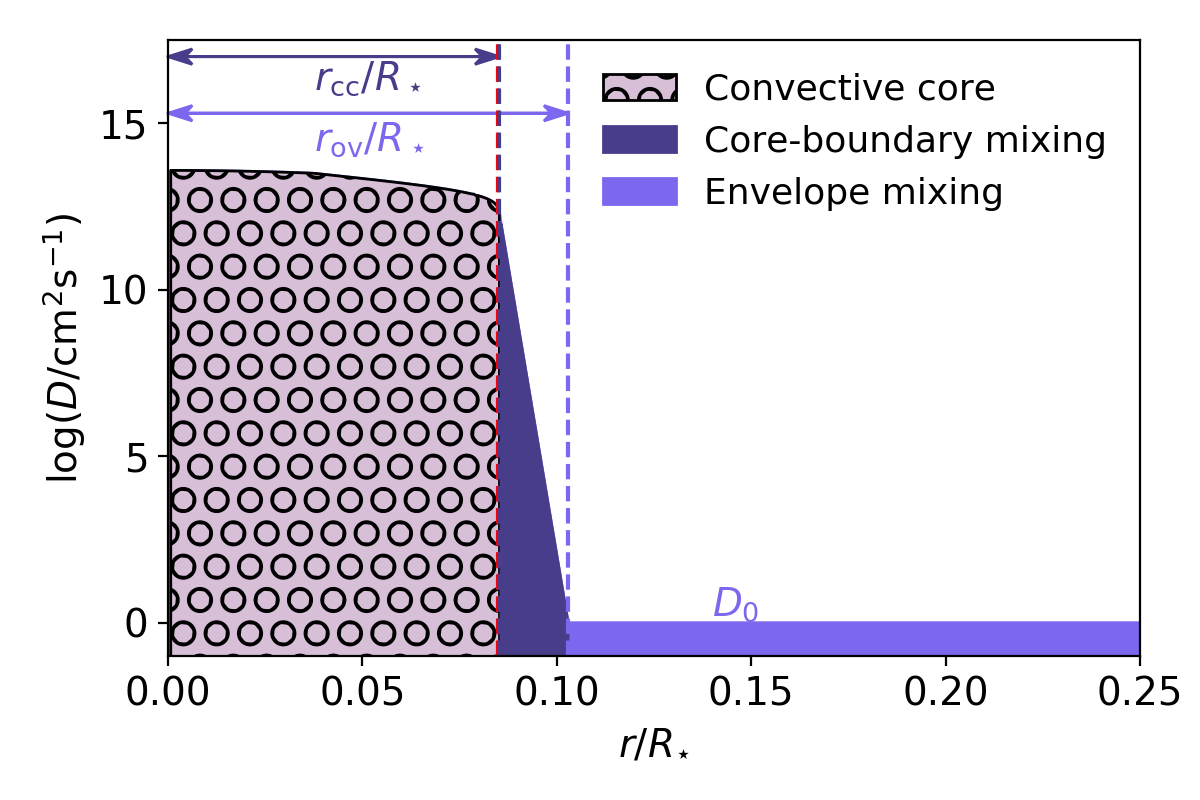}
    \caption{Chemical mixing profile showing the different mixing zones in a model for an intermediate-mass star. The ordinate shows the local efficiency of chemical (diffusive) mixing. The thin outer convection zone is not shown in the plot. The radii of the convective core boundary ($r_{\rm cc}$) and  overshoot zone ($r_{\rm ov}$) are also indicated. Mixing inside the convective regions 
    {is based on MLT and}
    occurs instantaneously. \edit{The red dashed line slightly to the left of $r_{\rm cc}/R_\star$ indicates the starting point of the overshoot zone ($r_0$ in Eq.~\ref{eq:ov}).}}
    \label{fig:Dmix_sketch}
\end{figure}

One of the biggest challenges in asteroseismic modelling  of period spacing patterns is that it requires a parameter search in a high-dimensional space. The most important parameters constitute the stellar mass ($M_\star$), the initial metallicity $Z$, and the hydrogen mass fraction in the convective core ($X_{\rm c}$; a proxy for the stellar age). In addition, the chemical mixing profile is typically split in two parts; the convective core overshooting (dark-coloured regions in Figure~\ref{fig:Dmix_sketch}) and the mixing efficiency in the radiative envelope (denoted as $D_0$ in Figure~\ref{fig:Dmix_sketch}). \edit{Given that the physical mechanisms at stake are still unknown, both} the overshoot and envelope mixing profile are parameterized by a function dependent on the local radius $r$, 
{where the exact functional description in real stars 
{is still a matter of debate \citep[e.g.,][for a general discussion and example profiles]{Aerts2021}. \citet{Mombarg2019} showed that diffusive exponentially decaying core overshooting cannot be distinguished from convective penetration in their modelling based on $\Pi_0$. As our main aim here is to evaluate our new deep learning modelling method in terms of capacity to estimate the global parameters and not to investigate the detailed morphology of the period spacing patterns,} we rely on an exponentially decaying diffusive  core-overshooting prescription. Our focus thus lies on the parameter estimation delivered by the individual period spacing values coupled to a neural network. 
The diffusive overshooting is expressed as a dimensionless parameter $f_{\rm ov}$ times the pressure scale height \citep{Freytag1996}:
\begin{equation} \label{eq:ov}
    D_{\rm CBM}(r) = D_{\rm CBM}(r_0)\exp\left( \frac{-2(r-r_0)}{f_{\rm ov}H_{P}(r_{\rm cc})}\right),
\end{equation}
where $r_{\rm cc}$ and $H_P$ are the distance from the stellar centre to the \edit{(Schwarzschild)} core boundary, and pressure scale height, respectively. For numerical purposes, the mixing efficiency at the stitching point between convective mixing in the core and the core overshoot mixing is evaluated at $r_{\rm cc} - f_0H_{P}(r_{\rm cc})$, for which we set $f_0 = 0.005$.} 
{As said, we do not focus on the shape of the mixing profile in the outer radiative envelope in this first application of the neural network. Hence we assume a constant mixing efficiency throughout the radiative zone for simplicity. 
}

{Aside from these most important parameters determining the stellar structure and evolution (SSE) models, 
$M_\star$, $Z$, $X_{\rm c}$, and $f_{\rm ov}$, 
we introduce the internal rotation 
frequency $f_{\rm rot}$ at the level of the pulsation computations.} 
Hence, asteroseismic modelling  of stars with a convective core is in general a +6D problem \citep{Aerts2018-apjs,Aerts2021}. 
This modelling scheme becomes rapidly computationally expensive with 
{the option to also assess mixing profiles via free parameters and for application to large \gDor sample sizes such as the one provided by \citet{Li2020}. To deal with such types of future applications,} grids of stellar or pulsation models can be approximated and replaced by statistical models (\cite{Mombarg2019, Pedersen2021}) or neural networks (this work). The latter method has been applied to non-rotating pulsation models for a benchmark sample of pulsating B-type stars by \cite{hendriks2019}, reaching an accuracy of $\sim$10\% on the pulsation frequencies predicted by deep learning models, compared to those from the pulsation models. {The observed values for \zini and \frot show that the stars in our sample cover a large range. Furthermore, the \gDor instability strip is narrow, compared to those of other pulsators. As such, properly covering the most important stellar parameters with an equally-spaced linear grid set-up, as is commonly done in asteroseismic modelling, would require a very large amount of equilibrium models. We seek to circumvent this by replacing the need of such extensive model grids by a neural network (NN). This study serves as an initial proof-of-concept study of the application of deep learning to asteroseismic modelling of g-mode pulsators, with the idea that future studies may rely on a larger set of grids with different input physics. }

\subsection{Predicting pulsation periods}
In this work, we train a neural network to predict g-mode periods of theoretical pulsation models for a given $M_\star$, \xc, \zini, \fov, $D_0$, and \frot. We have chosen this approach instead of using the asteroseismic data as input and the stellar parameters as output \citep[e.g.][]{Hon2020}, because this allows for more flexibility in the number of free parameters, \edit{as we can fix some of these parameters without having to retrain the NN.} {We keep this outlook in mind for future applications based on more complex envelope mixing profiles. In this initial study,  we compute a grid of stellar equilibrium models for the parameter ranges listed in Table~\ref{tab:train_grid}, based on the ranges used in \cite{Mombarg2019}.} \edit{We base the range of \fov on the typical values found by \cite{claret-torres2017} for the appropriate mass regime.} The parameter space is sampled both linear, and quasi random from a Sobol sequence, following \citet{Bellinger2016}. This way, we obtain a high sampling density for each parameter whilst also allowing the parameters to vary independently. The stellar equilibrium models have been computed with the SSE code \mesa (r11701) \citep{Paxton2011, Paxton2013, Paxton2015, Paxton2018, Paxton2019} for the computations of the stellar equilibrium models, using the same input physics as the models without atomic diffusion described in \cite{mombarg2020}. The stars in our sample are assumed to have a chemical mixture similar to the Sun, that is, we take the Solar abundances from \cite{Asplund2009}, and scale these according to the metallicity, \zini. The helium mass fraction of the models are set according to the enrichment rate found by \cite{Verma2019} and the initial hydrogen mass fraction is set by $X_{\rm ini} = 1 - Y_{\rm ini} - Z = 0.756 - 2.226Z$ with $^2{\rm H}/^1{\rm H}$ and $^3{\rm He}/^4{\rm He}$ isotope ratios set to the values measured by \cite{Asplund2009}.

From the stellar equilibrium models, the theoretically predicted pulsations are computed with the pulsation code \gyre \citep[v5.2;][]{Townsend2013, Townsend2018}, using the adiabatic framework and treating the Coriolis acceleration non-perturbatively. The equations of motion are decoupled in the angular and radial components by neglecting the latitudinal component of the rotation vector \citep[Traditional Approximation of Rotation; TAR;][]{eckart1960,Townsend2003}. 
Analyses of large samples of \gDor stars by \cite{VanReeth2016} and \cite{Li2019} reveal the prograde dipole mode ($\ell = 1, m = 1$) is by far the most observed mode geometry. Hence, for each stellar equilibrium model we compute the predicted pulsation frequencies for $(\ell, m) = (1,1)$ modes and radial orders $n \in [15, 91]$ (thus $\Delta P_{n}$ for $n \in [15, 90]$), using the pulsation code \gyre \citep{Townsend2003, Townsend2013}. The range of radial orders is based on the distribution found by \cite{Li2020}, taking the most ubiquitous ones. Again, we quasi-randomly sample \xc and \frot within the ranges listed in Table~\ref{tab:train_grid}. The rotation frequency range is based on the minimum and maximum observed values in the sample of \cite{VanReeth2016}. The total output data set contains \SI{38915} pulsation models, of which 70\% is used for training and 30\% for validation.  

\begin{table}[]
    \centering
    \begin{tabular}{lll}
        \hline \hline
        Parameter & Lower & Upper \\
                 & Boundary & Boundary \\
        \hline
        $M_\star\,({\rm M_\odot})$ & 1.3 & 2.0 \\
        $X_{\rm c}$ & 0.05 & 0.70 \\
        $Z$ & 0.011 & 0.023 \\
        $f_{\rm ov}$ & 0.010 & 0.030 \\
        $D_0\,({\rm cm^2s^{-1}})$ & 1 & 100 \\
        $f_{\rm rot}\,({\rm d^{-1}})$ & 0.0 & 2.3 \\
        \hline
    \end{tabular}
    \caption{Extent of the grid of stellar models used to train the \ctpo neural network.  }
    \label{tab:train_grid}
\end{table}

\begin{figure*}
    \centering
    \includegraphics[width = \textwidth]{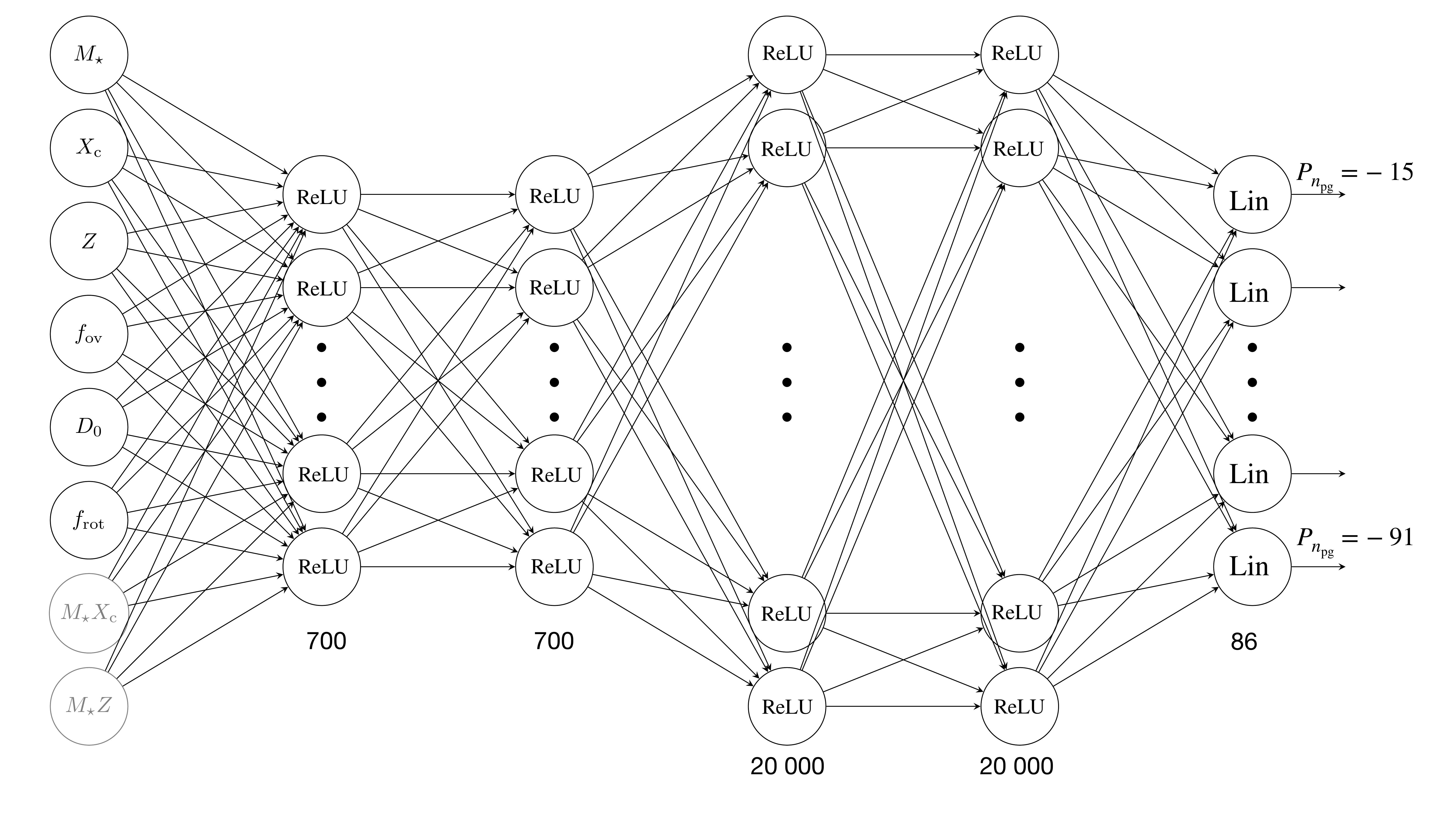}
    \caption{\edit{Schematic overview of the configuration of the dense neural network trained to predict the periods of the ($\ell, m$) = (1,1) modes for a given $\theta = (M_\star, X_{\rm c}, Z, f_{\rm ov}, D_{0}, f_{\rm rot}, M_\star \cdot X_{\rm c}, M_\star \cdot Z)$. The values of the input parameters are transformed as per Eq.~(\ref{eq:project}).}}
    \label{fig:DNN_P}
\end{figure*}

{Neural networks are powerful numerical methods suitable to treat multi-dimensional and complex regression problems \cite[see][to which we refer the reader for an introduction and general overview on deep learning]{Bishop1995,glorot2011}. }
We constructed a dense NN
comprising five layers, using the \texttt{KERAS} package for \texttt{Python} (with \texttt{Tensorflow} 1.14.0), for which the schematic overview of the NN is shown in Figure~\ref{fig:DNN_P}. A rectified linear unit (ReLU) was used as activation function for the neurons of the first four layers, and a linear activation function for the output layer, including a bias term for both activation functions. {We have tested several of the most commonly used activation functions (ReLU, sigmoid, tanh), for which we found that the ReLU activation functions yielded the best performance for our NN.} The output of the NN are the predicted periods of radials orders $n \in [15, 91]$. 

{Similarly to 
{the construction of}
statistical models {as a means to represent reality}, adding nonlinear combinations of the input parameters, may increase the performance of a NN.
{Inspired by the regressions performed in \citet{Mombarg2019}, we}
found the performance of our NN to be enhanced by adding two product terms of the most correlated parameters, the mass and age, and mass and metallicity, to the input vector $\theta = (M_\star, X_{\rm c}, Z, f_{\rm ov}, D_{0}, f_{\rm rot}, M_\star \cdot X_{\rm c}, M_\star \cdot Z)$}. Each component of $\theta$ is \edit{normalized} as per,
\begin{equation} \label{eq:project}
    \bar{\theta}^{i}_{j} = \frac{\theta^i_j - \min \theta^i}{\max \theta^i - \min \theta^i},
\end{equation}
where $i$ is the parameter index and $j$ the model index. The technique of `early stopping' is applied to prevent the loss from increasing again after several epochs, that is, the training is terminated when the validation loss has not decreased during three consecutive epochs. The weights are then restored to the configuration for which the validation loss was the lowest. 

The problem of overfitting is common when training a NN, that is, the NN achieves a high precision on the training set while it fails to generalize the correlations, and thus underperforms on the validation set. To remedy overfitting, a penalty term $\lambda \sum_j w_j^2$ is added to the loss function, where $w_j$ are the synaptic weights and $\lambda$ the control parameter, which we set to 0.01. This technique penalizes for large weights and is commonly referred to as L2 or ridge regularization \citep[cf.][Chap. 9.2]{Bishop1995}. As a loss function, the mean squared error is used. 
The final configuration of the weights is somewhat dependent on the initial weights, which introduces noise in the predictions. To remedy this, we train six NNs with different initial weights and average their predictions (this technique is often referred to as `ensemble learning'). 

\begin{figure}
    \centering
    \includegraphics[width = 0.48\textwidth]{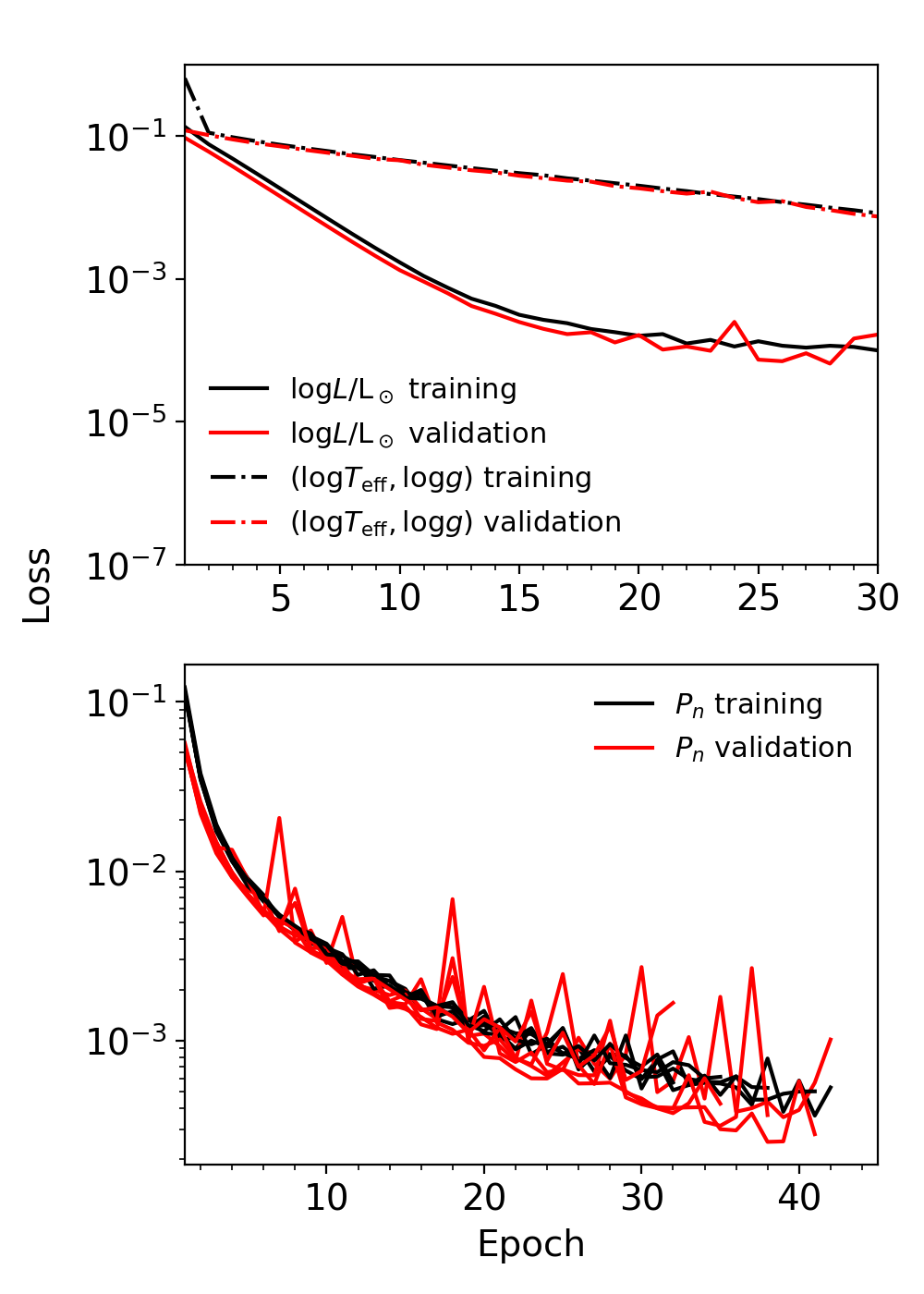}
    \caption{Loss (mean squared error + L2 regularization term) as a function of epoch. The NNs training on the pulsation models (bottom panel) reached an optimal solution within the maximum number of epochs. In all cases similar loss is obtained on the training and validation sets, indicating the NNs were able to generalize the synaptic weights. }
    \label{fig:lc}
\end{figure}

The so-called learning curve is a common diagnostic to assess whether a NN is able to make robust predictions. This curve shows the loss (which includes the L2 regularization term) as a function of epoch. In the top panel of Figure~\ref{fig:lc} we show the learning curves for all NNs trained to predict the pulsation periods. Ideally, a NN should obtain roughly the same loss on the training and validation set, which is indeed the case here. When ensemble learning is applied, a mean absolute error of $0.0024\,{\rm d}$ ($207\,{\rm s}$) on the pulsation periods is obtained over all models used to train and validate the NN. Typical observed spacings between periods of consecutive radial order vary from several hundreds to several thousands of seconds. It is more instructive to compare the accuracy of the NN with typical spacings rather than observational uncertainties, as the accuracy of the state-of-the-art pulsation models is typical orders of magnitude worse than the uncertainties on the measured periods from \textit{Kepler} light curves \cite[e.g.][]{buysschaert2018, mombarg2020}. 

\subsection{Lifting degeneracies}
\begin{figure}
    \centering
    \includegraphics[width = 0.50\textwidth]{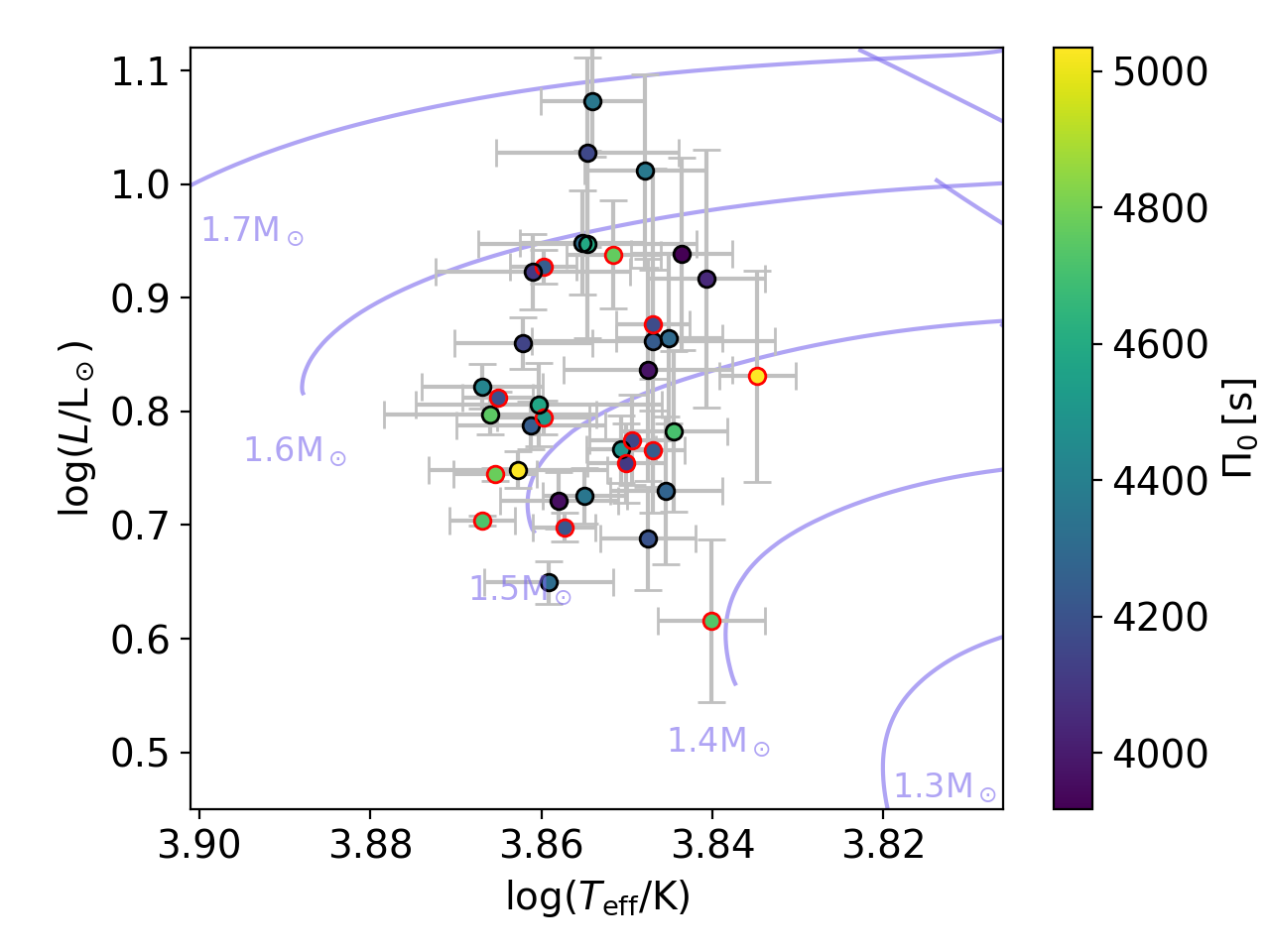}
    \caption{Positions in the Hertzsprung-Russel diagram of all stars in our sample. The values of \teff and \pin are taken from \cite{VanReeth2016}. The evolution tracks shown are for \zini = 0.014 and \fov = 0.0175. For the stars marked with red circles no reliable extinction estimate could be made, because these are too close to the Sun, therefore, their interstellar extinction values are close to zero.  }
    \label{fig:HRD}
\end{figure}

The modelling  of mode periods or period spacing patterns is prone the degeneracies, mainly between mass and age, and mass and metallicity \citep[e.g.][]{Moravveji2015, Mombarg2019, mombarg2020}. We apply the same methodology as \cite{mombarg2020}, where the best model is selected from models that are compatible with the effective temperature (\teff) and surface gravity (\logg), measured from spectroscopy. In addition to these two parameters, the luminosity (\logL) derived from the Gaia DR2 distances ($D$) measured by \cite{bailer-jones2018} is used:
\begin{equation}
    \log L/{\rm L_\odot} = -0.4\left(M_{V} - V_\odot - 31.572 + [{\rm BC}_V - {\rm BC}_{V, \odot}]\right),
\end{equation}
where $M_V = m_V\,-\,5 \log(D/10\,{\rm pc})\,-\,3.3E(B-V)$, $V_\odot = -26.76$, and ${\rm BC}_{V, \odot} = -0.080$.
Bolometric corrections (${\rm BC}_V$) are computed as per \cite{torres2010} and reddening corrections $E(B-V)$ are taken from the \texttt{Bayerstar2019} extinction map \citep{green2018}. From the extinction, the correction to the visual magnitude is calculated using $A_{V} = 3.3 E(B-V)$. Figure~\ref{fig:HRD} shows the position of all stars in our sample in a Hertzsprung-Russell diagram (see Table~\ref{tab:gaia} in Appendix). 

\begin{figure*}
    \centering
    \includegraphics[width = \textwidth]{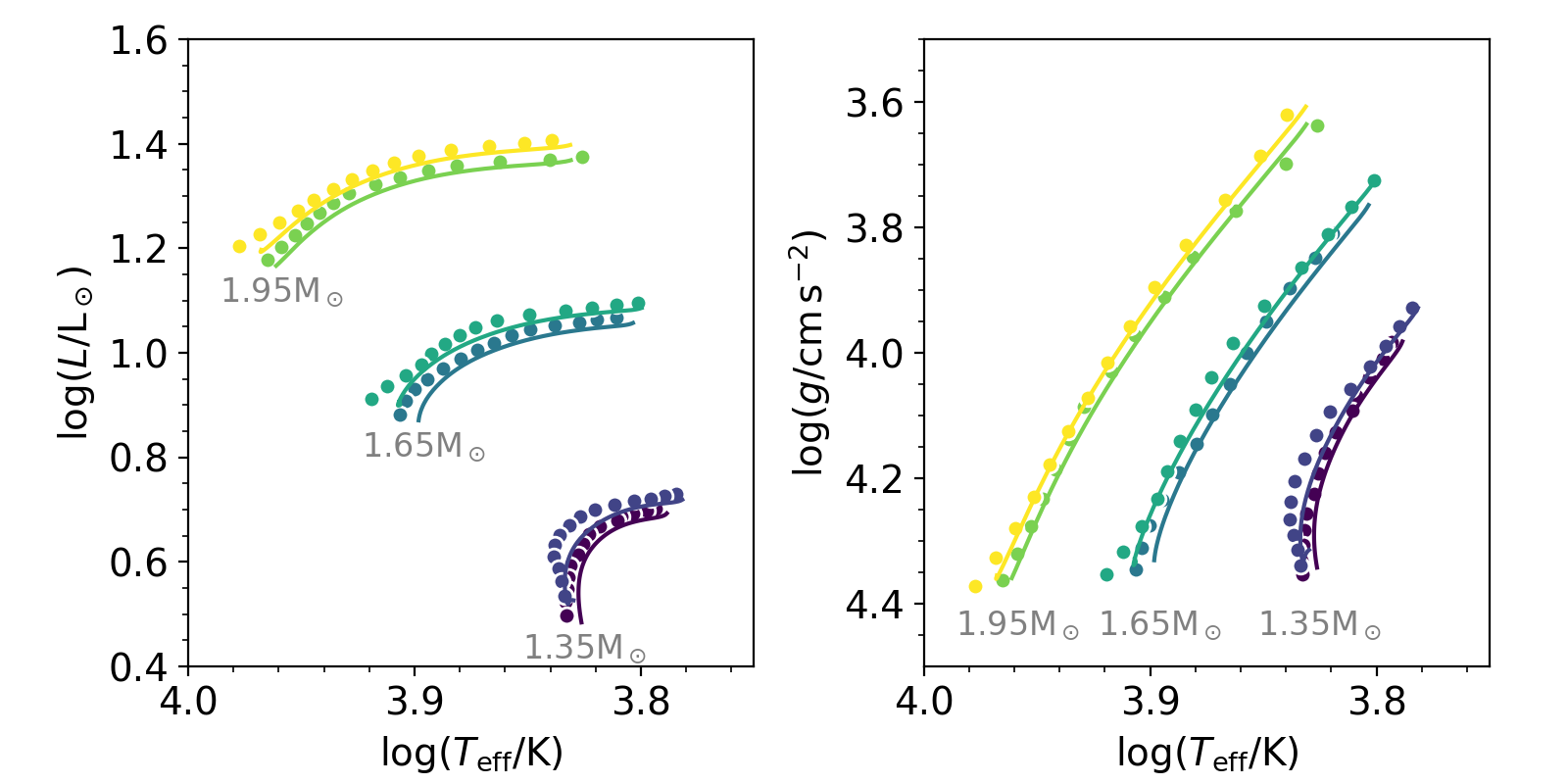}
    \caption{Comparison between the predictions of \ctpo (dots) and the \mesa benchmark evolution tracks (solid lines) for \zini = 0.012 and \zini = 0.023. The predictions are sampled from \xc = 0.70 to 0.05 with a step size of 0.05.  }
    \label{fig:HRD_predict}
\end{figure*}

In addition to the mode periods, we have trained two NNs to predict (\teff, \logg) and $\log L$, respectively, using the same configuration as the network shown in Figure~\ref{fig:DNN_P}. To train these NNs, we use all time steps (on the MS) computed by \mesa as opposed to using only the time step for which pulsation models were computed, that is, the training set comprises 423691 vectors and the validation set comprises 141231 vectors.  In Figure~\ref{fig:HRD_predict} we compare the predictions of the NNs of these photospheric observables to \mesa evolutionary tracks. We obtain mean absolute errors of $\bar{\epsilon}_{\log L/{\rm L_\odot}} = 0.010, \bar{\epsilon}_{\log T_{\rm eff}} = 0.004$, and $\bar{\epsilon}_{\log g} = 0.011$~dex on the complete data set (training plus validation).  The models shown in this figure were not included in the training or validation set. 
{There is no exact method to find the optimal configuration of a NN. We started from a similar network configuration as presented in \cite{hendriks2019}, and experimented with different numbers of hidden layers, and the number of neurons per layer, by means of trial and error.   } Table~\ref{tab:c3po} summarizes the ensemble of all the networks for Computing Pulsation Periods and Photospheric Observables, \ctpo. 

\begin{table}[]
    \centering
    \begin{tabular}{llll}
    \hline \hline
        Output & Configuration & Networks  \\
        \hline
        $P_{n}$, $n \in [15, 91]$ & 700, 700, 20000, 20000, 86 & 6 \\
        $\log L/{\rm L_\odot}$ & 700, 700, 20000, 20000, 1 & 1 \\
        $\log T_{\rm eff}, \log g$ & 700, 700, 20000, 20000, 2 & 1 \\
        \hline
    \end{tabular}
    \caption{Summary of the network configuration of \ctpo. The second column lists the number of neurons for each layer. }
    \label{tab:c3po}
\end{table}

\section{Asteroseismic modelling } \label{sec:modelling }
Until recently, g-mode modelling  of single \gDor stars relied on (\pin, \teff, \logg) as input \citep{Mombarg2019}. In the work by \cite{mombarg2020}, \pin is replaced with the individual mode periods as asteroseismic input for the modelling  of two slowly rotating \gDor stars. Furthermore, these authors investigated whether \teff and \logg should be added to the fit or used a posteriori to select models which are consistent within $n \sigma$ compared to the observations. As typically tens of excited modes of consecutive radial order are observed in a single star, the spectroscopic observables have little relative weight in the merit function. Therefore, we opt to use the \lteff and \logg to select the best-fitting model from a subset of models which are within the 2-$\sigma$ uncertainty ranges of these photospheric observables. 
\cite{Aerts2018-apjs} introduced the Mahalanobis distance (MD) as a merit function in asteroseismic modelling to account for the correlated nature of the observed periods, as well as the theoretical uncertainty of the modes. The MD takes the (co)variances of the input parameters into account such that the contours of equal MD are aligned with the principal components, whereas for the $\chi^2$ merit function contours of equal $\chi^2$ are aligned with the base vectors (free parameters). For a given vector $\bold{Y}^{\rm (th)}$, containing the periods of the identified radial orders, and a vector $\bold{Y}^{\rm (obs)}$ containing the observed periods, the MD is defined as
\begin{equation} \label{eq:MD}
    {\rm MD} = \left(\bold{Y}^{\rm (th)} - \bold{Y}^{\rm (obs)} \right)^\top \left(\bold{V} + \bold{\Sigma} \right)^{-1} \left(\bold{Y}^{\rm (th)} - \bold{Y}^{\rm (obs)} \right). 
\end{equation}

The matrix $\bold{V}$ is the (co)variance matrix and $\bold{\Sigma}$ is a diagonal matrix, constructed from the uncertainties on the predicted mode periods by the NN. These can be treated as aleatoric, as the residuals on the training/validation set follow a normal distribution (see Appendix\,\ref{app:res}). Therefore, each uncertainty $\sigma_{P_n^{\rm (th)}}$, with $n$ the radial order, is taken to be the standard deviation of the residuals over the complete grid of pulsation models. 
The uncertainties of the predictions are typically two orders of magnitude larger than those from the observations. \newline

The matching between the observed radial orders to those from a model is non trivial. {Observed period spacing patterns of \gDor stars do not always form a single sequence of consecutive radial orders, but may be interrupted by missing modes. Therefore,} the radial order matching in this paper is done as follows: 
\begin{enumerate}
    \item Start with the longest observed continuous sequence of $M$ periods and assign the shortest period $P_{0}^{\rm (obs)}$ to the best-matching period $P_{j, i}^{\rm (th)}$, as the shortest period has the smallest relative uncertainty.  
    \item Assign periods $P_{1}^{\rm (obs)}, \dotsc, P_{M}^{\rm (obs)}$ to the consecutive radial orders in the model.
    \item Repeat step 1 for the second longest sequence, and so on. When a radial order is selected which has already been assigned to an observed period in one of the longer sequences, omit the model. 
\end{enumerate}
{In this way, we account for the possibility of having multiple sequences of consecutive radial orders with missing modes in between those sequences}.
The construction of $\bold{V}$ can be done in two ways. 1: For each evaluated model $j$, each of the observed periods $P_i^{(\rm obs)}$ is matched with a period $P_i^{(\rm th)}$ corresponding to a radial order that may vary for different models. Matrix $V$ is given by, 
\begin{equation}
    \bold{V} = \frac{1}{q-1}\sum_{j}^{q} \left(\bold{Y}_j^{\rm (th)} - \bar{\bold{Y}}^{\rm (th)} \right) \left(\bold{Y}_j^{\rm (th)} - \bar{\bold{Y}}^{\rm (th)} \right)^\top,
\end{equation}
where $q$ is the number of evaluated models, and 
\begin{equation}
    \bar{\bold{Y}}^{\rm (th)} = \frac{1}{q}\sum_j^q \bold{Y}_j^{\rm (th)}. 
\end{equation}
2: Take $\bold{Y}_j^{\rm (th)}$ to contain the periods of all the radial orders predicted by the NN, that is, $n \in [15, 91]$, and construct $\bold{V}$, thereby yielding a matrix with dimensions 77\texttimes77. Consecutively, for each model with radial order identifications $n_{\rm min}, ..., n_{\rm max}$ corresponding to the $N$ observed mode periods, remove the rows and columns of $\bold{V}$ belonging to the radial orders that were not identified in model $j$, yielding an $N$\texttimes$N$ matrix, $\bold{V}^\prime_j$. 

In this paper, we have used method (2) to compute the MD, such that the (co)variances do not depend on the radial order identification. This is the case for method (1), because the lowest period is matched first and the consecutive periods are then matched to the consecutive radial orders of the model. Hence, when the observed $\Pi_0$ does not match the $\Pi_0$ of the model, the discrepancy between the model and observations is larger for higher radial orders. Therefore the variance across the grid will also be larger for these radial orders, and by extension will be given a smaller weight.
The maximum likelihood estimate (MLE) for our method is not equivalent to minimizing the merit function, as is the case for $\chi^2$. Instead, the likelihood of the observed data $\bm{D}$, given parameters $\bm{\theta} = (\theta^1, \theta^2, \theta^3)$, is given by,
\begin{equation}\label{eq:MLE}
\begin{split}
        L(\bm{\theta} | \bm{D}) &= P(\bm{D}|\bm{\theta}) \\ &= \exp\left(-\frac{1}{2}(\ln |\bm{V^\prime}| + {\rm MD} + k\ln(2\pi)) \right),
 \end{split}
\end{equation}
where $k = 3$ is the number of free parameters.
According to Bayes' theorem, the probability of component $\theta^k$ being within interval $[\theta_a^k, \theta_b^k]$ is given by
\begin{equation} \label{eq:MLE_err}
\begin{split}
         P(\theta_a^k < \theta^k < \theta_b^k|\bm{D}) &= \\ &\frac{\sum_i^{q} P(\bm{D}|\bm{\theta}_i)P(\theta_i^1)P(\theta_i^2)P(\theta_i^3)}  {\sum_j^Q P(\bm{D}|\bm{\theta}_j)P(\theta_j^1)P(\theta_j^2)P(\theta_j^3)}.
\end{split}         
\end{equation}
The sum over index $i$ is taken over the $q$ models with the highest likelihood such that $P(\theta_a^k < \theta^k < \theta_b^k|\bm{D}) = 0.68$, where the sum over index $j$ is taken over all of the models that are consistent with the observed photospheric observables. 
\newline  

The modelling of \gmodes in \gDor stars is a high-dimensional, degenerate problem. Optimizing all six aforementioned parameters at once will lead to noisy solution spaces, making uncertainty determination unwieldy. Yet, some parameters are more dominant than others, and thus, we first focus on the parameters which have the largest influence on the mode periods. Since the metallicity and near-core rotation rate have already been determined by \cite{VanReeth2016}, we fix these two parameters to the measured values. 
The influence of \fov, and specifically $D_0$ on the period spacings are mostly seen in the mode trapping. In this work, our aim is not to precisely model the trapped modes, but rather get an accurate fit to the global pattern of all observed modes. Accurately modelling trapped modes adds additional dimensionality to the problem in the form of structured (non-constant) mixing profiles \citep{Pedersen2021}, which we do not consider here. For each star in our sample, a best-fitting model is found as follows:
\begin{enumerate}
    \item Randomly sample 5000 models in ($M_\star, X_{\rm c}, f_{\rm ov}$) with \ctpo, where $Z$ and $f_{\rm rot}$ are fixed to the measured values, and $D_0 = 1\,{\rm cm^2\,s^{-1}}$.  
    \item If some observed periods cannot be matched in 90\% of the models (because the radial order is outside the range of NN) remove either the shortest or the longest period depending on which one is outside the range. Repeat until a model can be matched. 
    \item From these 5000 models, get the minimum and maximum values of $M_\star$ and  $X_{\rm c}$ such that the predicted $\log (L/{\rm L_\odot}), \log T_{\rm eff}$, and \logg are consistent within 2-$\sigma$ of the respective uncertainties. 
    \item Randomly sample 15000 models, but now within the mass and $X_{\rm c}$ ranges obtained in the previous step. The solution and uncertainties are computed according to Eqs~\ref{eq:MLE} and \ref{eq:MLE_err}. 
    
    \item Since the exact periods of trapped modes are difficult for a NN to predict, we optimize the overshoot parameter with \gyre models once we have a good estimate of ($M_\star, X_{\rm c}, f_{\rm ov}$) from the previous steps. We compute \gyre models for $f_{\rm ov} \in [0.005, 0.010, 0,015, 0.020, 0.025, 0.030, 0.035 ]$ for the MLEs for $M_\star$ and $X_{\rm c}$ and for the upper/lower limits of these two parameters (63 models per star in total). Models rotating faster than the critical Roche frequency, $\Omega_{\rm crit, Roche} = \sqrt{(8 G M_\star) /( 27 R_\star^3)}$, are not taken into account. 

\end{enumerate}

In the latter step, a $\chi^2$ merit function is used instead of the MD, as scanning such a small parameter range results in (co)variance matrices with very high condition numbers. The uncertainties on \fov are determined by,
\begin{equation}
    \chi^2_{1\sigma} = \chi^2_{\rm min} \left(1 + \sqrt{\frac{2}{N - k}}\right).
\end{equation}
For stars where an additional mode geometry has been observed, the corresponding pulsation models are also computed and fitted simultaneously with the prograde dipole modes in Step 5. 

\subsection{Theoretical benchmark}
\begin{figure*}[th!]
    \centering
    \includegraphics[width = 0.97\textwidth]{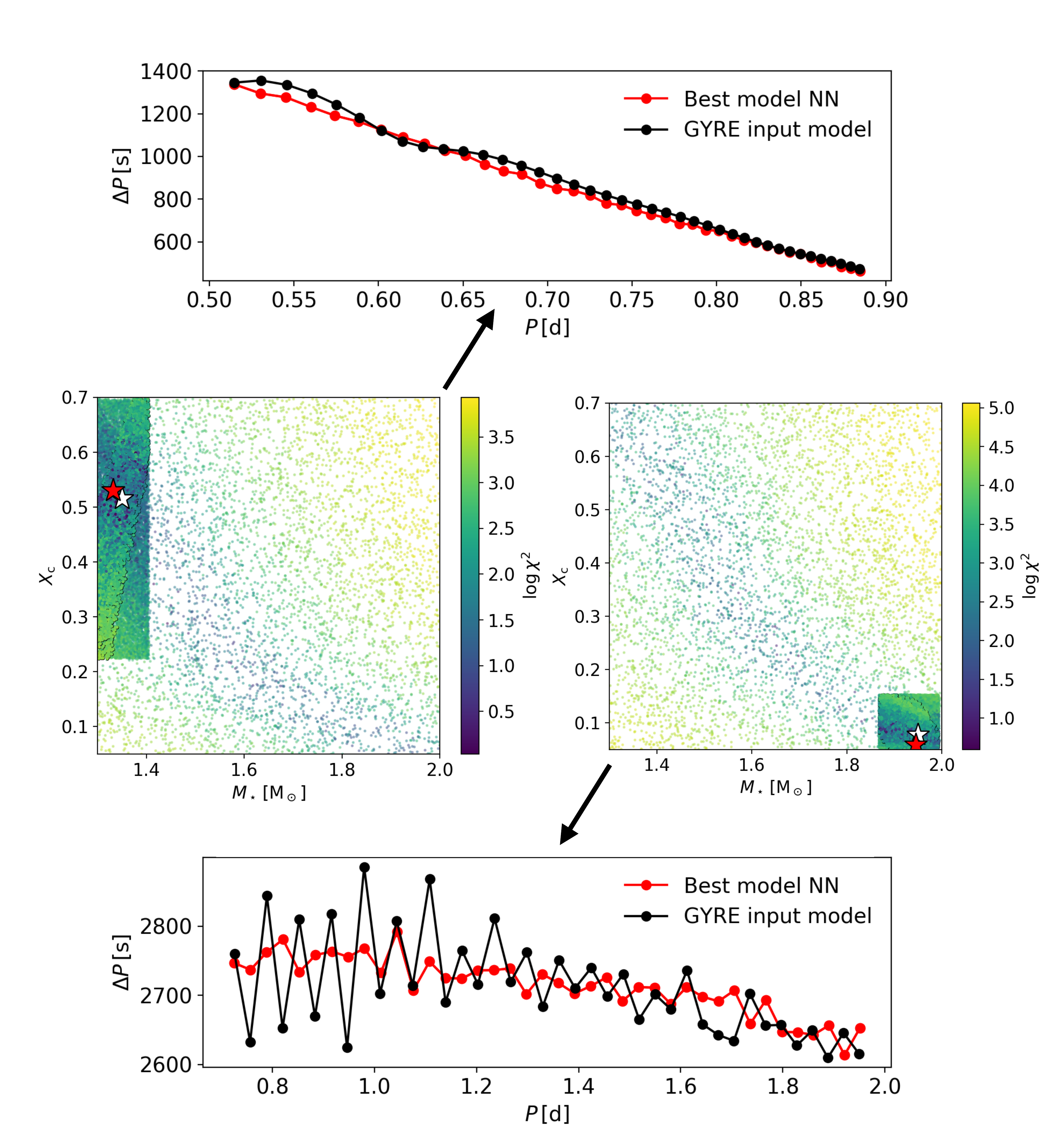}
    \caption{Examples of the recovered mass and \xc of two benchmark models (white stars: input, red stars: best model predicted by NN). Although the period spacings of trapped modes are more difficult to predict (bottom panel), the inclusion of the photospheric observables yields accurate predictions of the mass and \xc. The middle panels show the sampling density is increased in the parameter space where the predicted photospheric observables comply with those of the input model, as explained in Section~\ref{sec:modelling }. Model 1: $(M_\star, X_{\rm c}, Z, f_{\rm ov}, D_{0}, f_{\rm rot}) = (1.35\,{\rm M_\odot}, 0.515, 0.014, 0.0225, 1\,{\rm cm^2\,s^{-1}}, 0.7457\,{\rm d^{-1}})$. Model 2: $(M_\star, X_{\rm c}, Z, f_{\rm ov}, D_{0}, f_{\rm rot}) = (1.95\,{\rm M_\odot}, 0.08, 0.014, 0.0225, 1\,{\rm cm^2\,s^{-1}}, 0.0449\,{\rm d^{-1}})$. Typical uncertainties on $\Delta P$ are of the order of several tens of seconds.    }
    \label{fig:bm_example}
\end{figure*}

\begin{figure}
    \centering
    \includegraphics[width = 0.5\textwidth]{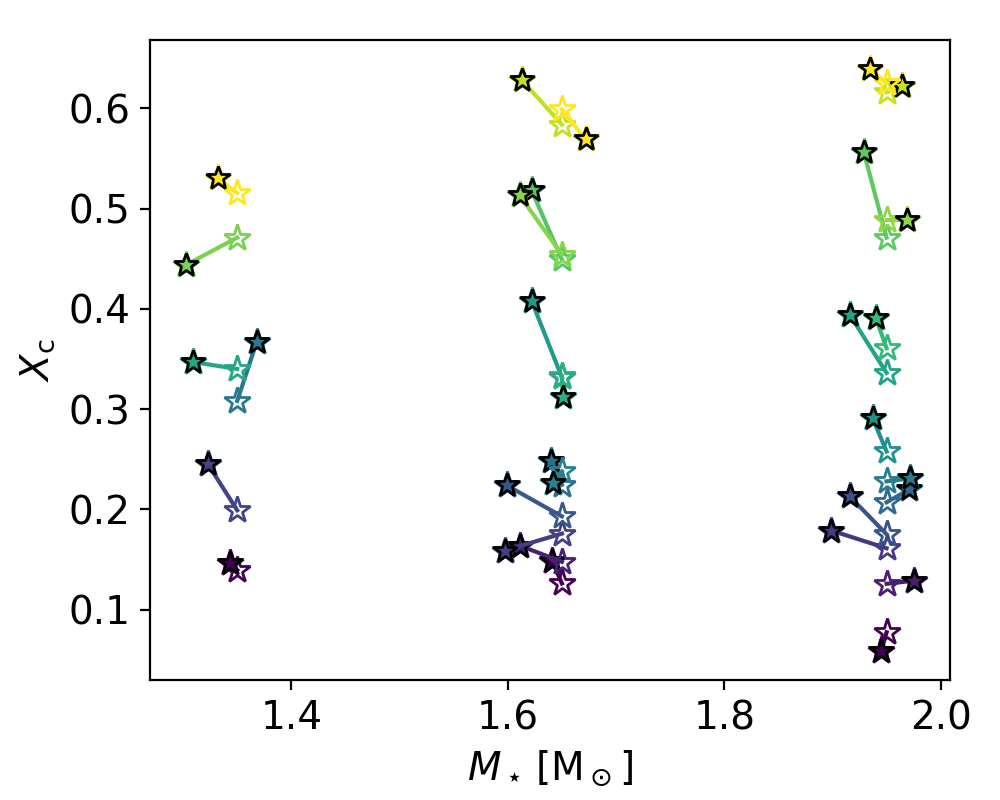}
    \caption{Mass and \xc values of the benchmark models (open symbols) and the corresponding retrieved values from the best-fitting model computed with \ctpo (filled symbols). For the benchmark models $Z = 0.014$ and $f_{\rm ov} \in [0.0175, 0.0225]$.}
    \label{fig:benchmark}
\end{figure}

First, we demonstrate the capabilities of \ctpo by applying the NN to a set of benchmark models (1.35\Msun, 1.65\Msun, 1.95\Msun) which were not included in the training/validation set. We use radial orders $n \in [20, 60]$, and the photospheric observables from the \mesa model as input, where we assume the following uncertainties: $\sigma_{\log\,{L/{\rm L_\odot}}} = 0.05\,{\rm dex}, \sigma_{\log\,{T_{\rm eff}}} = 0.015\,{\rm dex} $, and $\sigma_{\log\,g} = 0.6\,{\rm dex}$. The metallicity and rotation frequency are fixed to the values of the models in the fit, since these are also known for the observations. Two examples are shown in Figure~\ref{fig:bm_example} for a young star and an old star. The accuracy of the NN is less for patterns with clear mode trapping, yet in both cases the periods are still predicted with enough accuracy such that, with the inclusion of the photospheric observables, $M_\star$ and \xc can be recovered. In Figure~\ref{fig:benchmark}, we show the mass and \xc values of all the inputs models, and the obtained estimates from \ctpo. On average, the input mass, \xc, and \fov are recovered within 2\%, 15\%, and 16\%, respectively. It should also be noted that, as mentioned before, degeneracies exist between young lower-mass stars and old higher-mass stars. Thus, even if the NN would be able to perfectly predict the pulsation periods, exactly recovering a (perturbed) input model would still be impossible (see middle panels of Figure~\ref{fig:bm_example}).   

\subsection{Observational benchmark}
One of the stars in our sample, \kn, has been analysed by \cite{mombarg2020} who modelled the star using models both with and without atomic diffusion (including radiative levitation). As mentioned before, in this paper we use the same input physics in our models compared to their grid without atomic diffusion. It should be noted, however, that in this work only the periods of the prograde mode are modelled, whereas in the work by \cite{mombarg2020} the periods of the retrograde, zonal, and prograde are modelled simultaneously. 
 
\begin{figure}
    \centering
    \includegraphics[width = 0.5\textwidth]{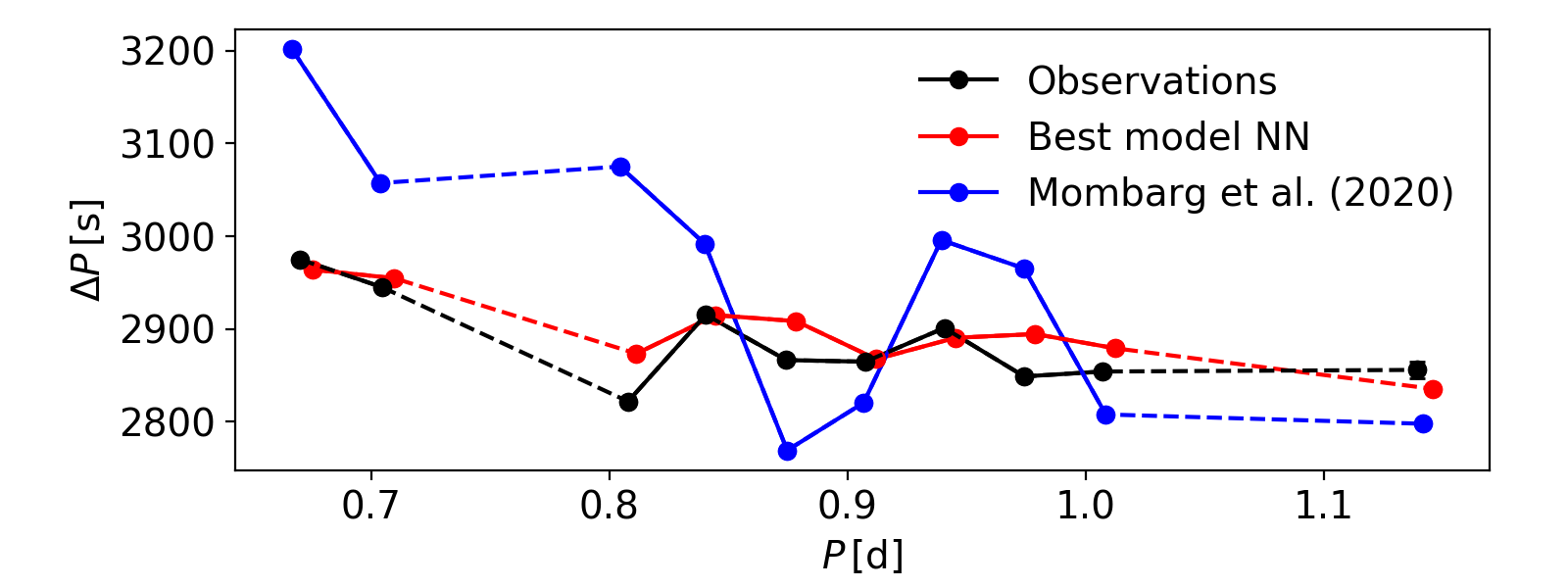}
    \caption{Observed period spacing pattern of \kn from \cite{VanReeth2016} (black dots, prograde dipole mode only) and the best-fitting pattern predicted with \ctpo (red dots). For comparison, the solution from \cite{mombarg2020} without atomic diffusion (their model M01) is also shown (blue dots). The observational uncertainties are typically smaller than the symbol size. }
    \label{fig:BM_975}
\end{figure}

  This yields $M_\star = 1.740_{-0.169}^{+0.009}\,{\rm M_\odot}, X_{\rm c} = 0.241_{-0.040}^{+0.147}$, and $f_{\rm ov} = 0.0124_{-0.0024}^{+0.0121}$, all of which are consistent with the values found by \cite{mombarg2020}. The predicted period spacing pattern of this best-fitting model from \ctpo is shown in Figure~\ref{fig:BM_975}.

\section{Application to the sample}
For all 37 stars in our sample, we apply the modelling  scheme discussed in Section~\ref{sec:modelling } to constrain $M_\star$, \xc, and \fov, 
{in order to compare the results with those based on $\Pi_0$ 
obtained by \citet{Mombarg2019}. We recall that our aim is not to achieve good fits to the dips in the patterns, as this requires stratified envelope mixing profiles \citep{Pedersen2021}. Here, we merely wish to compare the capacity of the NN with previous modelling based on actual equilibrium models rather than a numerical NN approximation thereof.}

\begin{figure*}
    \centering
    \includegraphics[width = \textwidth]{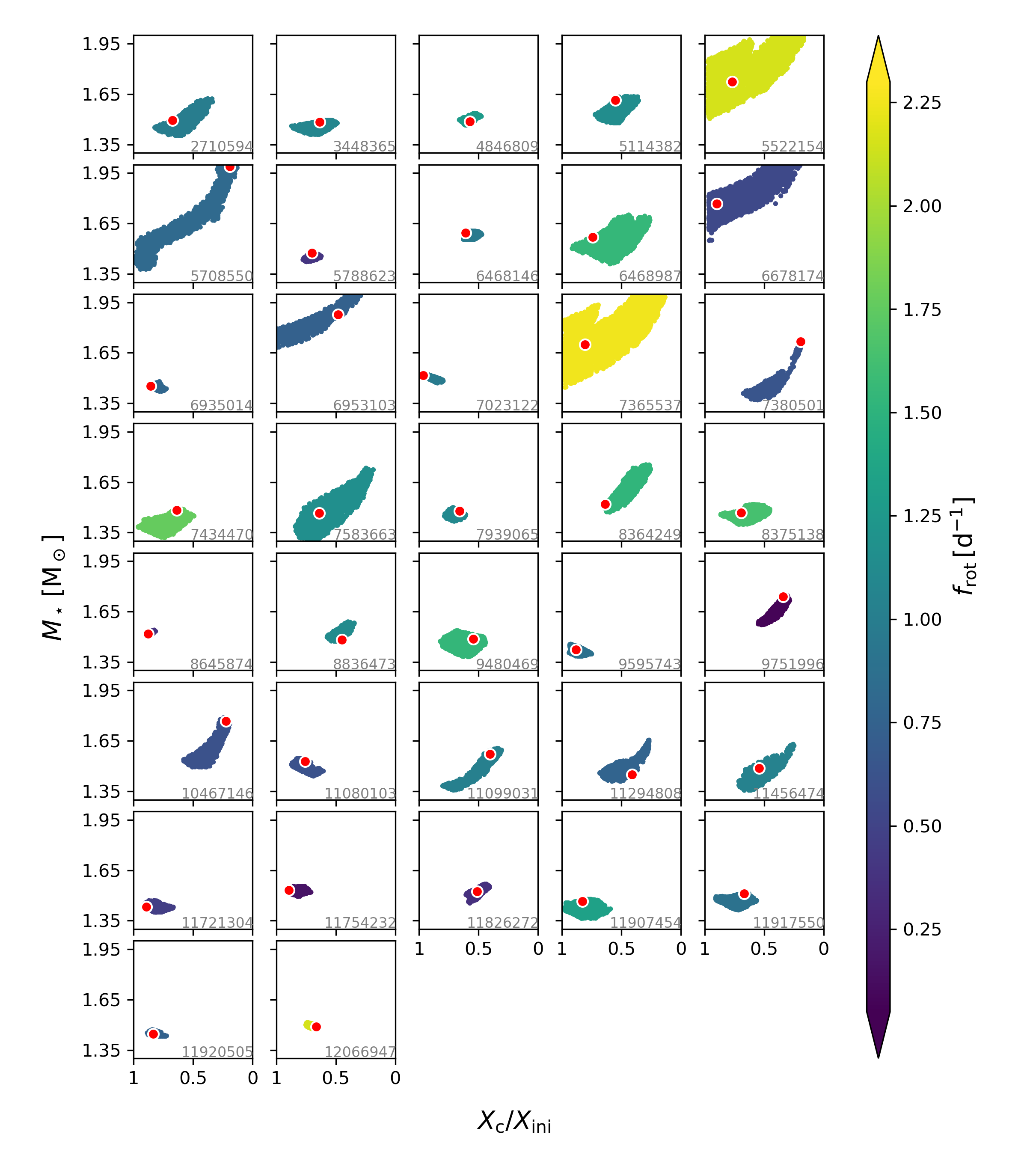}
    \caption{Correlation structure of the 68\%-confidence intervals of $M_\star$ and $X_{\rm c}/X_{\rm ini}$ for all 37 stars in our sample from modelling  the dipole period spacing patterns with \ctpo. The MLE is indicated by the red dot. The KIC number is indicated in the bottom right corner of each plot.}
    \label{fig:M-Xc_sample}
\end{figure*}

Figure~\ref{fig:M-Xc_sample} shows the mass and \xc of the best model predicted by \ctpo of each star. {As is expected from the position of the \gDor instability strip \citep{dupret2004}, there is a correlation between the estimated mass and age, i.e., no young, more massive or old, less massive stars are observed,
{as already found by \citet{Mombarg2019}.}}
KIC\,2710594, KIC\,3448365, KIC\,6678174, KIC\,6953103, KIC\,7380501, KIC\,8645874, KIC\,11099031, and KIC\,11294808 are stars for which a particularly complicated period spacing pattern is observed, \edit{i.e., large deviations in the observed morphology occur, compared to what we expect from the pulsation models limited to the physics described in this work.} 
For these stars, part of the pattern has not been taken into account in the fitting. In the case of KIC\,5522154, KIC\,5708550, KIC\,6678174, KIC\,6953103, KIC\,7365537, the best solutions to the observed mode periods do not seem to be in agreement with the photospheric observables. The former and latter are fast rotators ($f_{\rm rot} > 2.15\,{\rm d}^{-1}$, \cite{VanReeth2016}), and hence the inferred \teff could suffer a larger systemic error due to the larger uncertainty of the spectrum normalization and gravity darkening. For the other three stars, it is primarily the luminosity which does not agree with the observed pulsations. In any case, for these five stars, no $2-\sigma$ cutoff in $L, T_{\rm eff}$, and $\log g$ is imposed, resulting in a larger uncertainty on $M_\star$ and \xc, as can be seen in Figure~\ref{fig:M-Xc_sample}.

We have investigated the effect of adding the rotation frequency (within the observational uncertainties) as an additional free parameter in the fitting on the MLEs. For both the fastest rotating star (KIC\,12066947) and the star with the least precisely determined rotation rate (KIC\,6678174), fixing the rotation rate in the modelling yields consistent estimates of the varied stellar parameters, compared to when we allow it to vary.

In addition to $(\ell, m) = (1,1)$ modes, $(\ell, m) = (1,0)$ modes for KIC\,4846809 and KIC\,9595743, and an $(\ell, m) = (2,2)$ mode for KIC\,11294808 have been observed by \cite{VanReeth2016}. The additional period spacing patterns of these three stars have also been taken into account in Step 5 of the modelling  scheme (Section~\ref{sec:modelling }).

\begin{figure}
    \centering
    \includegraphics[width = 0.5\textwidth]{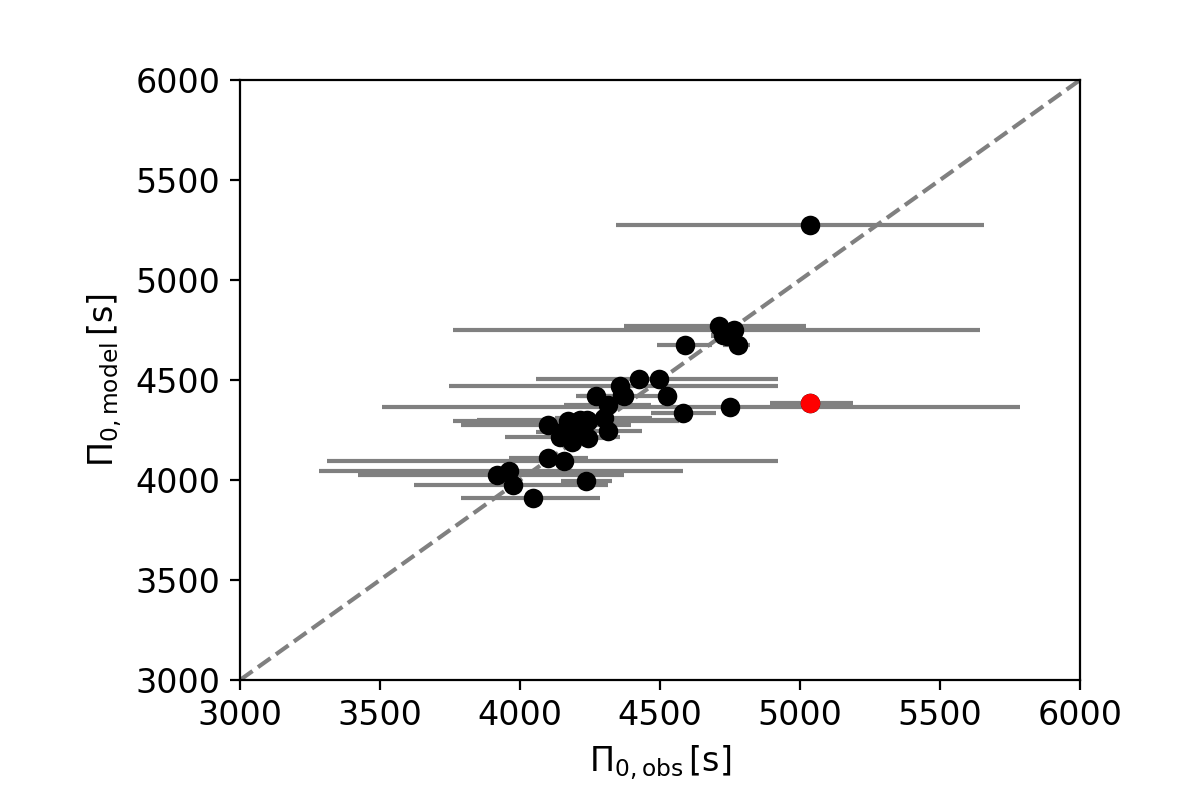}
    \caption{{Comparison between the \pin values derived by \cite{VanReeth2016, VanReeth2018} ($\Pi_{\rm 0, obs}$, simultaneously with \frot) versus the values found from the best models presented in Table\,\ref{tab:theta_sample} of this paper ($\Pi_{\rm 0, model}$). The grey dashed line indicates perfect correspondence. The outlier KIC\,11099031 is indicated in red. }}
    \label{fig:Pi0_comparision}
\end{figure}

Table~\ref{tab:theta_sample} lists the parameters of the best models found by \ctpo and of the best-fitting \gyre model. In Appendix\,\ref{app:psp}, we show the theoretically predicted period spacing patterns for all stars in our sample, as well as the distribution of radial orders, similar to the result from \cite{Li2020}. 
{Moreover, the Brunt-V\"ais\"al\"a frequency profiles of the best-fitting models are shown in Fig.~\ref{fig:BV_profiles}. 
\edit{In many cases, we are not able to reproduce the wiggly characteristics of the observed patterns, indicating that the physics used in this work is still incomplete and requires improvement.
Therefore, these} deviations between the patterns from our best solution and the observed ones offer a fruitful guide to future studies with the aim to explain the morphology of the patterns, by means of \edit{different core-boundary mixing prescriptions and/or} introducing stratified envelope mixing profiles. In order to guide such future studies, we focus on one
of the stars in our sample which shows clear `wiggles' in its observed period spacing patterns, i.e., KIC\,11294808. With the overshoot prescription used in this work (Eq.~\ref{eq:ov}), such wiggles are only observed in our models at a level of $D_0 = 0.05\,{\rm cm^2 s^{-1}}$ (see Fig.~\ref{fig:psp_112_lowD}). The same conclusion is found for this star when we recompute the \mesa and \gyre models, but this time using a step-like {penetrative convection
prescription as convective-boundary mixing prescription,} where we explore $\alpha_{\rm ov} \in [0.05, 0.35; 0.05]$ (other parameter ranges kept the same). We obtain the same best model as for an exponential core overshoot in terms of mass and evolutionary stage. This corresponding best convective penetration model has a value of $\alpha_{\rm ov}$ that is ten times the value of $f_{\rm ov}$,
in agreement with previous studies where such comparisons were made \citep{Moravveji2016}. This solution is also shown in Fig.~\ref{fig:psp_112_step} in the Appendix (grey open circles). 

{The remaining differences between the observed and predicted mode periods via the NN seen in most of the stars is due to the lack of stratified envelope mixing. As shown by \cite{Pedersen2021}, the latter results in structure to occur in the patterns. Such structure can have various physical causes, for example, atomic diffusion with radiative levitation \citep[e.g.][]{Deal2020, mombarg2020}, magnetism \citep[e.g.][which causes saw-tooth like features]{prat2019, vanbeeck2020}, shear instabilities due to rotation or wave mixing \citep{Pedersen2021}, or nonlinear effects (discussed later on). 
Our results open the way to include those types of phenomena by modelling the residuals between the periods predicted by the current initial NN and the observed ones.}

As an extra sanity check 
{to assess the quality of the NN solution}, we show in Figure~\ref{fig:Pi0_comparision} the \pin value derived from our best model and compare it to the observational values from \cite{VanReeth2016, VanReeth2018} as derived from the mode identification along with the near-core rotation frequency. In general, we find adequate correspondence between the two ways of estimating \pin. The most significant discrepancy is found in KIC\,11099031 (data point indicated in red). This is one of stars for which \cite{christophe2018} found a much lower value of \pin, compared to \cite{VanReeth2016} (see Fig.~5 in \cite{ouazzani2018}).}

\longtab{
\begin{landscape}
\begin{table*}[]
    \centering
    \begin{tabular}{lllllllllllllll}
    \hline \hline
    KIC & $M_\star\,[{\rm M}_\odot]$ & $X_{\rm c}$ & $X_{\rm ini}$ & $Z$ & $f_{\rm ov}$ & $\log L/{\rm L_\odot}$ & $\log T_{\rm eff}$ & $\log g$ &  $\tau\,[{\rm Gyr}]$ & $m_{\rm cc}/M_\star$ & $r_{\rm cc}\,[R_\odot]$ & $\Omega/\Omega_{\rm crit, Roche}$ & $r_{\rm ov} - r_{\rm cc}\,[{\rm R_\odot}]$& $R_\star\,[R_\odot]$ \\
    \hline
2710594 &     1.495 &     0.482 &     0.718 &     0.017 & $    0.020_{-    0.020}^{+    0.005}$ &     0.811 &     3.853 &      4.17 &     1.304 &     0.109 &     0.141 &      0.37 &     0.023 &     1.671 \\
3448365 &     1.485 &     0.459 &     0.723 &     0.015 & $    0.005_{-    0.005}^{+    0.005}$ &     0.770 &     3.850 &      4.19 &     1.157 &     0.082 &     0.124 &      0.39 &     0.005 &     1.612 \\
4846809 &     1.536 &     0.409 &     0.711 &     0.020 & $    0.005_{-    0.005}^{+    0.005}$ &     0.863 &     3.858 &      4.15 &     1.148 &     0.099 &     0.137 &      0.50 &     0.006 &     1.732 \\
5114382 &     1.635 &     0.252 &     0.709 &     0.021 & $    0.010_{-    0.005}^{+    0.005}$ &     1.023 &     3.847 &      3.97 &     1.402 &     0.100 &     0.143 &      0.62 &     0.012 &     2.189 \\
5522154 &     1.504 &     0.700 &     0.720 &     0.016 & $    > 0.030$ &     0.716 &     3.865 &      4.31 &     0.166 &     0.112 &     0.150 &      0.63 &     0.044 &     1.420 \\
5708550 &     1.377 &     0.699 &     0.714 &     0.019 & $    > 0.030$ &     0.558 &     3.837 &      4.32 &     0.151 &     0.099 &     0.138 &      0.23 &     0.040 &     1.342 \\
5788623 &     1.474 &     0.505 &     0.720 &     0.016 & $    0.005_{-    0.005}^{+    0.010}$ &     0.747 &     3.852 &      4.22 &     0.981 &     0.080 &     0.124 &      0.14 &     0.005 &     1.557 \\
6468146 &     1.594 &     0.444 &     0.729 &     0.012 & $    0.015_{-    0.005}^{+    0.005}$ &     0.913 &     3.866 &      4.15 &     1.228 &     0.106 &     0.143 &      0.38 &     0.018 &     1.766 \\
6468987 &     1.568 &     0.522 &     0.705 &     0.023 & $    0.025_{-    0.005}^{+    0.005}$ &     0.911 &     3.878 &      4.19 &     0.945 &     0.128 &     0.159 &      0.57 &     0.033 &     1.669 \\
6678174 &     2.000 &     0.144 &     0.725 &     0.014 & $    > 0.030$ &     1.498 &     3.839 &      3.55 &     1.297 &     0.100 &     0.164 &      0.65 &     0.049 &     3.932 \\
6935014 &     1.452 &     0.617 &     0.716 &     0.018 & $    > 0.030$ &     0.711 &     3.854 &      4.26 &     0.783 &     0.113 &     0.145 &      0.25 &     0.043 &     1.484 \\
6953103 &     1.877 &     0.339 &     0.705 &     0.023 & $    > 0.030$ &     1.347 &     3.898 &      3.91 &     1.101 &     0.133 &     0.185 &      0.46 &     0.056 &     2.519 \\
7023122 &     1.519 &     0.697 &     0.723 &     0.015 & $    0.030_{-    0.005}^{+    0.005}$ &     0.734 &     3.868 &      4.31 &     0.191 &     0.108 &     0.148 &      0.29 &     0.039 &     1.429 \\
7365537 &     1.444 &     0.700 &     0.718 &     0.017 & $    > 0.030$ &     0.642 &     3.850 &      4.31 &     0.165 &     0.106 &     0.144 &      0.66 &     0.042 &     1.393 \\
7380501 &     1.718 &     0.138 &     0.716 &     0.018 & $    0.005_{-    0.005}^{+    0.010}$ &     1.097 &     3.840 &      3.89 &     1.352 &     0.084 &     0.132 &      0.40 &     0.006 &     2.460 \\
7434470 &     1.497 &     0.449 &     0.705 &     0.023 & $    0.020_{-    0.005}^{+    0.005}$ &     0.849 &     3.854 &      4.13 &     1.322 &     0.116 &     0.146 &      0.70 &     0.024 &     1.736 \\
7583663 &     1.467 &     0.450 &     0.705 &     0.023 & $    0.015_{-    0.010}^{+    0.005}$ &     0.802 &     3.849 &      4.15 &     1.295 &     0.107 &     0.140 &      0.45 &     0.017 &     1.680 \\
7939065 &     1.479 &     0.465 &     0.705 &     0.023 & $    0.005_{-    0.005}^{+    0.005}$ &     0.793 &     3.856 &      4.19 &     1.025 &     0.096 &     0.134 &      0.40 &     0.005 &     1.616 \\
8364249 &     1.519 &     0.466 &     0.711 &     0.020 & $    0.020_{-    0.005}^{+    0.005}$ &     0.857 &     3.858 &      4.15 &     1.259 &     0.115 &     0.147 &      0.59 &     0.024 &     1.719 \\
8375138 &     1.469 &     0.499 &     0.720 &     0.016 & $    0.020_{-    0.020}^{+    0.005}$ &     0.768 &     3.849 &      4.18 &     1.300 &     0.104 &     0.137 &      0.60 &     0.022 &     1.622 \\
8645874 &     1.540 &     0.639 &     0.729 &     0.012 & $    0.020_{-    0.005}^{+    0.005}$ &     0.777 &     3.869 &      4.28 &     0.517 &     0.095 &     0.139 &      0.12 &     0.024 &     1.489 \\
8836473 &     1.589 &     0.243 &     0.705 &     0.023 & $    0.005_{-    0.005}^{+    0.005}$ &     0.966 &     3.844 &      4.00 &     1.392 &     0.095 &     0.136 &      0.58 &     0.006 &     2.085 \\
9480469 &     1.487 &     0.579 &     0.705 &     0.023 & $    0.005_{-    0.005}^{+    0.015}$ &     0.767 &     3.866 &      4.26 &     0.567 &     0.092 &     0.135 &      0.49 &     0.005 &     1.493 \\
9595743 &     1.456 &     0.672 &     0.716 &     0.018 & $    > 0.000$ &     0.673 &     3.855 &      4.30 &     0.228 &     0.065 &     0.120 &      0.26 &     0.005 &     1.410 \\
9751996 &     1.740 &     0.201 &     0.705 &     0.023 & $    0.015_{-    0.005}^{+    0.005}$ &     1.170 &     3.848 &      3.85 &     1.322 &     0.100 &     0.150 &      0.05 &     0.019 &     2.580 \\
10467146 &     1.768 &     0.161 &     0.720 &     0.016 & $    0.010_{-    0.005}^{+    0.005}$ &     1.157 &     3.846 &      3.86 &     1.338 &     0.089 &     0.140 &      0.41 &     0.012 &     2.574 \\
11080103 &     1.528 &     0.545 &     0.718 &     0.017 & $    0.015_{-    0.005}^{+    0.005}$ &     0.817 &     3.867 &      4.22 &     0.869 &     0.105 &     0.142 &      0.21 &     0.017 &     1.580 \\
11099031 &     1.572 &     0.290 &     0.716 &     0.018 & $    0.030_{-    0.015}^{+    0.005}$ &     1.002 &     3.830 &      3.90 &     1.970 &     0.110 &     0.146 &      0.61 &     0.038 &     2.313 \\
11294808 &     1.655 &     0.190 &     0.711 &     0.020 & $    0.005_{-    0.005}^{+    0.010}$ &     1.033 &     3.842 &      3.95 &     1.384 &     0.090 &     0.135 &      0.43 &     0.006 &     2.265 \\
11456474 &     1.630 &     0.178 &     0.716 &     0.018 & $    0.010_{-    0.005}^{+    0.005}$ &     1.016 &     3.830 &      3.91 &     1.608 &     0.088 &     0.132 &      0.63 &     0.011 &     2.350 \\
11721304 &     1.431 &     0.633 &     0.709 &     0.021 & $    0.025_{-    0.005}^{+    0.005}$ &     0.679 &     3.852 &      4.28 &     0.544 &     0.102 &     0.139 &      0.14 &     0.029 &     1.440 \\
11754232 &     1.554 &     0.641 &     0.716 &     0.018 & $    0.010_{-    0.005}^{+    0.005}$ &     0.809 &     3.880 &      4.29 &     0.353 &     0.097 &     0.142 &      0.05 &     0.012 &     1.472 \\
11826272 &     1.524 &     0.364 &     0.709 &     0.021 & $    0.015_{-    0.005}^{+    0.005}$ &     0.888 &     3.846 &      4.07 &     1.502 &     0.107 &     0.141 &      0.16 &     0.018 &     1.885 \\
11907454 &     1.465 &     0.597 &     0.720 &     0.016 & $    0.015_{-    0.005}^{+    0.005}$ &     0.714 &     3.854 &      4.26 &     0.706 &     0.088 &     0.131 &      0.43 &     0.016 &     1.486 \\
11917550 &     1.512 &     0.486 &     0.729 &     0.012 & $    0.025_{-    0.005}^{+    0.005}$ &     0.821 &     3.853 &      4.16 &     1.434 &     0.109 &     0.141 &      0.34 &     0.029 &     1.692 \\
11920505 &     1.471 &     0.637 &     0.720 &     0.016 & $    0.010_{-    0.005}^{+    0.005}$ &     0.700 &     3.856 &      4.28 &     0.445 &     0.076 &     0.126 &      0.23 &     0.011 &     1.448 \\
12066947 &     1.513 &     0.464 &     0.714 &     0.019 & $    0.005_{-    0.005}^{+    0.005}$ &     0.818 &     3.859 &      4.19 &     1.032 &     0.094 &     0.134 &      0.78 &     0.006 &     1.638 \\
\hline
    \end{tabular}
    \caption{Parameters of the best-fitting \mesa/\gyre models. }
    \label{tab:theta_sample}
\end{table*}
\end{landscape}
}

\subsection{\edit{On the possible origins of CBM}}
\begin{figure}
    \centering
    \includegraphics[width = 0.5\textwidth]{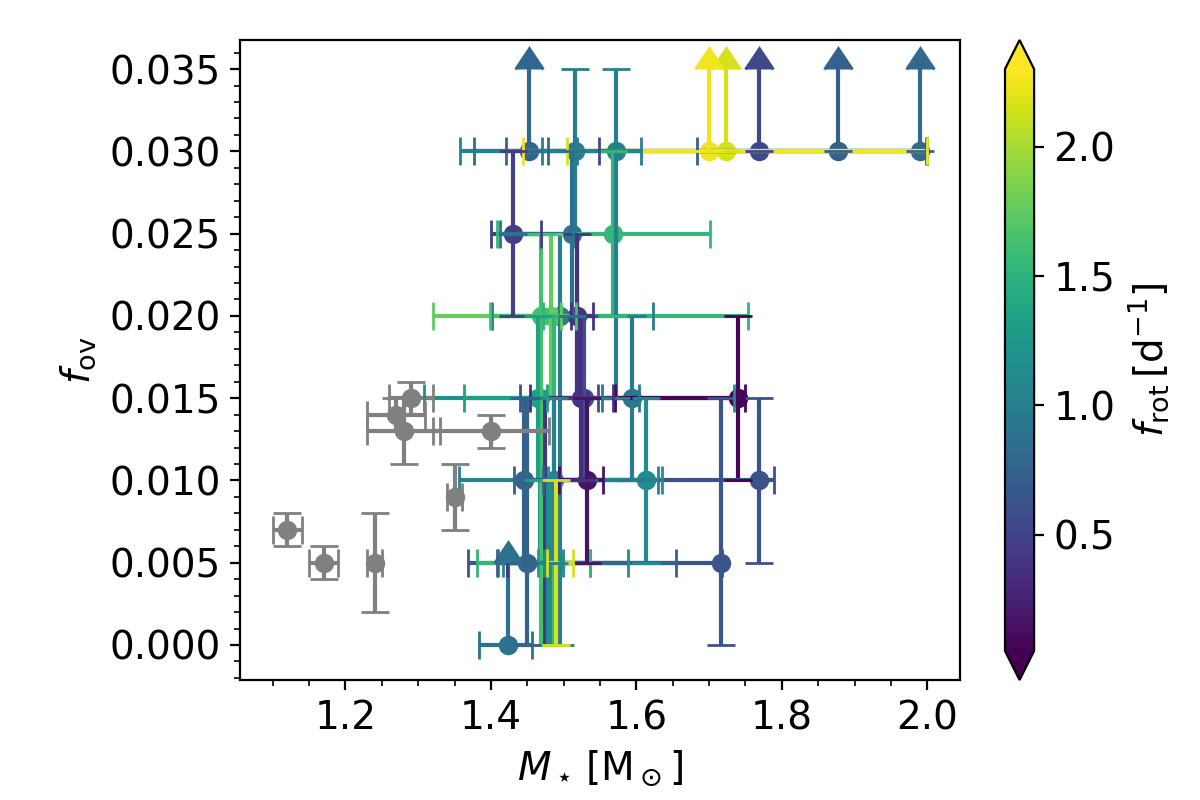}
    \caption{Derived masses and overshoot parameters for all stars in our sample from \gyre models. For comparison, inferred overshoot parameters of low-mass stars by \citet{Deheuvels2016} (solutions with microscopic diffusion, where we have used $\alpha_{\rm ov} \approx 10f_{\rm ov}$) are plotted in grey.  }
    \label{fig:M-fov-frot}
\end{figure}

The parameters of the best-fitting models discussed in this section are those extracted from the \mesa and \gyre models, computed in Step 5 of the modelling  scheme. As illustrated in Figure~\ref{fig:M-fov-frot}, the modelling  of individual pulsations instead of the method employed by \cite{Mombarg2019} allows for a better constraint on \fov, \edit{as the individual mode periods lead to less degeneracy with respect to the stellar mass and age. In addition, the luminosity is in general more precisely determined than \logg for F-type stars, which further reduces the degeneracy between the stellar mass and age. }   
The overshoot parameter has been asteroseismically calibrated for eight solar-like oscillators with a convective core by \cite{Deheuvels2016}, covering a mass range from roughly 1.1 to 1.45\Msun. They observe an increase of the overshoot parameter with increasing mass, although this trend is much less outspoken for $M_\star > 1.25{\rm M_\odot}$. The results from \cite{Deheuvels2016} for models with microscopic diffusion are shown in Figure~\ref{fig:M-fov-frot} in grey. To convert from a step-like overshoot ({penetrative}, $\alpha_{\rm ov}$) to the exponential-like overshoot ({diffusive}) parameterizaton used in this work, we have used the approximation $\alpha_{\rm ov} \approx 10f_{\rm ov}$ { as inferred by \cite{claret-torres2017}. A similar correlation between overshooting and the mass was found by these authors} from isochrone fitting of binary systems, plateauing to a value of $\sim$0.0175 for $M_\star \gtrsim 2.0{\rm M_\odot}$.

\begin{figure}
    \centering
    \includegraphics[width = 0.5\textwidth]{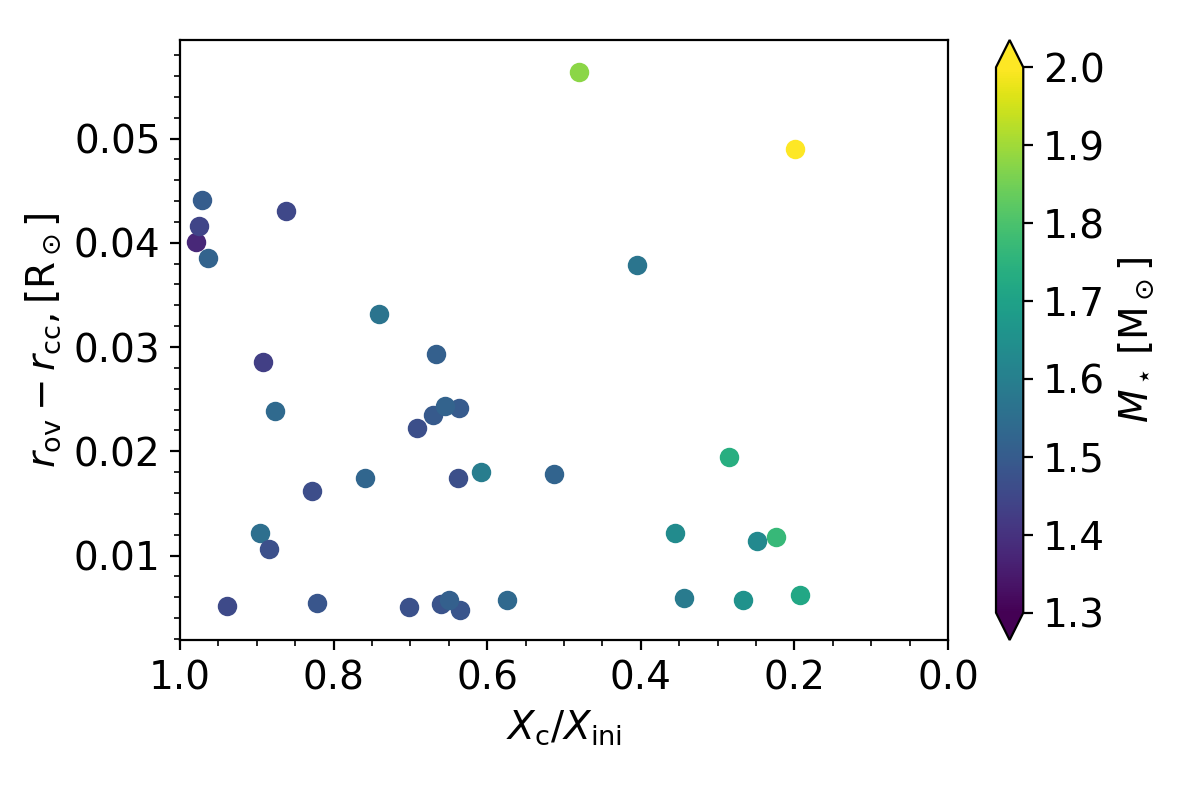}
    \caption{Extent of the overshoot zone plotted against $X_{\rm c}/X_{\rm ini}$. }
    \label{fig:Xc-rov-M}
\end{figure}

\begin{figure}
    \centering
    \includegraphics[width = 0.5\textwidth]{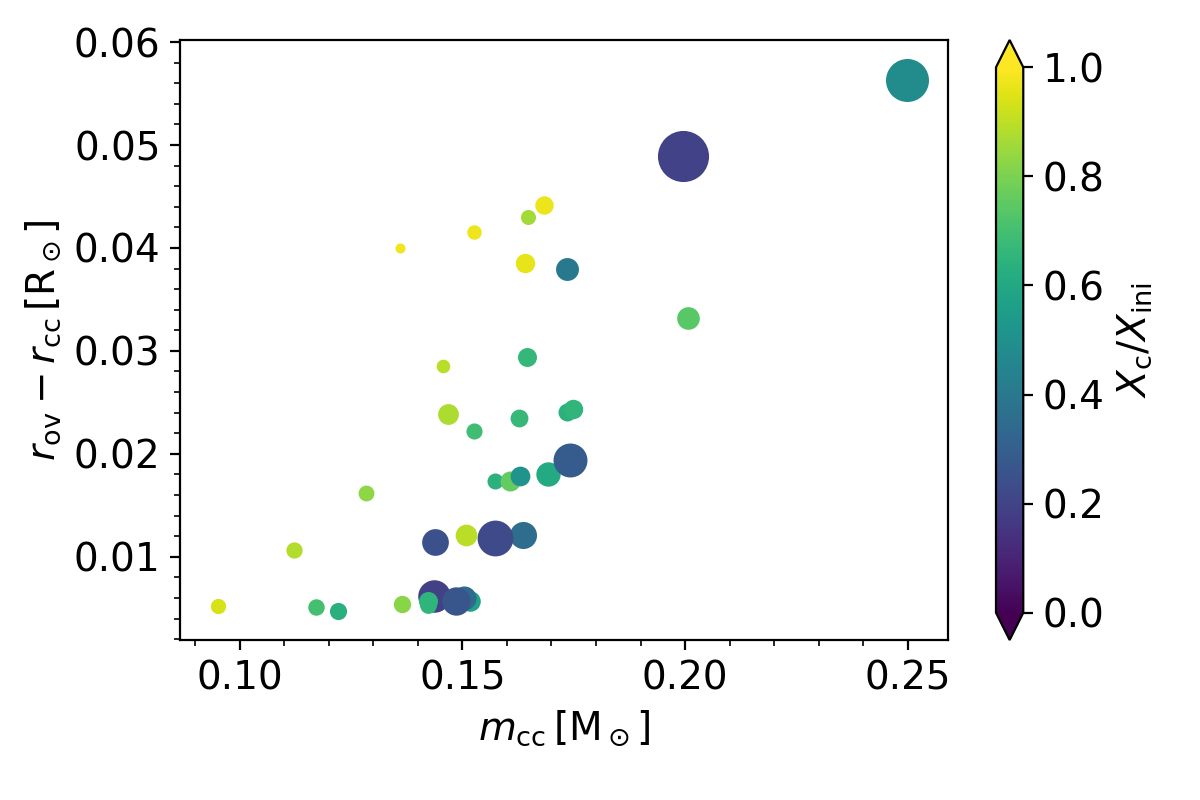}
    \caption{Correlation between the convective core mass and extent of the overshoot zone. The symbol size is indicative of the mass.}
    \label{fig:mcc-rov}
\end{figure}

\begin{figure}
    \centering
    \includegraphics[width = 0.5\textwidth]{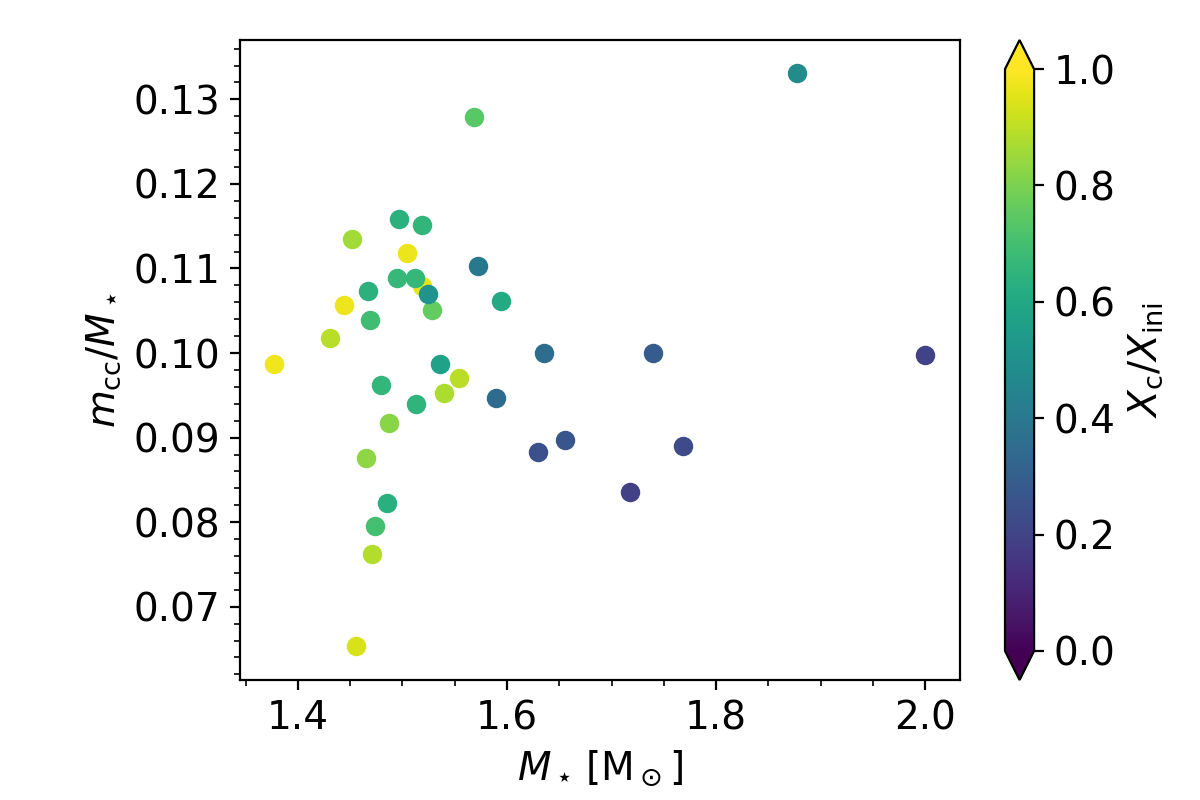}
    \caption{Derived fractional convective core mass versus stellar mass. }
    \label{fig:M-qcc}
\end{figure}

Yet for the single stars modelled in this work, we do not observe such a mass-overshoot relation, albeit a much smaller range in mass is probed. Furthermore, for a significant portion of the sample, we obtain higher values for \fov than those from \cite{claret-torres2017}. Caution is advised when comparing overshoot values between different studies, as the exact location in the star where the overshoot (core-boundary mixing; CBM) profile starts is set by an additional parameter, $f_{\rm 0}$, normalized by the local pressure scale height, for which we use 0.005 in this work. In Figure~\ref{fig:Xc-rov-M}, the extent of the overshoot zone is plotted against $X_{\rm c}/ X_{\rm ini}$. The size of the overshoot region is defined as the difference between $r_{\rm ov}$ and $r_{\rm cc}$, where $r_{\rm cc}$ is defined as the radius where the temperature gradient transitions from adiabatic to radiative, and $r_{\rm ov}$ is the radius where the overshoot zone ends
{(cf.\,Figure\,\ref{fig:Dmix_sketch})}. 
No evident correlation is observed between the extent of the overshoot zone and the evolutionary stage, although the most evolved stars in our sample typically have smaller overshoot zones. It should be noted that if the value of \fov is dependent on the stellar age, this is not taken into account in current SSE models. The modelling  of period spacing patterns does not probe the value of \fov at the current age of the star, but rather the constant value that is needed to have the correct core properties at the age when the observations were taken. As illustrated in Figure~\ref{fig:mcc-rov}, we observe a correlation between the core mass, $m_{\rm cc}$, and the extent of the overshoot region. An increase in $r_{\rm ov} - r_{\rm cc}$ with increasing $m_{\rm cc}$ is observed, where there is an offset between younger and older stars. For the less massive stars, the convective core grows in mass at first, whilst for the more massive stars, the core recedes throughout the MS \citep[cf. Figure 2 of][]{Mombarg2019}. This suggests that the correlation between $r_{\rm ov} - r_{\rm cc}$ and $m_{\rm cc}$ is dependent on the core density. The fractional core mass versus the stellar mass is shown in Figure~\ref{fig:M-qcc}.  

\begin{figure}
    \centering
    \includegraphics[width = 0.5\textwidth]{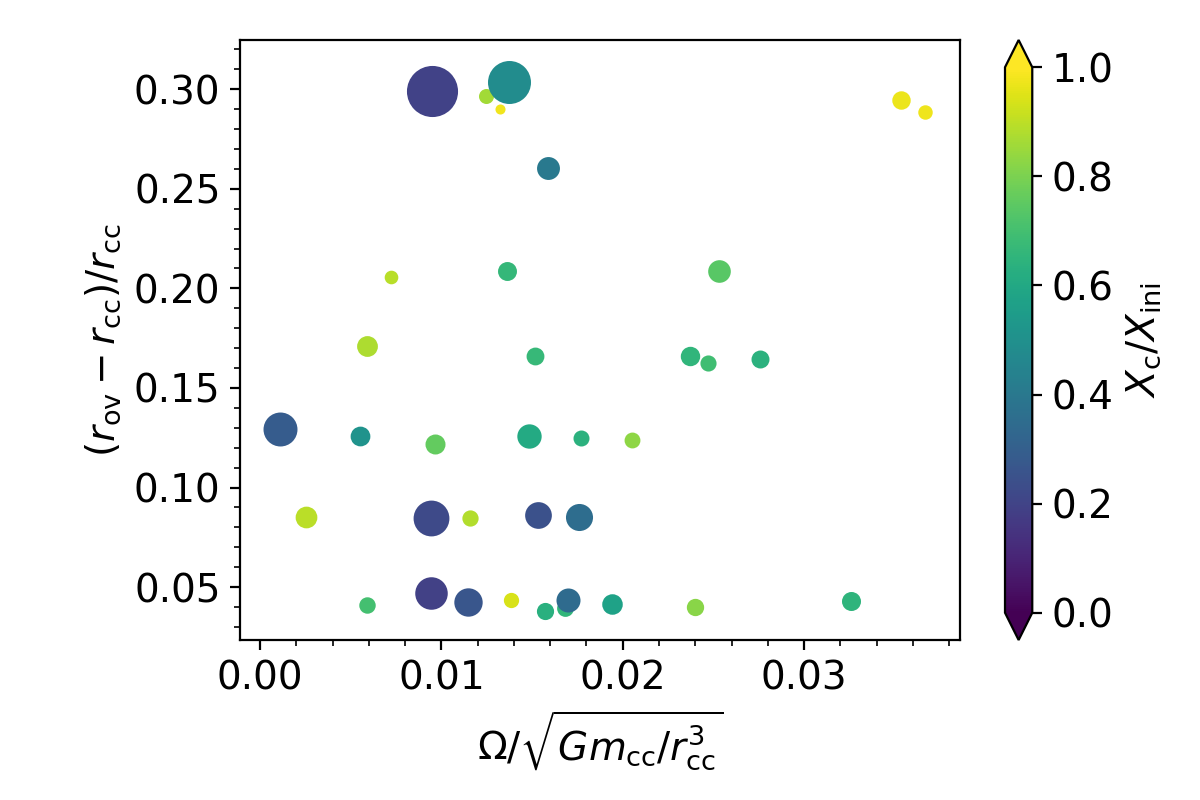}
    \caption{Extent of the overshoot region with respect to the radius of the convective core as a function of the angular rotation rates from \cite{VanReeth2016}, scaled with frequency $\sqrt{Gm_{\rm cc}/r_{\rm cc}^3}$, where $m_{\rm cc}$ and $r_{\rm cc}$ are the core mass and radius, respectively. The symbol size is indicative of the stellar mass.}
    \label{fig:omega-rov}
\end{figure}

Although we refer to $f_{\rm ov}$ as the convective core overshoot parameter, $f_{\rm ov}$ encompasses all forms of mixing occurring close to the core boundary. If CMB is induced via rotational shears between the core and the envelope, a dependence between the amount of CMB and the rotation rate is expected. In Figure~\ref{fig:omega-rov} the inferred extent of the overshoot zone, relative to the core size is plotted as a function of the near-core rotation rate from \cite{VanReeth2016} ($\Omega = 2 \pi f_{\rm rot}$), scaled by $\sqrt{G m_{\rm cc}/ r_{\rm cc}^3}$, where $G$ is the gravitational constant\footnote{No scaling of $\Omega$ does not reveal any correlation either.}. No evidence is found that the CMB increases with a faster rotating near-core region. To fully rule out the connection between rotation and the extent of the overshoot zone, the rotation of the convective core itself needs to be measured. \edit{\cite{Saio2021} have derived the rotation of the convective core in 16 rapidly rotating \gDor stars from the sample of \cite{Li2020} by studying the coupling between inertial modes in the convective core and \gmodes in the radiative envelope. These authors found only small differences with the rotation rates derived from \gmodes in the TAR framework for the majority of stars.}

\subsection{Angular momentum of the sample}
\begin{figure}
    \centering
    \includegraphics[width = 0.5\textwidth]{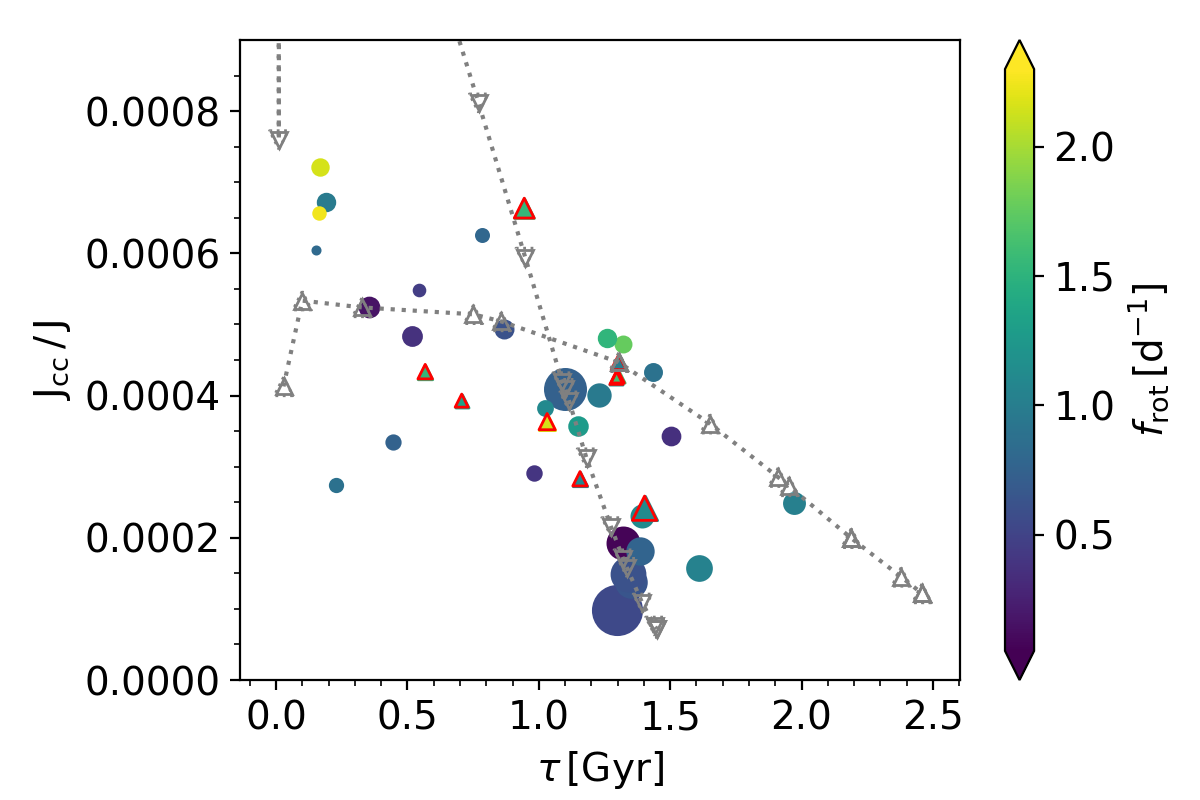}
    \caption{Fractional AM of the convective core to total AM, assuming rigid rotation throughout the MS. Stars with detected Rossby modes are plotted as triangles and highlighted in red. The symbol size is indicative of the mass. The grey symbols indicate the predicted evolution of $J_{\rm cc}/J$ (assuming rigid rotation) for KIC\,2710594 (upward triangles) and KIC\,6953103 (downward triangles). }
    \label{fig:tau-JccJ}
\end{figure}

The AM of a star, $J$, is defined as
\begin{equation} \label{eq:AM}
    J = \int r^2\Omega(r) {\rm d}m.
\end{equation}
For eight stars in our sample, \cite{VanReeth2018} have measured the ratio between the near-core rotation rate and the surface rotation rate, suggesting all eight stars are quasi-rigidly rotating. \edit{In addition, \cite{Saio2021} found the \gDor stars in their sample to be rotating nearly uniformly as well.}
Therefore, assuming $\Omega(r)$ is constant throughout the star and that typical mass loss in F-type stars ($\dot{M} \simeq 10^{-13}\,{\rm M_\odot/yr}$) is too small to carry away significant amounts of AM from the star, we can infer the rotation rate at any point in time, since $J$ is constant.
 Figure~\ref{fig:tau-JccJ} shows the fraction of AM in the convective core, compared to the total AM of the whole star, $J$, as a function of stellar age, $\tau$. Furthermore, two examples of the evolution of the AM of the convective core\footnote{\edit{$J_{\rm cc}$ is computed by discretizing the integral in Eq.~(\ref{eq:AM}), i.e. $J_{\rm cc} = \sum_i r_{i}^2 {\rm d}m_i$, where $i$ runs over the cells in the model below the convective core, and $r$ and ${\rm d}m$ are the central radius and enclosed mass of a cell, respectively.}}, $J_{\rm cc}/J$ for KIC\,2710594 ($\sim$1.5\Msun) and KIC\,7434470 ($\sim$1.9\Msun) when rigid rotation is assumed, are shown.     \newline

\begin{figure}
    \centering
    \includegraphics[width = 0.5\textwidth]{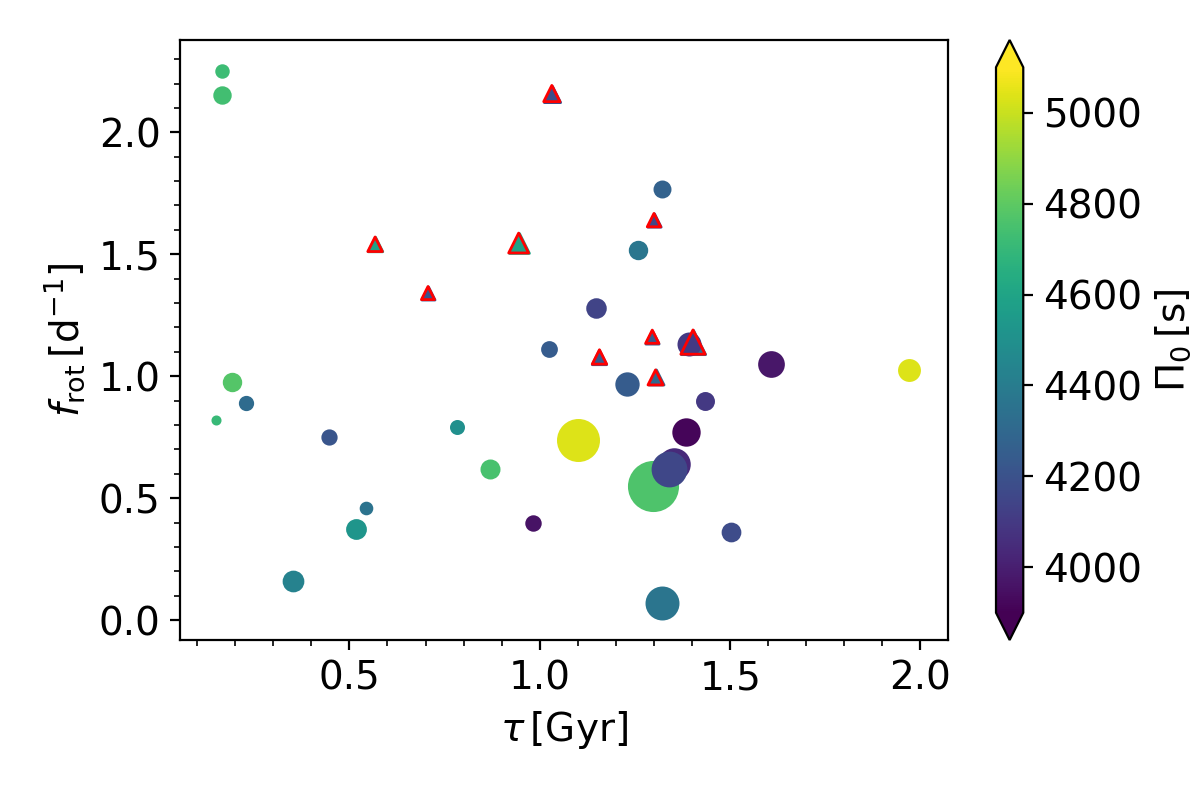}
    \caption{Near-core rotation rates (and $\Pi_0$) from \cite{VanReeth2016, VanReeth2018} as a function of age $\tau$ (this work). Stars with detected Rossby modes are highlighted as red triangles. The symbol size is indicative of the mass.}
    \label{fig:tau-frot}
\end{figure}

\cite{Mombarg2019} observed that stars with detected Rossby modes are situated across the entire MS (based on $X_{\rm c}/X_{\rm ini}$), which is in line with our findings. In Figure~\ref{fig:tau-frot}, we show the near-core rotation as a function of absolute age, $\tau$, and highlighted stars in red for which \cite{VanReeth2016} detected Rossby modes. Again, we do not find any correlation between a star's age and the presence of Rossby modes. 
 
\begin{figure}
    \centering
    \includegraphics[width = 0.5\textwidth]{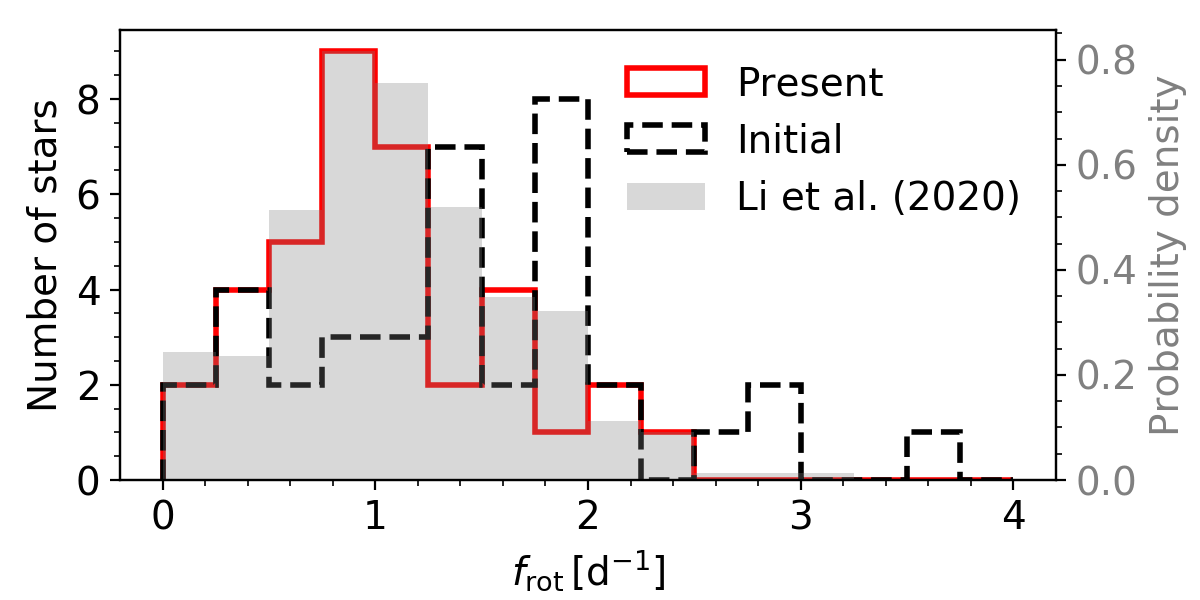}
    \caption{Red: Distribution of the present near-core rotation frequencies from \cite{VanReeth2016, VanReeth2018}. Black: Distribution of the rotation rate near the ZAMS, assuming rigid rotation throughout the MS. Grey: Normalized distribution from \cite{Li2020}.}
    \label{fig:hist_frot}
\end{figure}

When rigid rotation is assumed, the initial rotation frequency at the start of the MS can be estimated. Figure~\ref{fig:hist_frot} shows the distribution of the rotation frequencies near ZAMS (i.e. not more than 3\% of the initial hydrogen in the core burnt) as well as the distribution of the present rotation frequency \citep{VanReeth2016, VanReeth2018}. While the distribution of the present-day near-core rotation frequencies corresponds to the one found by \cite{Li2020}, we find a broader distribution for the near-ZAMS rotation rates, which peaks around 1.8\,${\rm d}^{-1}$. 

\subsection{Mode interaction in KIC\,12066947}
\begin{figure}[htb]
    \centering
    \includegraphics[width = 0.5\textwidth]{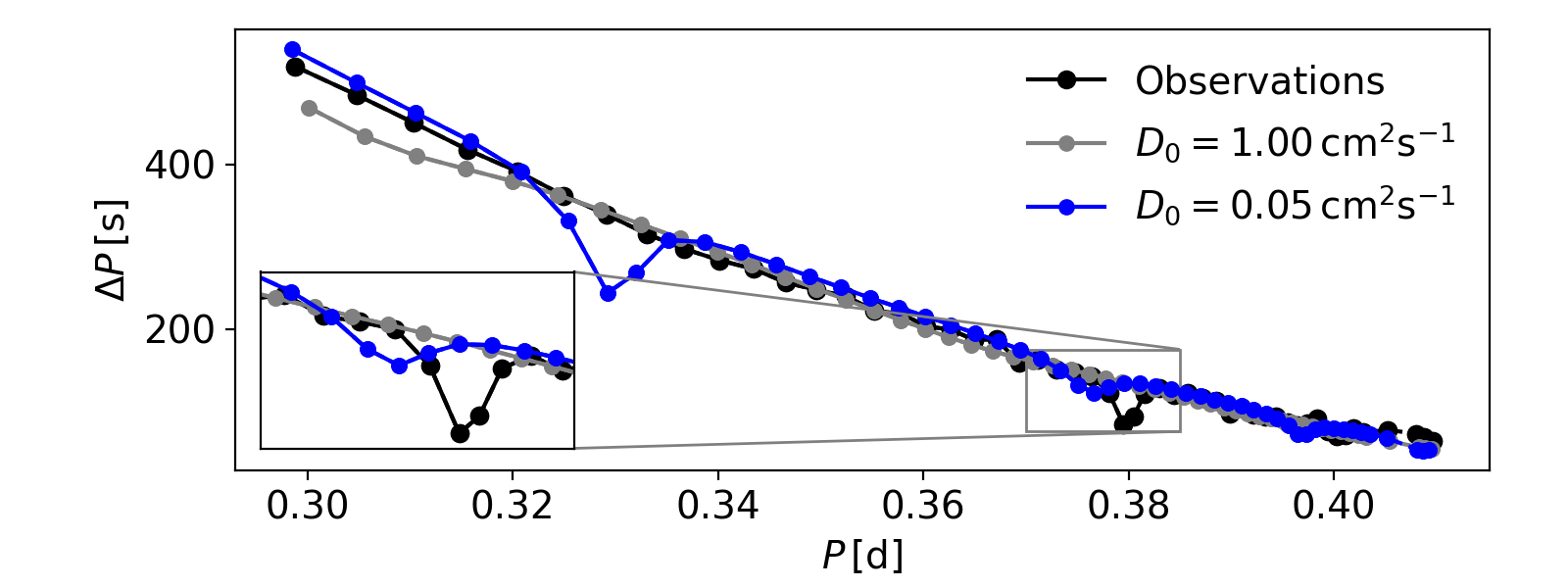}
    \caption{Best-fitting model with $D_0 = 0.05\,{\rm cm^2}{\rm s^{-1}}$ to the observed period spacing pattern of KIC\,12066947 \citep[black dots,][]{VanReeth2016}. For reference the best-fitting with the standard $D_0 = 1.0\,{\rm cm^2}{\rm s^{-1}}$ used in this work is shown in grey.}
    \label{fig:psp_120_lowD}
\end{figure}

One particularly interesting star in our sample is KIC\,12066947. It shows just one characteristic dip in the period spacing pattern at 0.38~days. 
{Such a single dip is not expected to be caused by a chemical gradient because this introduces multiple dips \citep{miglio2008}. \edit{Recently, \cite{Saio2021} were able to reproduce the sharp dip observed in the period spacing pattern of this star, using the formalism by \cite{LeeBaraffe1995} to compute the coupling between \gmodes in the radiative envelope and an inertial mode in the convective core.} To exclude the scenario of constant envelope mixing combined with a chemical gradient in the core boundary layer \edit{as a possible second explanation},} we again repeat Step 5 (Section~\ref{sec:modelling }), but with extremely inefficient envelope mixing, that is, $D_0 = 0.05\,{\rm cm^2\,s^{-1}}$.  As can be seen from the best pulsation model selected among those computed from the best stellar models according to \ctpo, see Figure~\ref{fig:psp_120_lowD}, a low $D_0$ does introduce a dip in the pattern around the observed dip in period spacing close to 0.38~days. However, the additional theoretically predicted dips are not observed. A rotation rate of $25^{+0.09}_{-0.10}\,\mu{\rm Hz}$ for KIC\,12066947 has been measured by \cite{VanReeth2016}. Assuming a uniform rotation profile at $25\,\mu{\rm Hz}$, \cite{ouazzani2020} have computed the radial order at which the interaction 
{between inertial and gravito-inertial modes}
occurs. According to the parameters derived for KIC\,12066947 in this work, the star is most akin to the mid-MS model from \cite{ouazzani2020}, for which the mode interaction is estimated to occur at $n_{\rm pg} = -44$. The dip in the period spacing pattern occurs at $n_{\rm pg} = -44$, therefore, if the dip is caused by interaction between gravito-inertial and pure inertial modes, it should occur close to the dip. Hence, the observed dip in the period spacing pattern of KIC\,12066947 is most likely caused by mode interaction,
confirming the results by \cite{ouazzani2020} \edit{and \cite{Saio2021}.}

\begin{figure*}[th!]
    \centering
    \includegraphics[width = 0.90\textwidth]{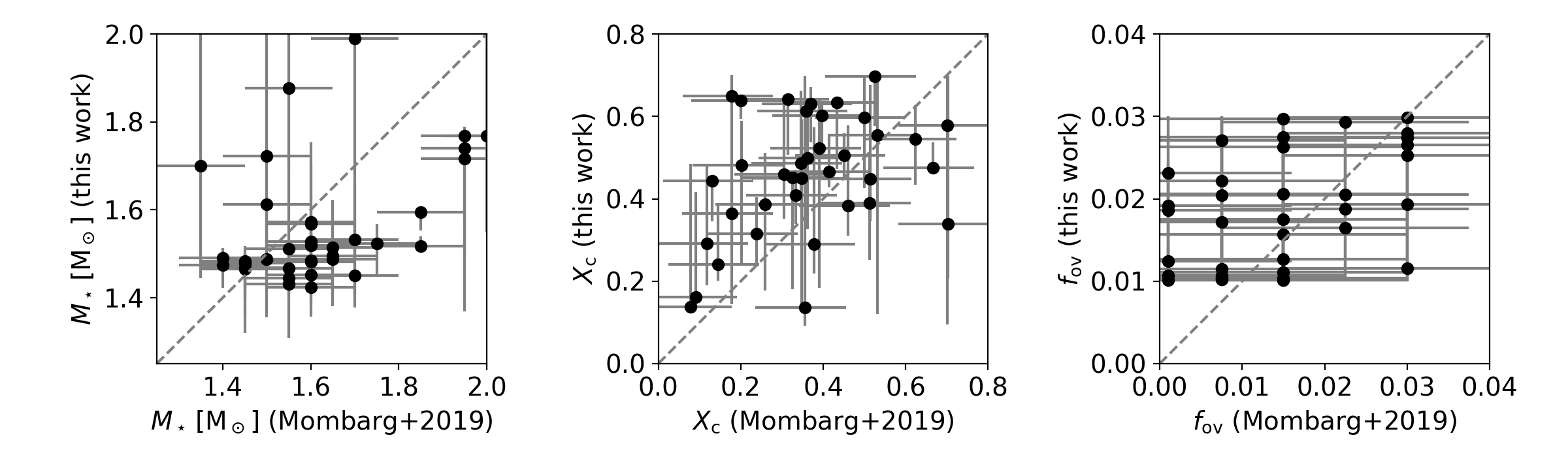}
    \caption{Predictions of the \ctpo NN versus the results from \cite{Mombarg2019}. The grey dashed lines indicate perfect concordance.}
    \label{fig:compare_NN_M2019}
\end{figure*}

\subsection{Comparison with \cite{Mombarg2019}}

In Figure~\ref{fig:compare_NN_M2019}, we compare the MLE for $M_\star, X_{\rm c}$, and $f_{\rm ov}$ from \ctpo (this work) with those from \cite{Mombarg2019}. 
Several key differences between 
{the NN methodology in this work and theirs should be noted while evaluating the comparison, as summarised in Table~\ref{tab:method_M2019}. Keeping these differences in mind, the agreement among the estimated parameters is adequate, given the uncertainties in the parameter estimation. 
This is not the case for all stars in terms of their 
masses, however. The mass estimates from the NN methodology are in general lower compared to those of \cite{Mombarg2019}. This is not a consequence of the more complex NN modelling strategy adopted here, however, but rather due to the inclusion of the stellar luminosities calculated from the Gaia DR2 data as a constraint imposed in the modelling.} 
{Figure\,\ref{fig:M_GaiaL} in the appendix shows the effect  of including the Gaia luminosity as an extra constraint in the modelling on the inferred stellar mass.} 
{A 
systematic underestimation of the stellar masses 
deduced from Gaia DR2 data 
was also found by \cite{Pedersen2021} in their
asteroseismic modelling of B-type stars. 
We conclude that the NN itself performs well as an asteroseismic modelling strategy and opens the way forward for applications to large samples of $\gamma\,$Dor pulsators and for the use of more complex stellar models, including parametrized stratified envelope mixing profiles.}

\begin{table*}[]
    \centering
    \begin{tabular}{c|cccc}
    \hline \hline
         & Asteroseismic & Spectroscopic & $Z$ & Gaia $L$  \\ 
          & input & $T_{\rm eff}$ and $\log g$ &  &   \\ \hline
        This work & $P_{n}$ & A posteriori & Fixed to value  & A posteriori  \\
         &  & cutoff & from spectroscopy &  cutoff \\
         &&&& \\
        \cite{Mombarg2019} & $\Pi_0$ & Fitted & Fixed to 0.014 & Not included \\
        \hline
    \end{tabular}
    \caption{Comparison between the method used in \cite{Mombarg2019} and this work. }
    \label{tab:method_M2019}
\end{table*}

\section{Conclusion \& Discussion}
In this paper, we have constructed a forward modelling  scheme for gravito-inertial modes in \gDor stars, \edit{and used it to estimate the stellar mass, the central hydrogen mass fraction, and the overshooting parameter}. We have eliminated the need of large grids of stellar models by training a dense neural network on a coarse grid of stellar evolution and pulsation models to predict the corresponding oscillation periods ($n \in [15, 91]$) and the luminosity, effective temperature, and surface gravity, given the mass, hydrogen mass fraction in the core, metallicity, core-boundary mixing and radiative envelope mixing efficiency, and rotation rate. All of the input parameters of the network have been varied within appropriate ranges of \gDor stars, making it a versatile tool for estimating stellar parameter ranges with minimal computational effort. The \ctpo neural network, comprising of eight different subnetworks has been applied to a sample of 37 \gDor stars for which period spacing patterns have been detected by \cite{VanReeth2016}. For \teff and \logg, we relied on the measurements from \cite{VanReeth2015b} and have derived luminosities for the sample, using Gaia DR2 distances from \cite{bailer-jones2018}.

Using the mass and $X_{\rm c}$ estimates of \ctpo, we have computed small grids of stellar pulsation models for each star, as to further constrain the overshoot $f_{\rm ov}$ as well. We find no evidence for the core boundary mixing efficiency to correlate with stellar mass, age, or rotation rate. However, we do observe stars with larger core masses tend to have larger overshoot regions. Furthermore, we find that Rossby modes are only detected in the more evolved stars in our sample, which is only seen when the actual stellar age is used instead of $X_{\rm c}/X_{\rm ini}$. Yet, a much larger sample is needed to conclude whether Rossby modes are indeed not observed for stars younger than about 1\,Gyr. \cite{Li2020} compiled a sample of 611 \gDor stars, 83 of which have observed Rossby modes, but many of these stars lack spectroscopic observations therefore the MLEs may in general be less constrained. 

\begin{figure}[th!]
    \centering
    \includegraphics[width = 0.5\textwidth]{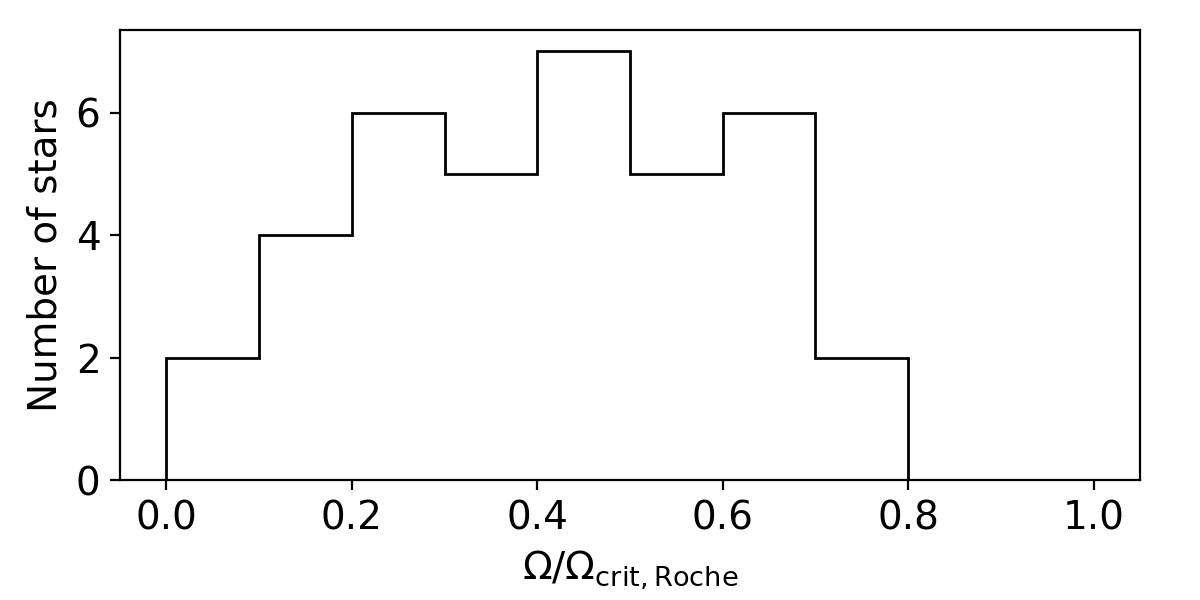}
    \caption{Distribution of the rotation frequency as a fraction of the critical rotation frequency in the Roche framework. Rotation frequencies are taken from \cite{VanReeth2016, VanReeth2018}.  }
    \label{fig:Omega_crit}
\end{figure}

To compute the mode periods, we have relied on the Traditional Approximation of Rotation (TAR), which assumes spherical symmetry and loses its validity for rapidly rotating stars deformed by the centrifugal force \citep[cf.\,][for an improved description of the TAR in that case]{MathisPrat2019}. In Figure~\ref{fig:Omega_crit}, the distribution of the critical Roche frequency for the stars in our sample is shown. Inspecting the quality of fit for the star with the largest fraction of $\Omega / \Omega_{\rm crit, Roche}$, namely KIC\,12066947 ($\sim$80\%), suggests the TAR is still able to reproduce the observed patterns to a satisfactorily level, in line with the findings by \edit{\cite{Henneco2021}}. However, our models are not able the reproduce the sharp dip in the observed pattern, suggesting its caused by resonances between gravito-inertial and pure-inertial modes \citep{ouazzani2020, Saio2021}. An improved formalism of the TAR, taking into account the centrifugal deformation, has been developed by \cite{MathisPrat2019}. 
{
The inclusion of the centrifugal force in the pulsation computations of \gDor stars has a smaller effect on the predicted values for the periods than the inclusion of stratified envelope mixing profiles due to radiative levitation \citep{mombarg2020} or other mixing phenomena, although for lower radial orders the effect of the centrifugal force should be observable in long time-base, space-based photometry \edit{\cite{Henneco2021}}.    } 

In the stellar structure and evolution models used in this work, the effect of atomic diffusion is neglected. \cite{mombarg2020} have demonstrated for two slowly-rotating \gDor stars that this process (including radiative levitations) can lead to significant differences in the derived mass, age, and near-core boundary mixing. The obtained period spacing pattern of the best-fitting model for KIC\,9751996 suggests that we are underestimating $D_0$. The need for more envelope mixing supports the conclusion of \cite{mombarg2020} who found that models without atomic diffusion better reproduce the observed mode periods for this particular star, if other forms of envelope mixing are able to counteract the effects of atomic diffusion. 
Even though the use of a neural network greatly decreases the number of required stellar models in the modelling, doing the full radiative levitation calculations in \mesa still requires a large amount of computation time. The modelling of the sample in this work with atomic diffusion (including radiative levitation), \edit{as well as a non-constant parametrised mixing profile in the radiative envelope}
will be taken up in a future paper.

\begin{acknowledgements}
  The research leading to these results has received funding from the European Research Council (ERC) under the European Unions Horizon 2020 research and innovation programme (grant agreement N$^\circ$670519: MAMSIE) and from the KU\,Leuven Research Council (grant C16/18/005: PARADISE). The computational resources and services used in this work were provided by the VSC (Flemish Supercomputer Center), funded by the Research Foundation - Flanders (FWO) and the Flemish Government department EWI. TVR gratefully acknowledges support from the Research Foundation Flanders (FWO) through grant 12ZB620N. We thank Dr. Dominic M. Bowman for his comments on the manuscript. 
\end{acknowledgements}

%
%
 \bibliographystyle{aa} 
\bibliography{main} 

\appendix
\section{luminosities from Gaia} \label{app:gaia}
\begin{table}[h!] \label{tab:gaia}
    \centering
    \begin{tabular}{lllllll}
    \hline \hline
    KIC & $m_{\rm V}$ & BC$_{\rm V}$ & $D\,$(pc) & $A_{\rm V}$ & Reliable
    $A_{\rm V}$? & $\log L/{\rm L_\odot}$ \\
    \hline

2710594 & 11.79 & 0.031 & $643_{-9}^{+9}$ & 0.21 & Y & $0.864_{-0.075}^{+0.075}$ \\
3448365 & 9.92 & 0.031 & $261_{-2}^{+2}$ & 0.05 & N & $0.766_{-0.035}^{+0.035}$ \\
4846809 & 12.36 & 0.035 & $852_{-22}^{+23}$ & 0.16 & Y & $0.86_{-0.023}^{+0.023}$ \\
5114382 & 11.55 & 0.034 & $618_{-10}^{+11}$ & 0.27 & Y & $0.948_{-0.046}^{+0.046}$ \\
5522154* & 10.44 & 0.028 & $285_{-2}^{+2}$ & 0.0 & N & $0.615_{-0.072}^{+0.072}$ \\
5708550* & 11.93 & 0.03 & $641_{-13}^{+13}$ & 0.15 & Y & $0.782_{-0.071}^{+0.071}$ \\
5788623 & 12.07 & 0.034 & $649_{-12}^{+12}$ & 0.12 & Y & $0.722_{-0.024}^{+0.025}$ \\
6468146 & 9.96 & 0.035 & $326_{-3}^{+3}$ & 0.02 & N & $0.927_{-0.015}^{+0.015}$ \\
6468987 & 12.67 & 0.034 & $910_{-18}^{+19}$ & 0.54 & Y & $0.947_{-0.082}^{+0.083}$ \\
6678174* & 11.74 & 0.033 & $686_{-12}^{+13}$ & 0.21 & N & $0.938_{-0.048}^{+0.048}$ \\
6935014 & 10.91 & 0.033 & $395_{-3}^{+3}$ & 0.15 & Y & $0.767_{-0.03}^{+0.03}$ \\
6953103* & 12.56 & 0.035 & $781_{-15}^{+16}$ & 0.27 & Y & $0.748_{-0.016}^{+0.016}$ \\
7023122 & 10.84 & 0.035 & $389_{-4}^{+4}$ & 0.05 & N & $0.744_{-0.006}^{+0.006}$ \\
7365537* & 9.19 & 0.035 & $178_{-1}^{+1}$ & 0.0 & N & $0.704_{-0.004}^{+0.004}$ \\
7380501 & 11.98 & 0.029 & $663_{-10}^{+11}$ & 0.46 & Y & $0.917_{-0.114}^{+0.114}$ \\
7434470 & 12.02 & 0.031 & $602_{-9}^{+9}$ & 0.25 & Y & $0.73_{-0.065}^{+0.065}$ \\
7583663 & 12.69 & 0.031 & $972_{-25}^{+26}$ & 0.21 & Y & $0.862_{-0.151}^{+0.151}$ \\
7939065 & 12.14 & 0.035 & $696_{-15}^{+15}$ & 0.2 & Y & $0.787_{-0.021}^{+0.021}$ \\
8364249 & 11.95 & 0.032 & $806_{-15}^{+16}$ & 0.25 & Y & $1.011_{-0.085}^{+0.085}$ \\
8375138 & 11.02 & 0.032 & $422_{-4}^{+4}$ & 0.13 & N & $0.775_{-0.04}^{+0.04}$ \\
8645874 & 9.92 & 0.035 & $274_{-2}^{+2}$ & 0.02 & N & $0.795_{-0.015}^{+0.015}$ \\
8836473 & 12.78 & 0.035 & $926_{-21}^{+22}$ & 0.56 & Y & $0.923_{-0.033}^{+0.033}$ \\
9480469 & 12.78 & 0.035 & $908_{-18}^{+18}$ & 0.31 & Y & $0.806_{-0.037}^{+0.037}$ \\
9595743 & 12.09 & 0.035 & $618_{-7}^{+8}$ & 0.06 & Y & $0.649_{-0.019}^{+0.019}$ \\
9751996 & 10.96 & 0.034 & $588_{-9}^{+9}$ & 0.1 & Y & $1.073_{-0.049}^{+0.049}$ \\
10467146 & 12.66 & 0.034 & $1114_{-41}^{+45}$ & 0.29 & Y & $1.027_{-0.082}^{+0.084}$ \\
11080103 & 12.94 & 0.035 & $1017_{-25}^{+27}$ & 0.2 & Y & $0.797_{-0.017}^{+0.018}$ \\
11099031 & 10.02 & 0.025 & $298_{-2}^{+2}$ & 0.02 & N & $0.831_{-0.093}^{+0.093}$ \\
11294808 & 11.73 & 0.03 & $679_{-8}^{+8}$ & 0.21 & Y & $0.939_{-0.085}^{+0.085}$ \\
11456474 & 12.49 & 0.032 & $864_{-17}^{+18}$ & 0.2 & Y & $0.836_{-0.098}^{+0.098}$ \\
11721304 & 11.71 & 0.034 & $550_{-7}^{+7}$ & 0.13 & Y & $0.726_{-0.024}^{+0.024}$ \\
11754232 & 12.24 & 0.035 & $799_{-21}^{+22}$ & 0.08 & Y & $0.821_{-0.019}^{+0.02}$ \\
11826272 & 10.21 & 0.031 & $341_{-5}^{+5}$ & 0.04 & N & $0.877_{-0.048}^{+0.048}$ \\
11907454 & 11.38 & 0.032 & $469_{-5}^{+5}$ & 0.05 & Y & $0.688_{-0.045}^{+0.045}$ \\
11917550 & 11.12 & 0.033 & $448_{-5}^{+5}$ & 0.05 & N & $0.754_{-0.035}^{+0.035}$ \\
11920505 & 9.88 & 0.034 & $243_{-1}^{+2}$ & 0.0 & N & $0.698_{-0.013}^{+0.013}$ \\
12066947 & 10.23 & 0.035 & $325_{-3}^{+3}$ & 0.0 & N & $0.812_{-0.006}^{+0.006}$ \\
\hline
    \end{tabular}
    \caption{Visual apparent magnitudes $m_{\rm V}$, bolometric corrections BC$_{\rm V}$, distances $D$ \citep{bailer-jones2018}, and extinctions $A_{\rm V}$ used to derive the luminosity. The `Reliable $A_{\rm V}$' flag indicates whether a reliable estimation of the extinction could be made with the \texttt{Bayerstar2019} extinction map. Stars for which no reliable estimation could be made according to the \texttt{Bayerstar2019} map are relatively close by and therefore the extinction is expected to be minimal. For the stars marked with an asterisk (*) the luminosity is inconsistent with the pulsations and is therefore not used in the modelling.  }
    \label{tab:gaia_L}
\end{table}
\clearpage

\section{Best-fitting parameters from neural network}
\begin{table}[h!]
    \centering
    \begin{tabular}{llll}
    \hline \hline
    KIC & $M_\star\,[\rm M_\odot]$ & $X_{\rm c}$ & $f_{\rm ov}$ \\
    \hline
2710594 & $1.495_{-0.093}^{+0.128}$ & $0.482_{-0.241}^{+0.108}$ & $0.0265_{-0.0164}^{+0.0035}$ \\
3448365 & $1.485_{-0.075}^{+0.010}$ & $0.459_{-0.107}^{+0.171}$ & $0.0271_{-0.0171}^{+0.0029}$ \\
4846809 & $1.487_{-0.022}^{+0.049}$ & $0.409_{-0.070}^{+0.062}$ & $0.0206_{-0.0106}^{+0.0037}$ \\
5114382 & $1.613_{-0.135}^{+0.022}$ & $0.390_{-0.138}^{+0.121}$ & $0.0222_{-0.0122}^{+0.0078}$ \\
5522154 & $1.723_{-0.219}^{+0.276}$ & $0.554_{-0.433}^{+0.146}$ & $0.0253_{-0.0153}^{+0.0047}$ \\
5708550 & $1.990_{-0.613}^{+0.010}$ & $0.137_{-0.045}^{+0.562}$ & $0.0165_{-0.0065}^{+0.0135}$ \\
5788623 & $1.474_{-0.052}^{+0.001}$ & $0.505_{-0.057}^{+0.054}$ & $0.0102_{-0.0002}^{+0.0056}$ \\
6468146 & $1.594_{-0.041}^{+0.010}$ & $0.444_{-0.099}^{+0.014}$ & $0.0205_{-0.0104}^{+0.0095}$ \\
6468987 & $1.568_{-0.159}^{+0.133}$ & $0.522_{-0.339}^{+0.122}$ & $0.0274_{-0.0174}^{+0.0026}$ \\
6678174 & $1.768_{-0.220}^{+0.232}$ & $0.650_{-0.506}^{+0.050}$ & $0.0157_{-0.0057}^{+0.0143}$ \\
6935014 & $1.452_{-0.031}^{+0.027}$ & $0.613_{-0.095}^{+0.004}$ & $0.0293_{-0.0120}^{+0.0007}$ \\
6953103 & $1.877_{-0.194}^{+0.121}$ & $0.339_{-0.133}^{+0.360}$ & $0.0280_{-0.0180}^{+0.0020}$ \\
7023122 & $1.515_{-0.045}^{+0.004}$ & $0.697_{-0.121}^{+0.002}$ & $0.0101_{-0.0001}^{+0.0199}$ \\
7365537 & $1.700_{-0.256}^{+0.300}$ & $0.578_{-0.483}^{+0.122}$ & $0.0188_{-0.0088}^{+0.0112}$ \\
7380501 & $1.717_{-0.348}^{+0.001}$ & $0.138_{-0.005}^{+0.346}$ & $0.0107_{-0.0007}^{+0.0135}$ \\
7434470 & $1.483_{-0.163}^{+0.014}$ & $0.449_{-0.103}^{+0.228}$ & $0.0263_{-0.0163}^{+0.0037}$ \\
7583663 & $1.467_{-0.159}^{+0.267}$ & $0.450_{-0.320}^{+0.143}$ & $0.0204_{-0.0104}^{+0.0096}$ \\
7939065 & $1.479_{-0.062}^{+0.020}$ & $0.465_{-0.037}^{+0.090}$ & $0.0193_{-0.0069}^{+0.0093}$ \\
8364249 & $1.519_{-0.048}^{+0.235}$ & $0.452_{-0.271}^{+0.014}$ & $0.0275_{-0.0174}^{+0.0025}$ \\
8375138 & $1.469_{-0.070}^{+0.048}$ & $0.499_{-0.173}^{+0.124}$ & $0.0116_{-0.0016}^{+0.0184}$ \\
8645874 & $1.518_{-0.008}^{+0.022}$ & $0.639_{-0.045}^{+0.012}$ & $0.0297_{-0.0057}^{+0.0002}$ \\
8836473 & $1.482_{-0.009}^{+0.107}$ & $0.315_{-0.072}^{+0.089}$ & $0.0111_{-0.0011}^{+0.0142}$ \\
9480469 & $1.487_{-0.106}^{+0.049}$ & $0.383_{-0.072}^{+0.196}$ & $0.0231_{-0.0131}^{+0.0069}$ \\
9595743 & $1.424_{-0.040}^{+0.032}$ & $0.630_{-0.093}^{+0.042}$ & $0.0107_{-0.0007}^{+0.0193}$ \\
9751996 & $1.740_{-0.169}^{+0.009}$ & $0.241_{-0.040}^{+0.147}$ & $0.0124_{-0.0024}^{+0.0121}$ \\
10467146 & $1.768_{-0.274}^{+0.021}$ & $0.161_{-0.023}^{+0.256}$ & $0.0115_{-0.0015}^{+0.0157}$ \\
11080103 & $1.528_{-0.088}^{+0.019}$ & $0.545_{-0.111}^{+0.082}$ & $0.0175_{-0.0075}^{+0.0125}$ \\
11099031 & $1.572_{-0.215}^{+0.034}$ & $0.290_{-0.071}^{+0.284}$ & $0.0299_{-0.0140}^{+0.0001}$ \\
11294808 & $1.450_{-0.041}^{+0.205}$ & $0.291_{-0.101}^{+0.194}$ & $0.0101_{-0.0001}^{+0.0198}$ \\
11456474 & $1.487_{-0.131}^{+0.143}$ & $0.387_{-0.209}^{+0.124}$ & $0.0127_{-0.0027}^{+0.0134}$ \\
11721304 & $1.431_{-0.031}^{+0.038}$ & $0.633_{-0.161}^{+0.015}$ & $0.0104_{-0.0004}^{+0.0196}$ \\
11754232 & $1.532_{-0.038}^{+0.022}$ & $0.641_{-0.134}^{+0.012}$ & $0.0102_{-0.0001}^{+0.0196}$ \\
11826272 & $1.524_{-0.070}^{+0.044}$ & $0.364_{-0.075}^{+0.066}$ & $0.0172_{-0.0060}^{+0.0119}$ \\
11907454 & $1.465_{-0.102}^{+0.012}$ & $0.597_{-0.171}^{+0.102}$ & $0.0104_{-0.0004}^{+0.0196}$ \\
11917550 & $1.512_{-0.099}^{+0.001}$ & $0.486_{-0.076}^{+0.176}$ & $0.0192_{-0.0092}^{+0.0108}$ \\
11920505 & $1.445_{-0.013}^{+0.026}$ & $0.602_{-0.082}^{+0.035}$ & $0.0101_{-0.0001}^{+0.0126}$ \\
12066947 & $1.490_{-0.013}^{+0.023}$ & $0.475_{-0.011}^{+0.062}$ & $0.0186_{-0.0079}^{+0.0044}$ \\
\hline
    \end{tabular}
    \caption{Best-fitting parameters of the model predicted by the \ctpo neural network. }
    \label{tab:theta_NN}
\end{table}

\section{Distribution of the residuals} \label{app:res}
\begin{figure}[h!]
    \centering
    \includegraphics[width = 0.55\textwidth]{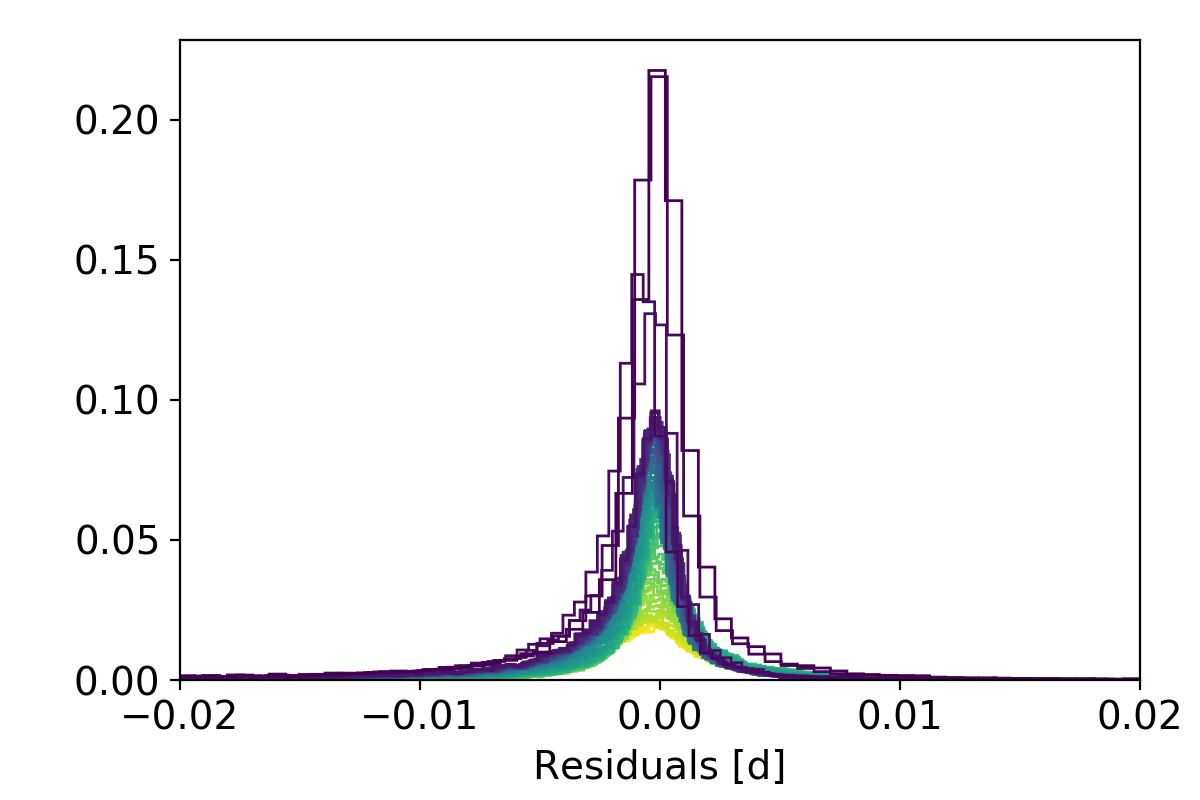}
    \caption{Normalized distributions of the residuals of the NN on the grid of pulsation models used for training and validation. Darker colors indicate higher radial orders. }
    \label{fig:hist_res}
\end{figure}

In Figure~\ref{fig:hist_res} we show the residuals ($P_{\rm true} - P_{\rm pred}$) -- per radial order -- of the NN on the grid of \SI{38915} stellar pulsation models used for training and validation.

\section{Best-matching period spacing patterns} \label{app:psp}
\begin{figure}
    \centering
    \includegraphics[width = 0.55\textwidth]{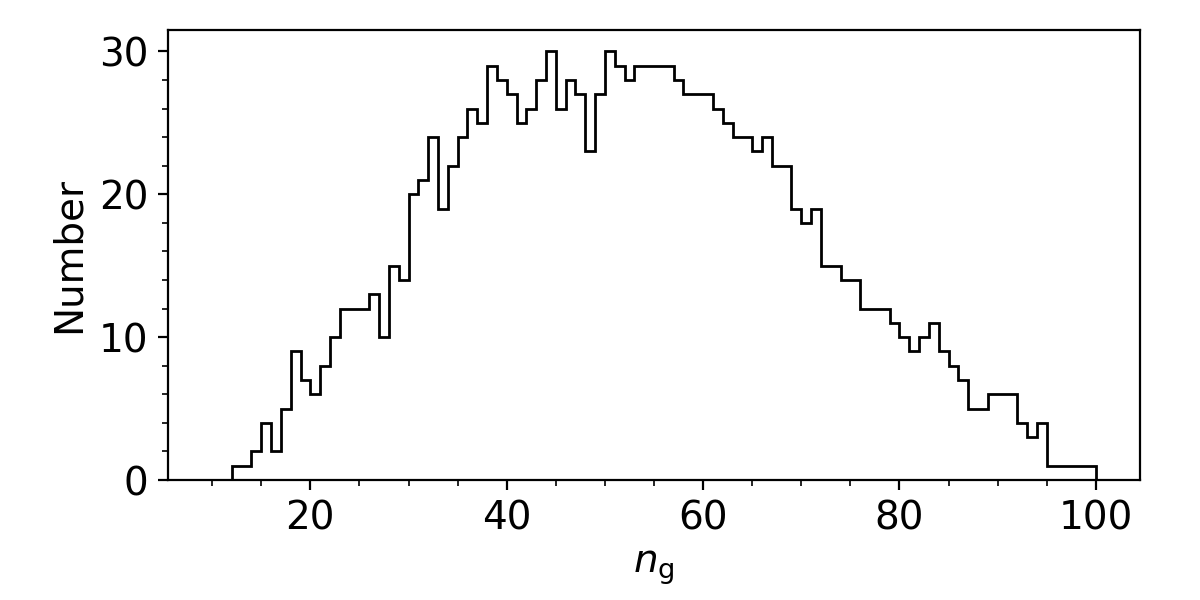}
    \caption{The distribution of the identified radial orders for all stars in our sample. }
    \label{fig:hist_n}
\end{figure}

In this appendix we present the best-fitting period spacing patterns found with \ctpo (in red) and \gyre (in blue) for all stars in our sample. The observed period spacing patterns are taken from \cite{VanReeth2015b}, where the uncertainties are typically smaller than the symbol size. Missing radial orders are indicated by dashed lines. \edit{As only the mode periods are fitted, these values are also indicated by dashes at the bottom/top of each panel.} Furthermore, in Fig~\ref{fig:hist_n} we show the distribution of radial orders. A similar distribution is found compared to the one from \cite{Li2020} (their Figure 17).

\clearpage
\begin{figure*}
    \centering
    \includegraphics[width = 0.9\textwidth]{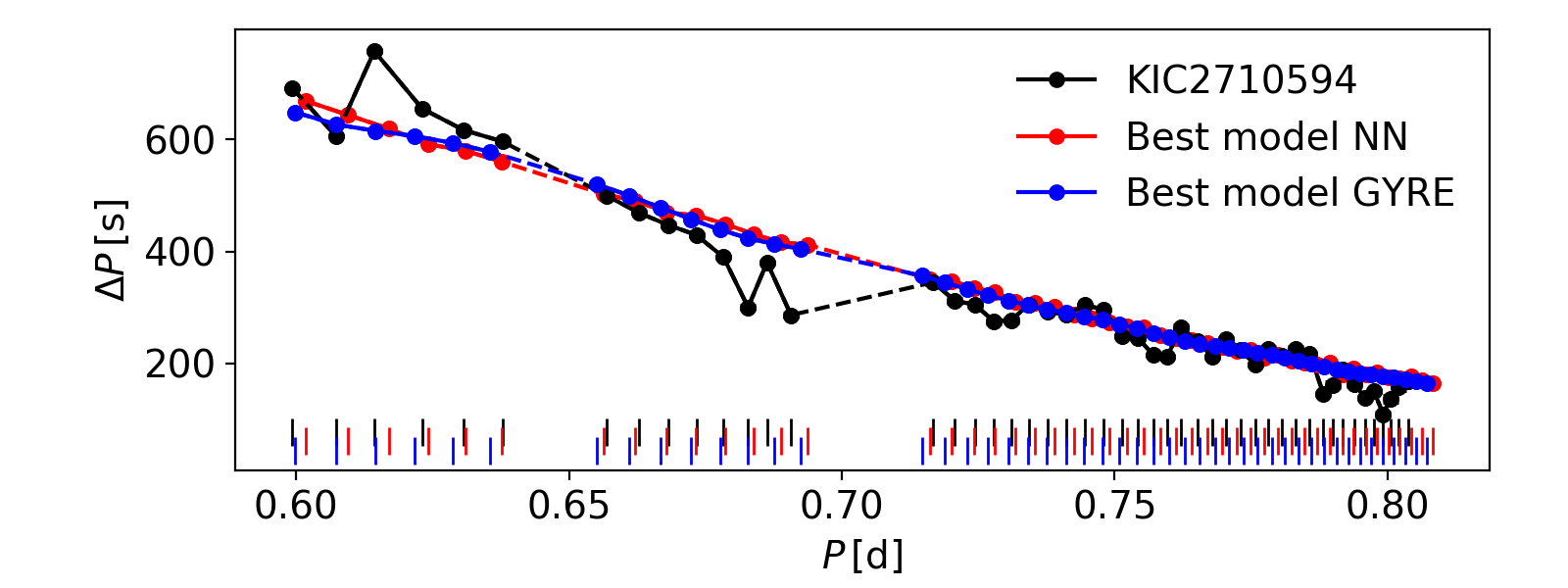}
    \caption{Best-fitting models from the neural network and \gyre for KIC\,2710594.}
    \label{fig:my_label}
\end{figure*}

\begin{figure*}
    \centering
    \includegraphics[width = 0.9\textwidth]{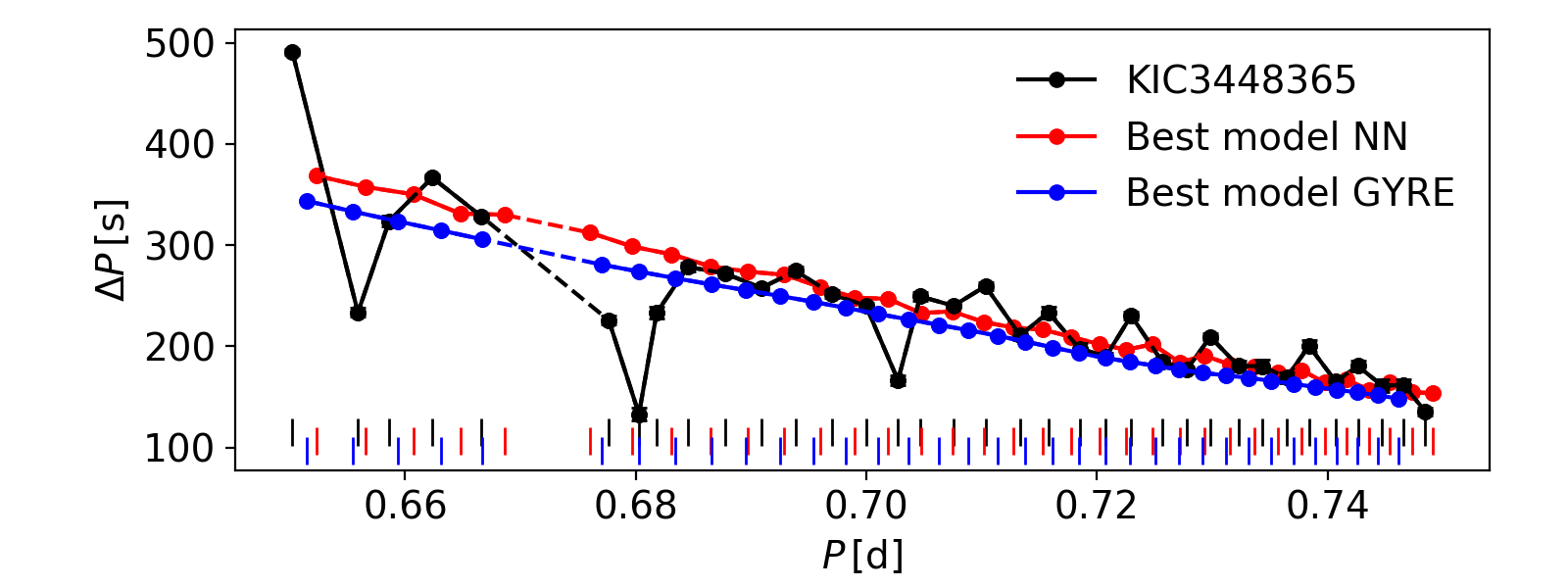}
    \caption{Best-fitting models from the   neural network and \gyre for KIC\,3448365.}
\end{figure*}

\begin{figure*}
    \centering
    \includegraphics[width = 0.9\textwidth]{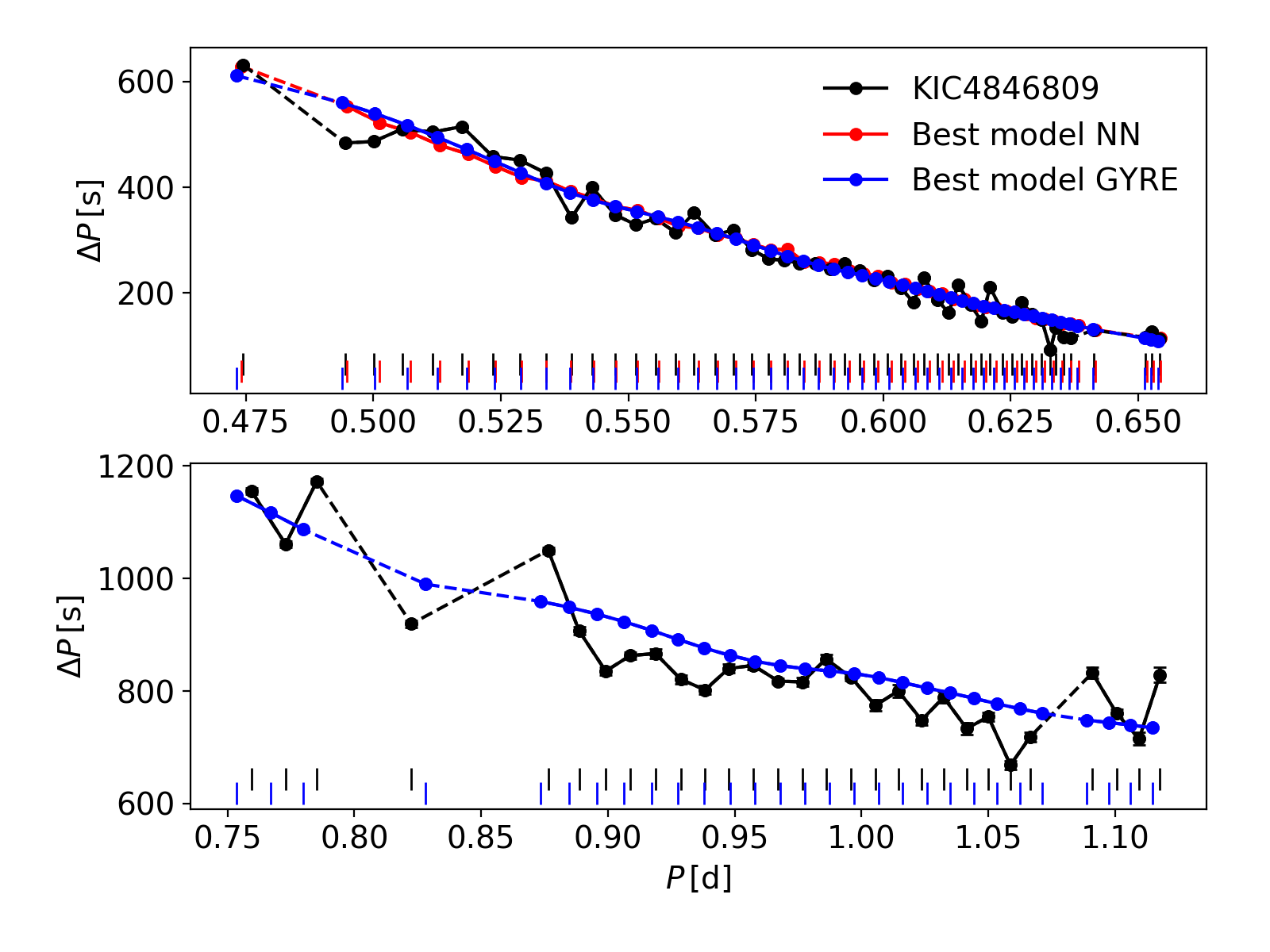}
    \caption{Best-fitting models from the   neural network and \gyre for KIC\,4846809. {Top: $(\ell, m) = (1,1)$, bottom $(\ell, m) = (1,0)$.}}
\end{figure*}

\begin{figure*}
    \centering
    \includegraphics[width = 0.9\textwidth]{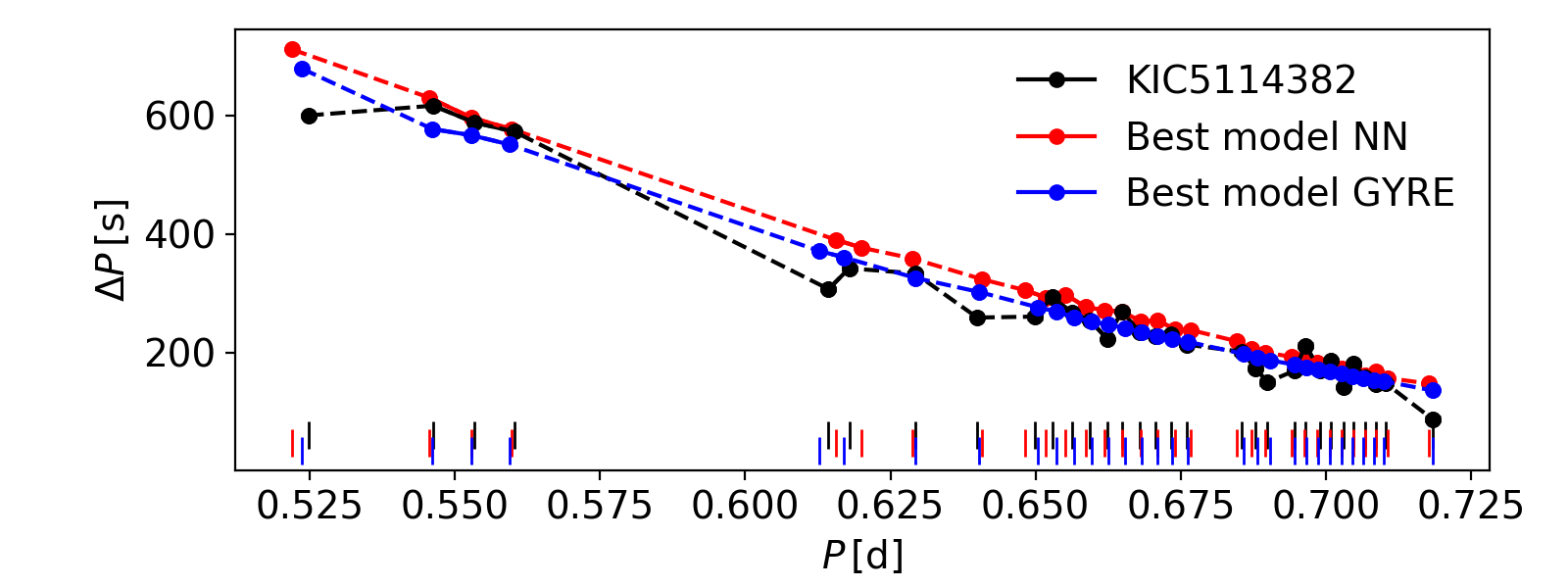}
    \caption{Best-fitting models from the   neural network and \gyre for KIC\,5114382.}
\end{figure*}

\begin{figure*}
    \centering
    \includegraphics[width = 0.9\textwidth]{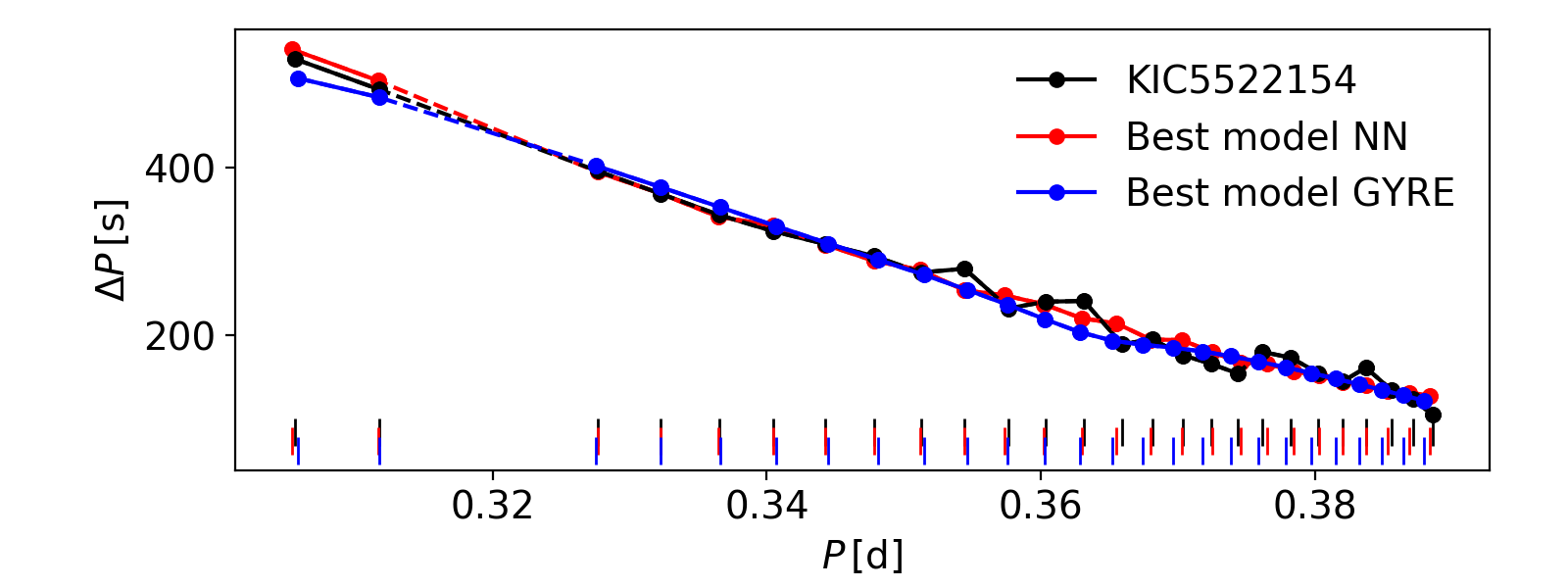}
    \caption{Best-fitting models from the   neural network and \gyre for KIC\,5522154.}
\end{figure*}

\begin{figure*}
    \centering
    \includegraphics[width = 0.9\textwidth]{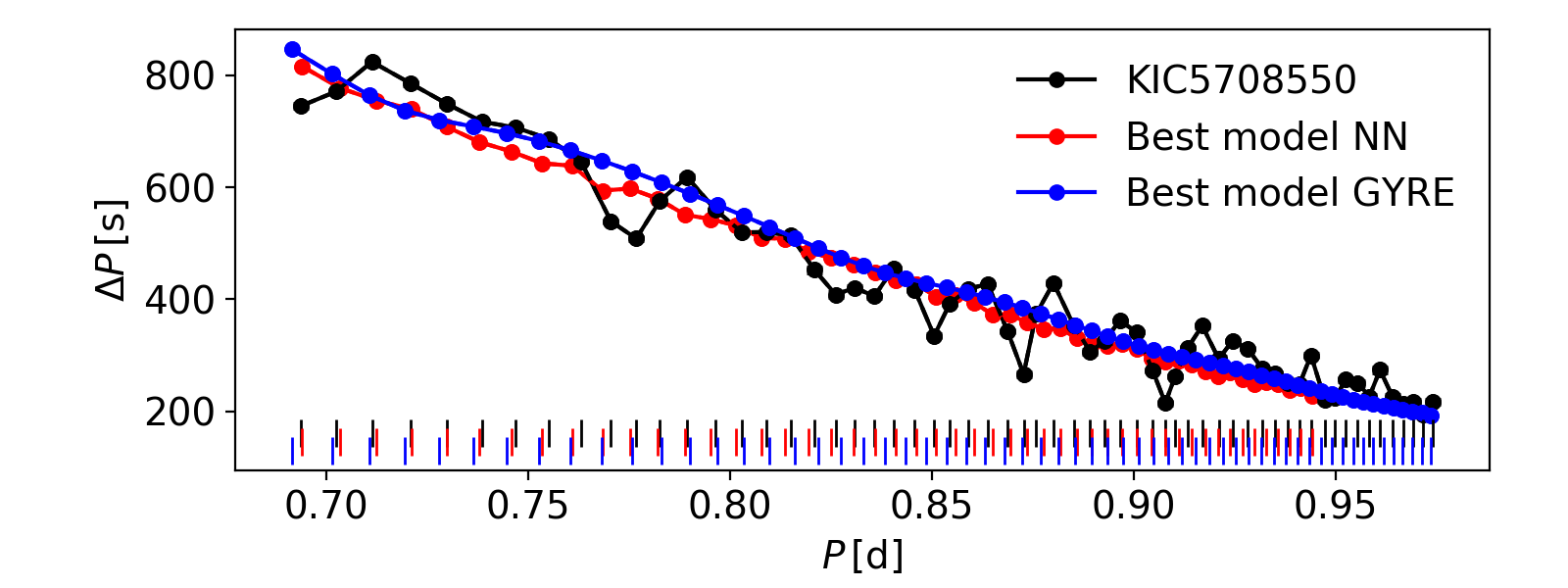}
    \caption{Best-fitting models from the   neural network and \gyre for KIC\,5708550.}
\end{figure*}

\begin{figure*}
    \centering
    \includegraphics[width = 0.9\textwidth]{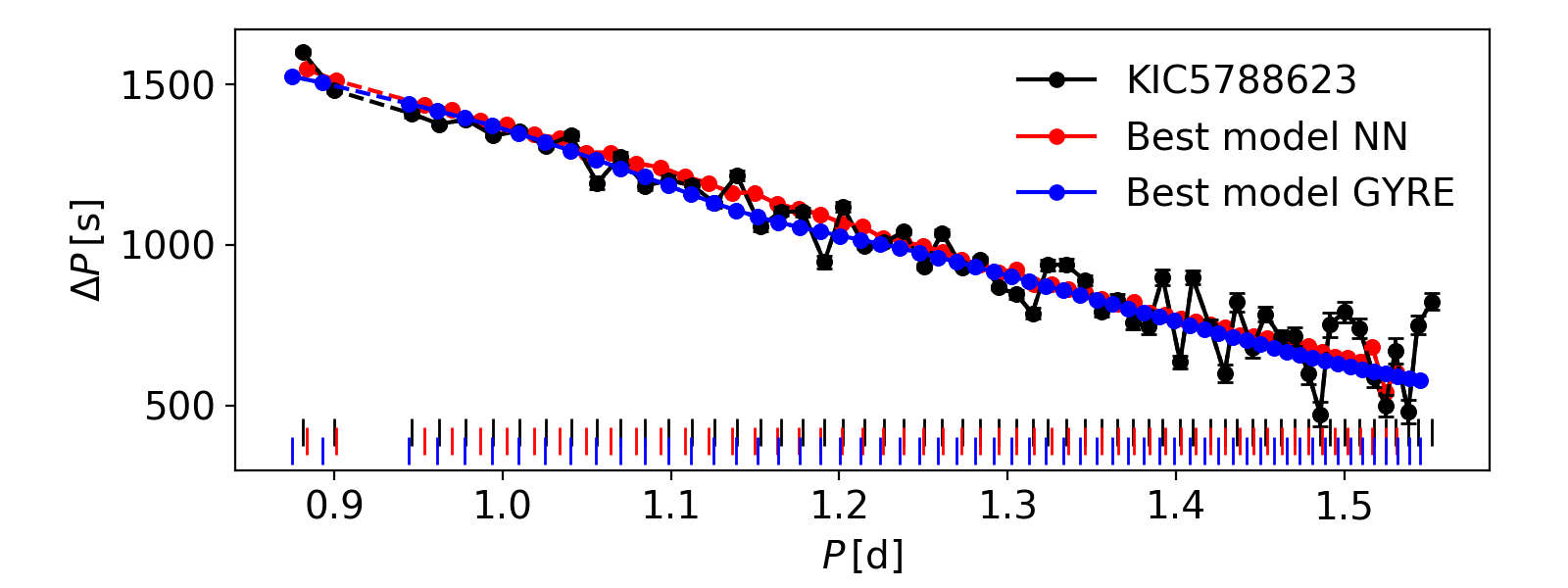}
    \caption{Best-fitting models from the   neural network and \gyre for KIC\,5788623.}
\end{figure*}

\begin{figure*}
    \centering
    \includegraphics[width = 0.9\textwidth]{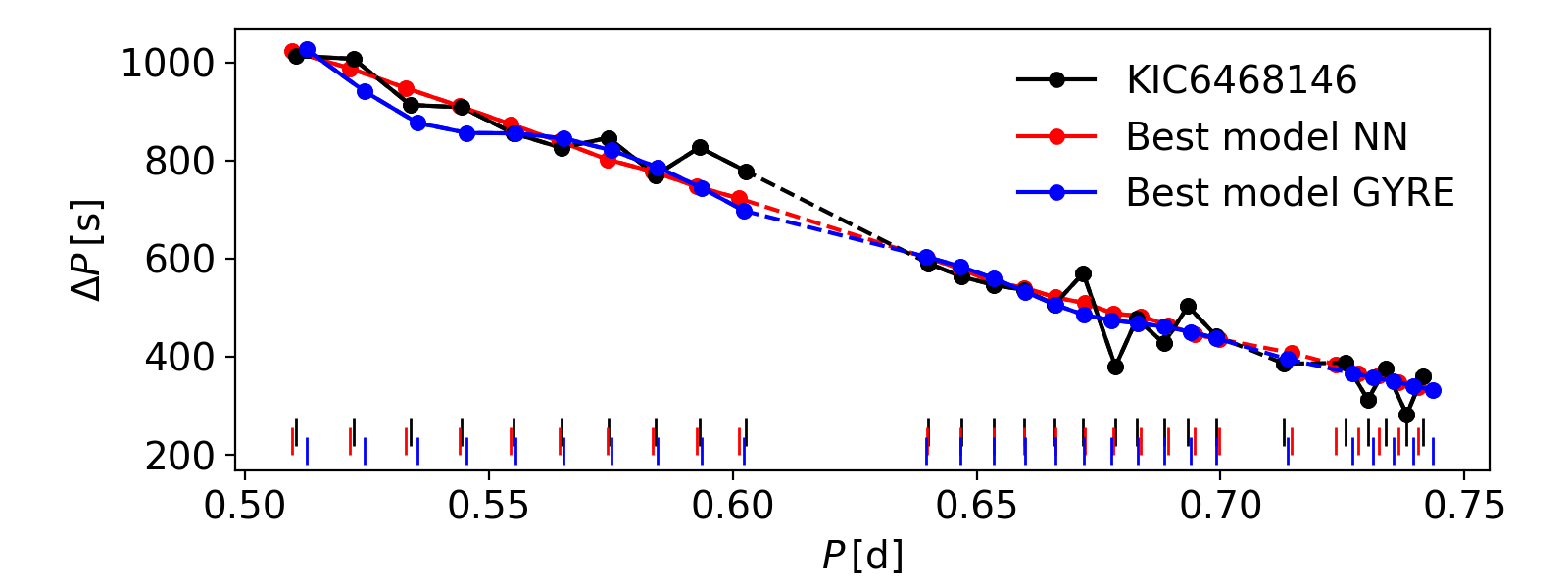}
    \caption{Best-fitting models from the   neural network and \gyre for KIC\,6468146.}
\end{figure*}

\begin{figure*}
    \centering
    \includegraphics[width = 0.9\textwidth]{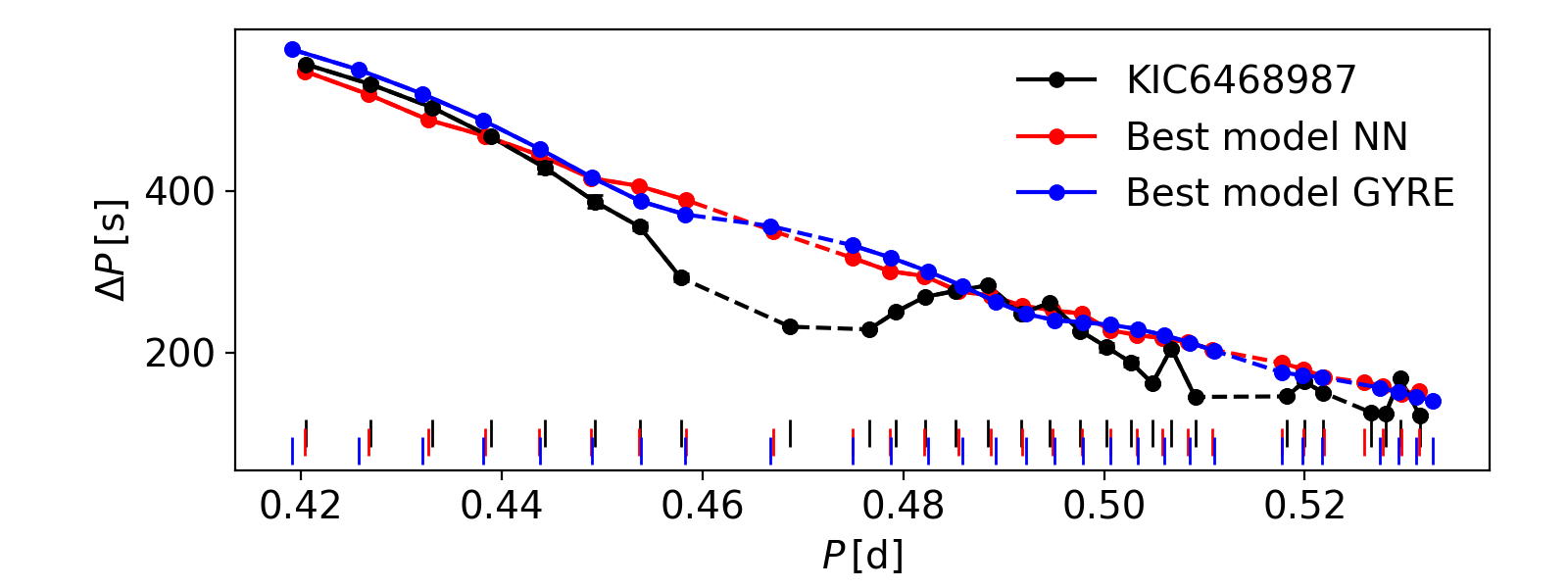}
    \caption{Best-fitting models from the   neural network and \gyre for KIC\,6468987.}
\end{figure*}

\begin{figure*}
    \centering
    \includegraphics[width = 0.9\textwidth]{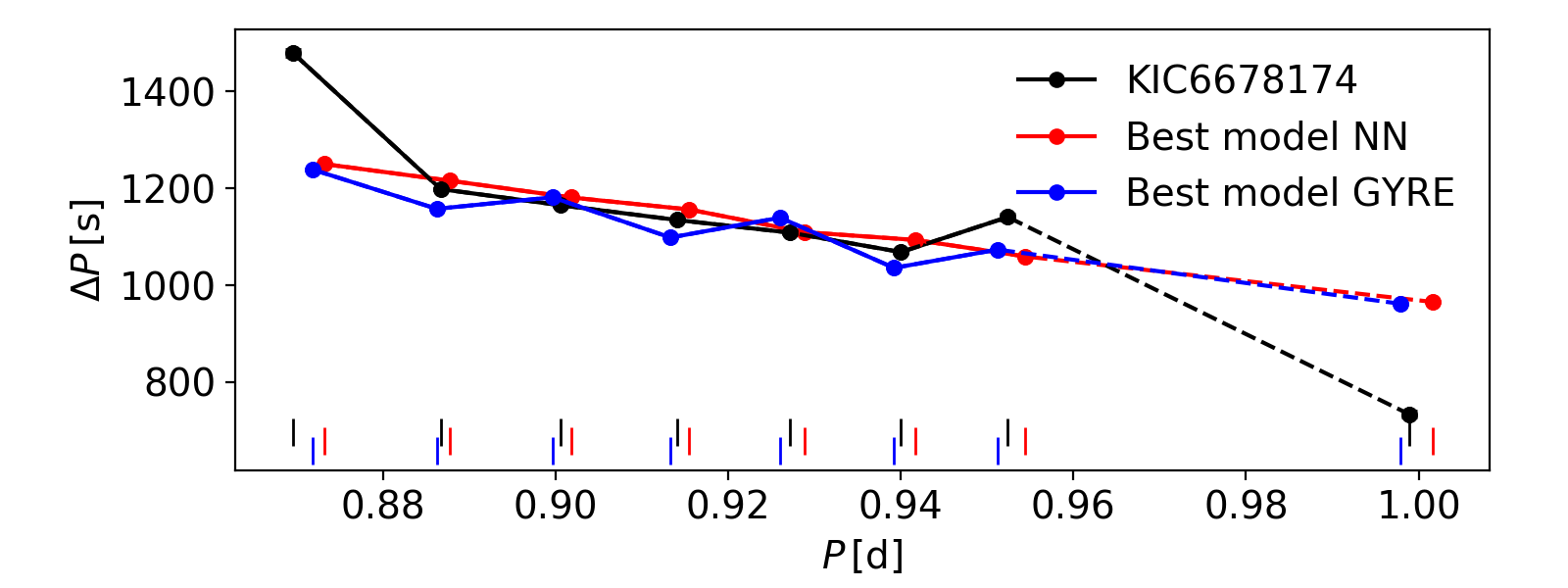}
    \caption{Best-fitting models from the   neural network and \gyre for KIC\,6678174.}
\end{figure*}

\begin{figure*}
    \centering
    \includegraphics[width = 0.9\textwidth]{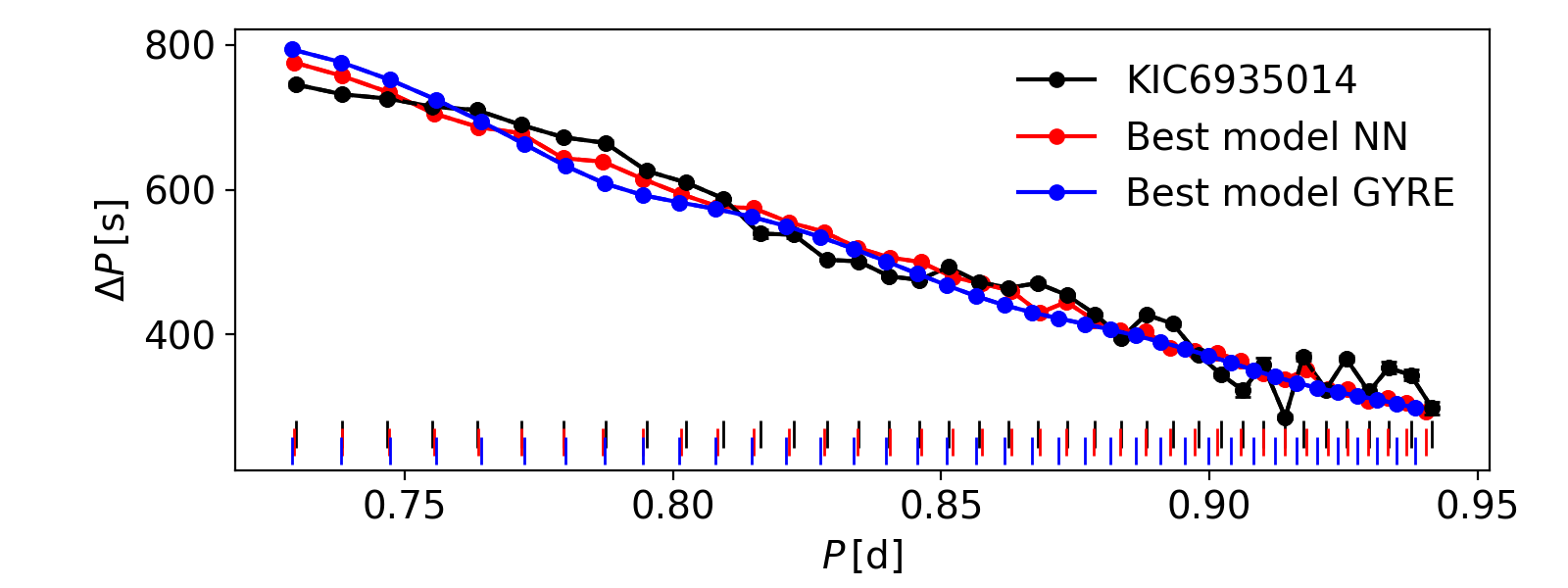}
    \caption{Best-fitting models from the   neural network and \gyre for KIC\,6935014.}
\end{figure*}

\begin{figure*}
    \centering
    \includegraphics[width = 0.9\textwidth]{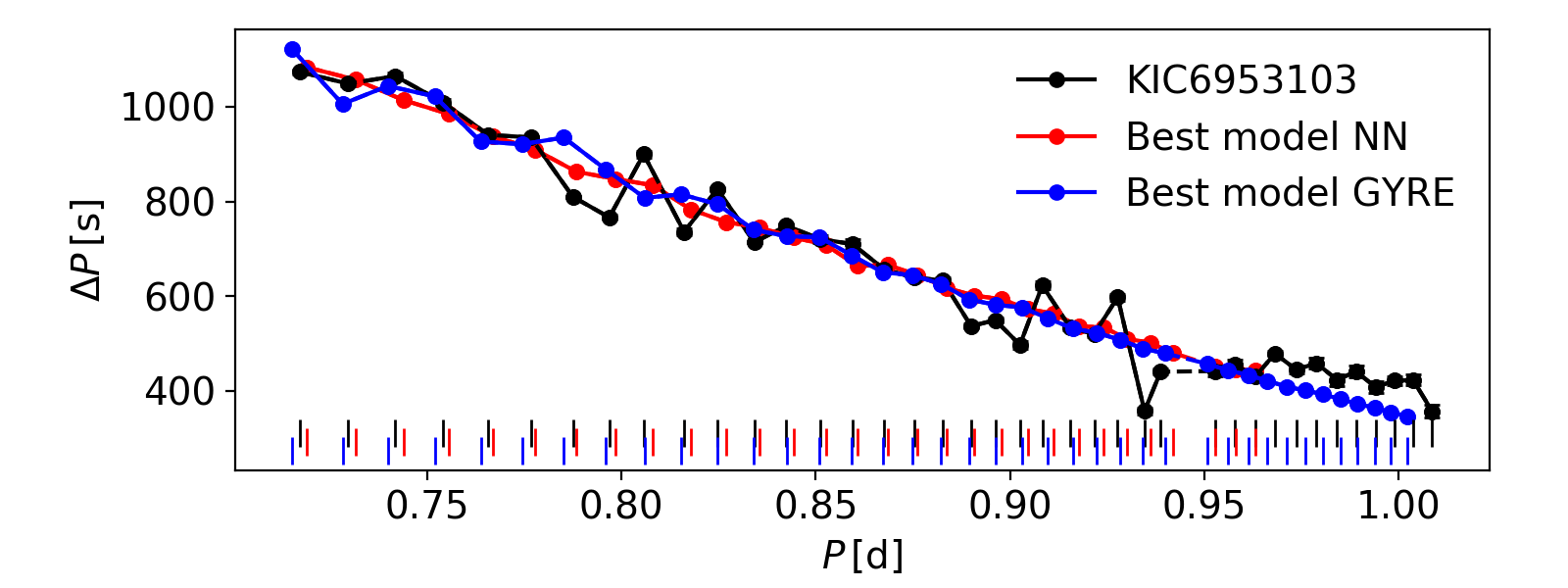}
    \caption{Best-fitting models from the   neural network and \gyre for KIC\,6953103.}
\end{figure*}

\begin{figure*}
    \centering
    \includegraphics[width = 0.9\textwidth]{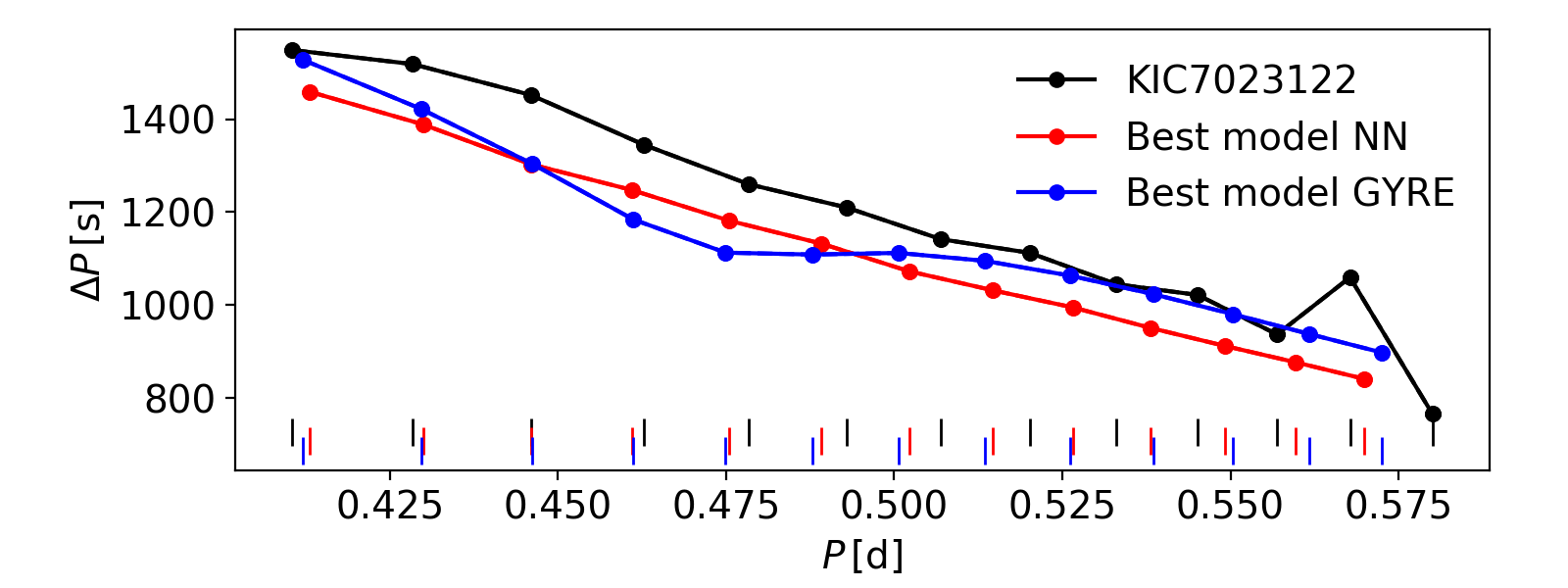}
    \caption{Best-fitting models from the   neural network and \gyre for KIC\,7023122.}
\end{figure*}

\begin{figure*}
    \centering
    \includegraphics[width = 0.9\textwidth]{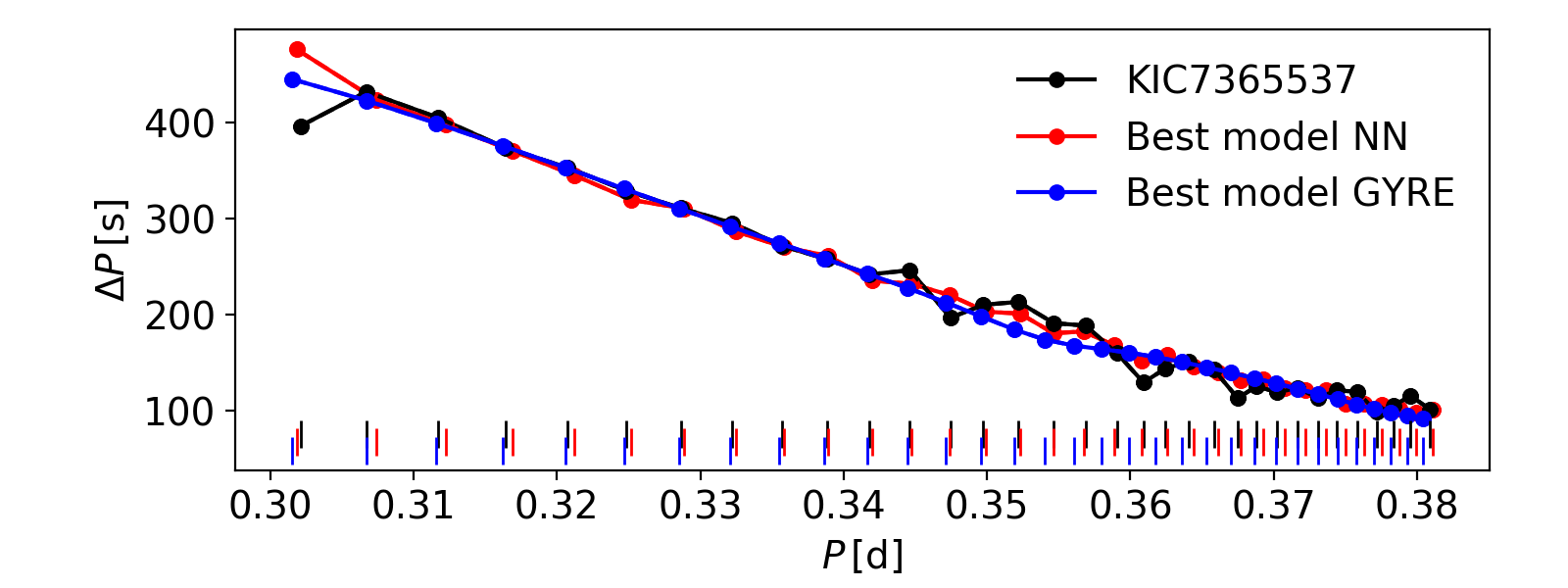}
    \caption{Best-fitting models from the   neural network and \gyre for KIC\,7365537.}
\end{figure*}

\begin{figure*}
    \centering
    \includegraphics[width = 0.9\textwidth]{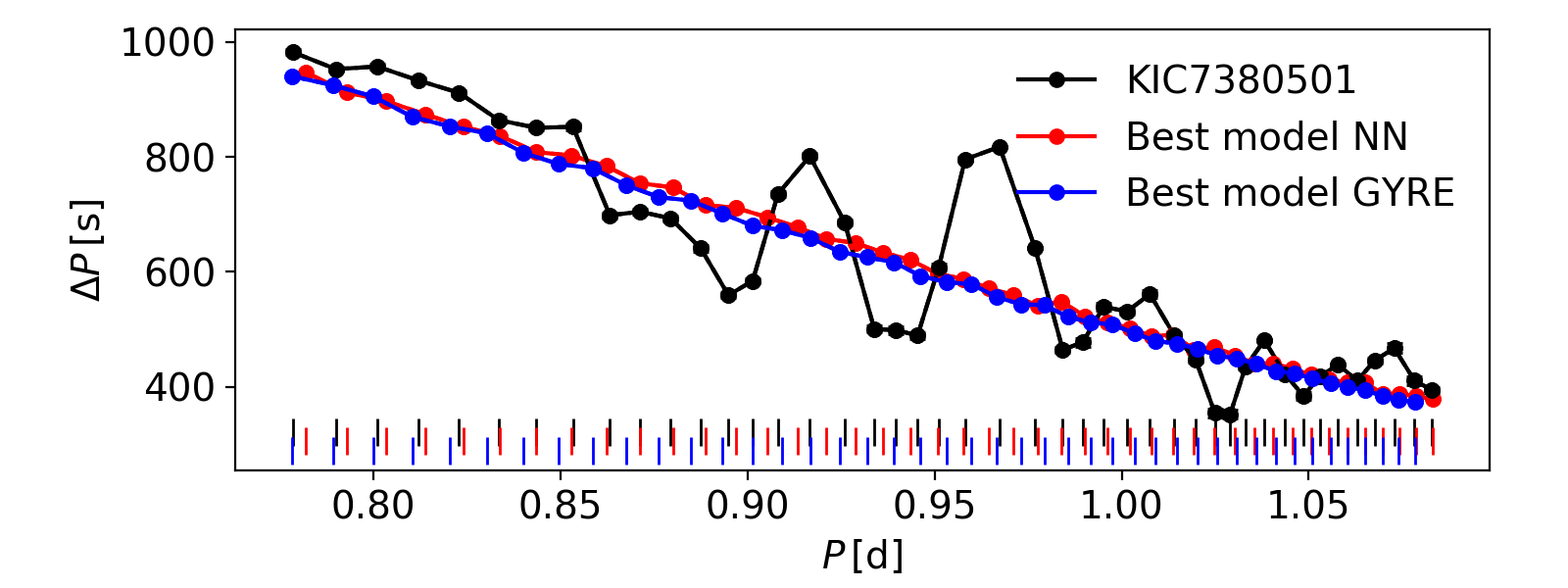}
    \caption{Best-fitting models from the   neural network and \gyre for KIC\,7380501.}
\end{figure*}

\begin{figure*}
    \centering
    \includegraphics[width = 0.9\textwidth]{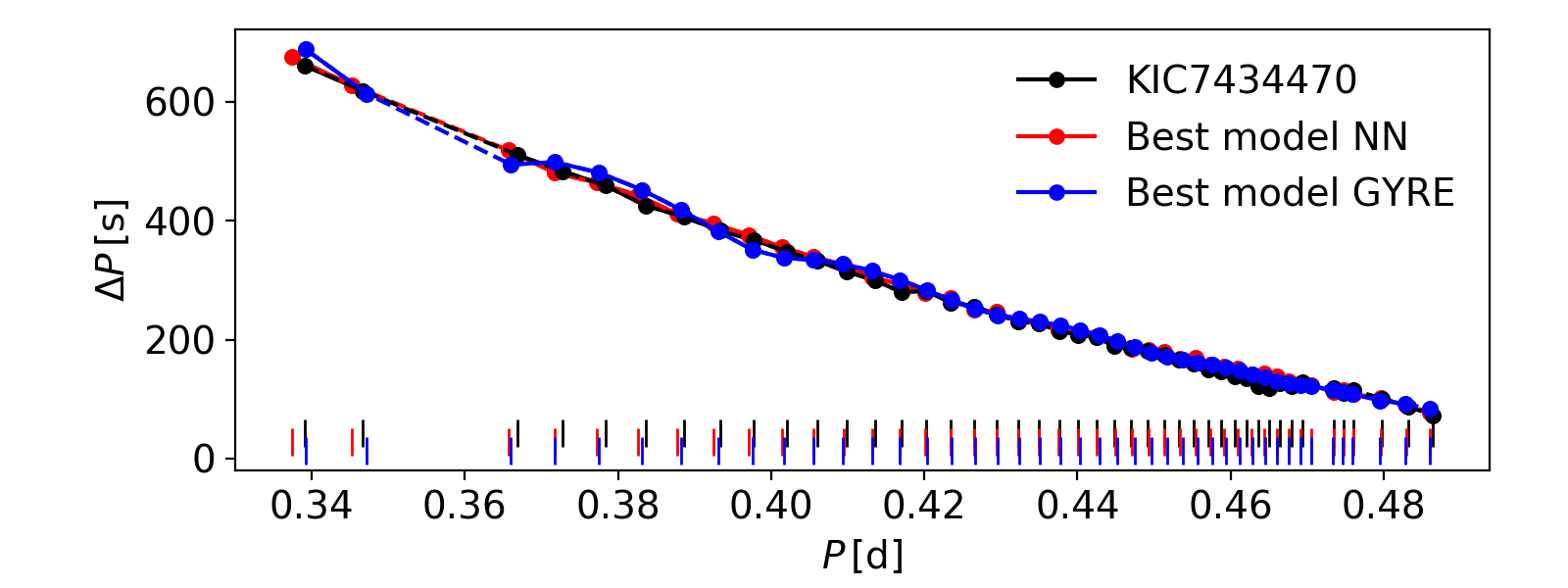}
    \caption{Best-fitting models from the   neural network and \gyre for KIC\,7434470.}
\end{figure*}

\begin{figure*}
    \centering
    \includegraphics[width = 0.9\textwidth]{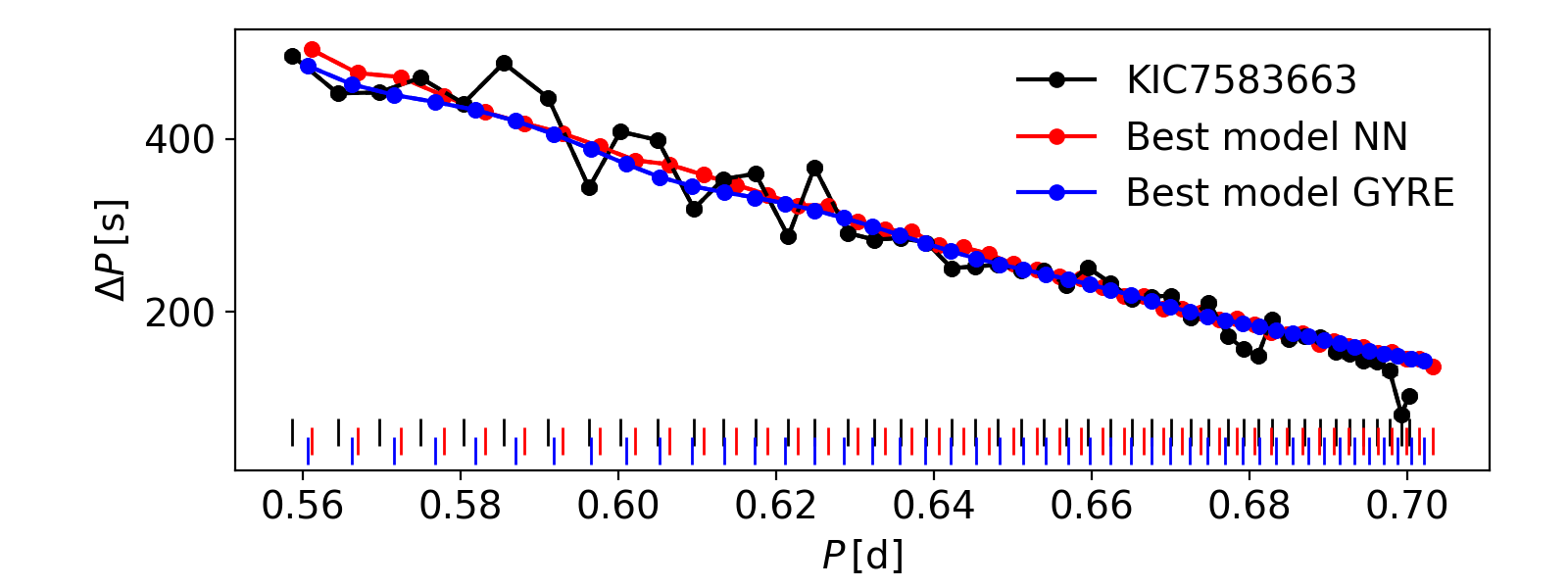}
    \caption{Best-fitting models from the   neural network and \gyre for KIC\,7583663.}
\end{figure*}

\begin{figure*}
    \centering
    \includegraphics[width = 0.9\textwidth]{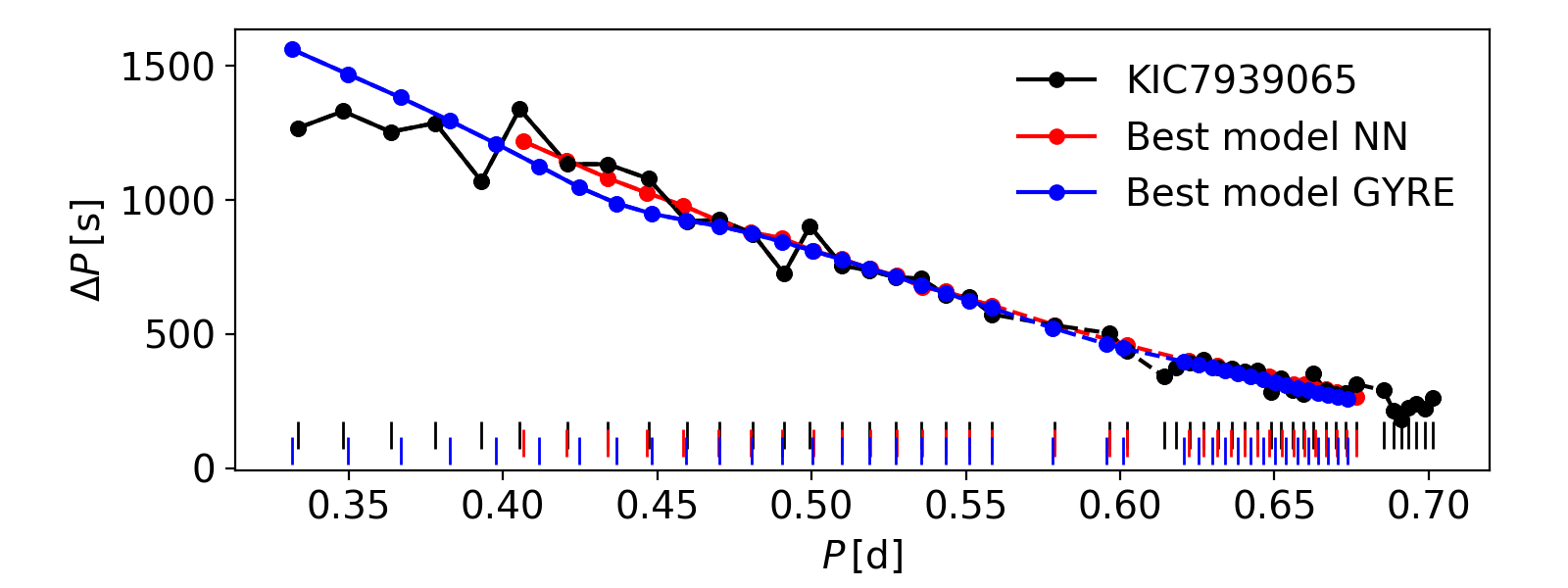}
    \caption{Best-fitting models from the   neural network and \gyre for KIC\,7939065.}
\end{figure*}

\begin{figure*}
    \centering
    \includegraphics[width = 0.9\textwidth]{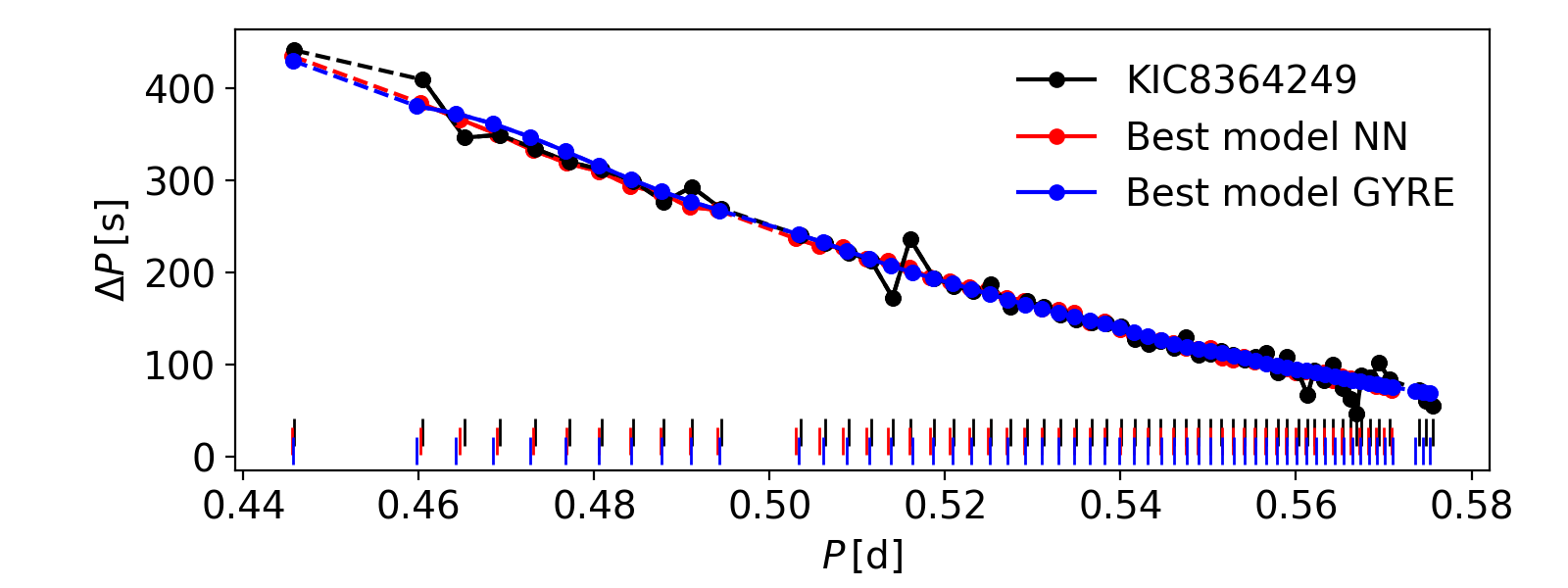}
    \caption{Best-fitting models from the   neural network and \gyre for KIC\,8364249.}
\end{figure*}

\begin{figure*}
    \centering
    \includegraphics[width = 0.9\textwidth]{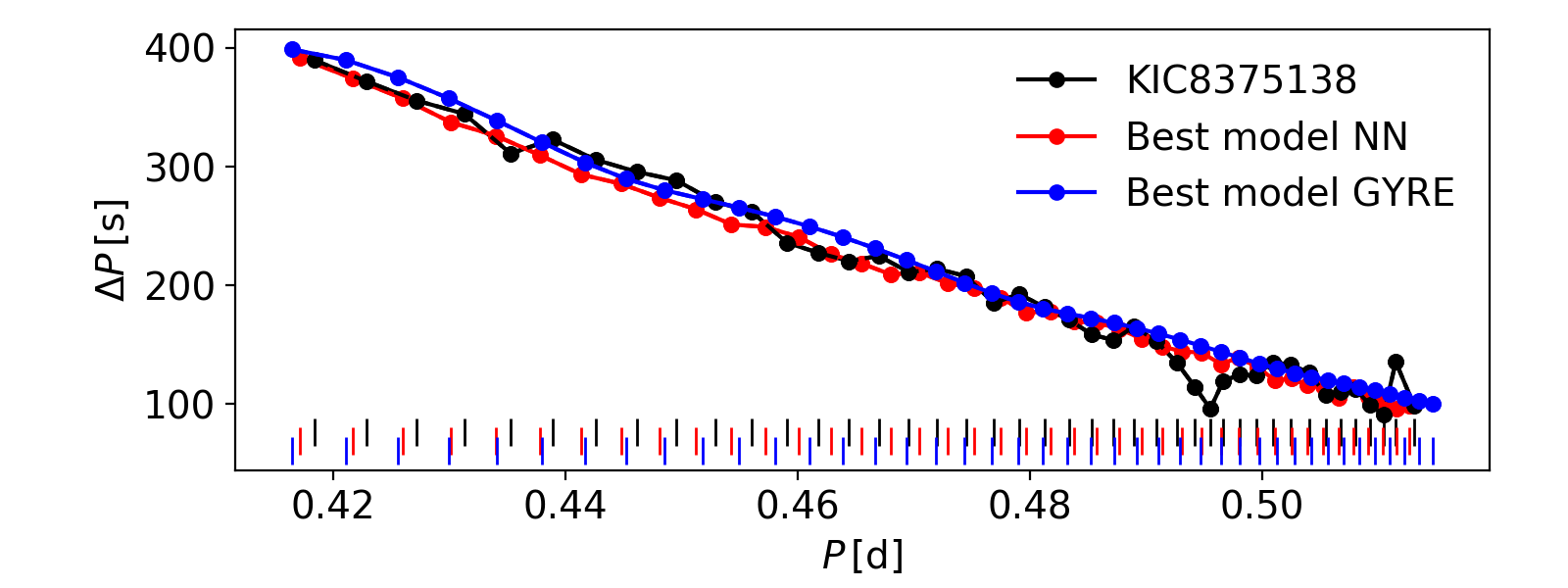}
    \caption{Best-fitting models from the   neural network and \gyre for KIC\,8375138.}
\end{figure*}

\begin{figure*}
    \centering
    \includegraphics[width = 0.9\textwidth]{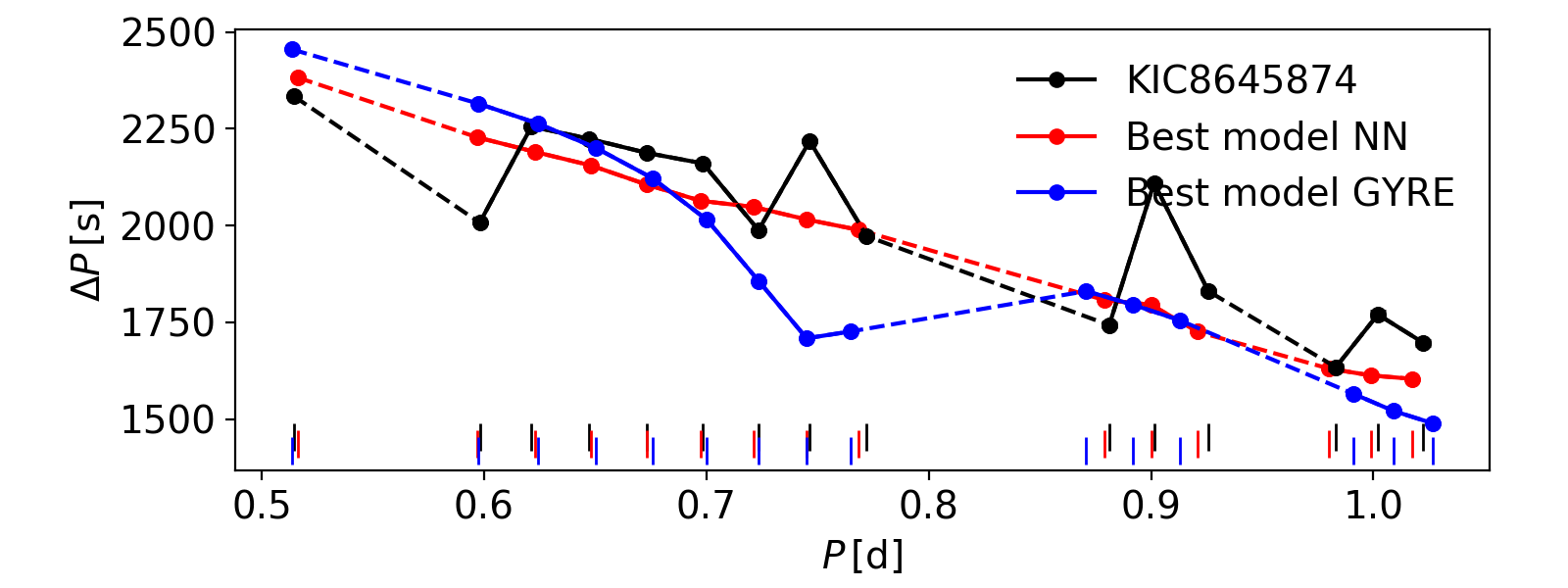}
    \caption{Best-fitting models from the   neural network and \gyre for KIC\,8645874.}
\end{figure*}

\begin{figure*}
    \centering
    \includegraphics[width = 0.9\textwidth]{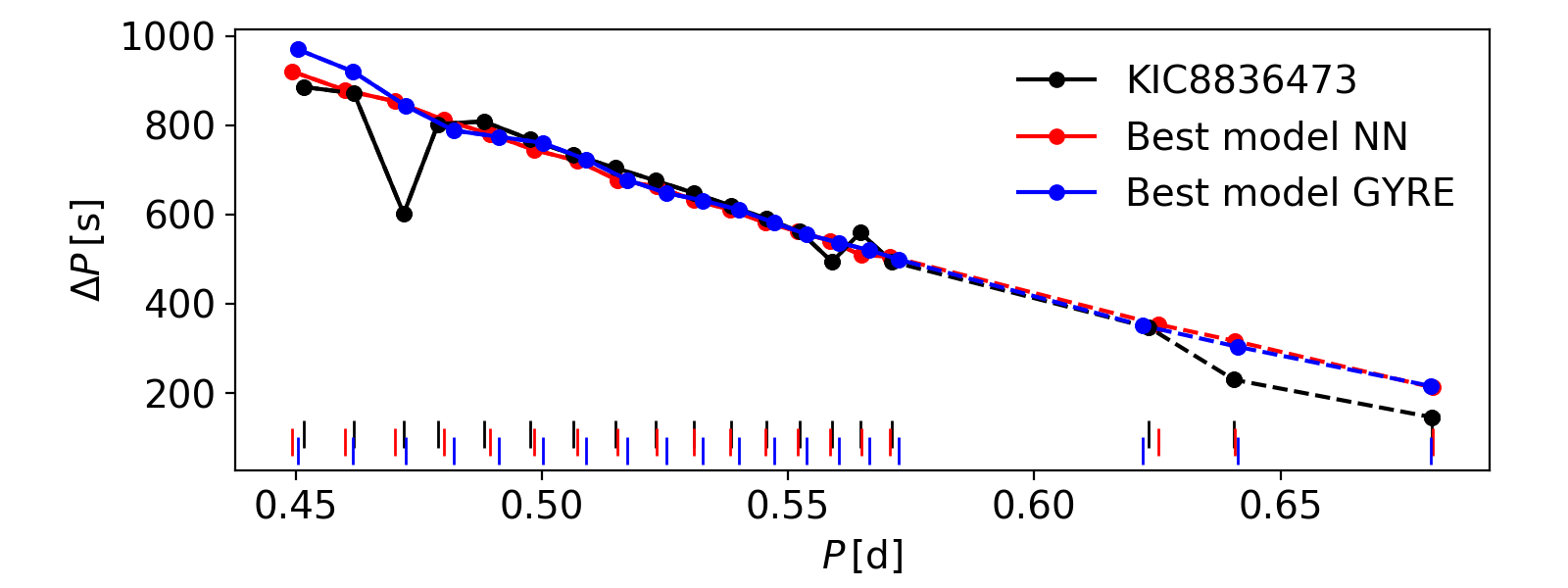}
    \caption{Best-fitting models from the   neural network and \gyre for KIC\,8836473.}
\end{figure*}

\begin{figure*}
    \centering
    \includegraphics[width = 0.9\textwidth]{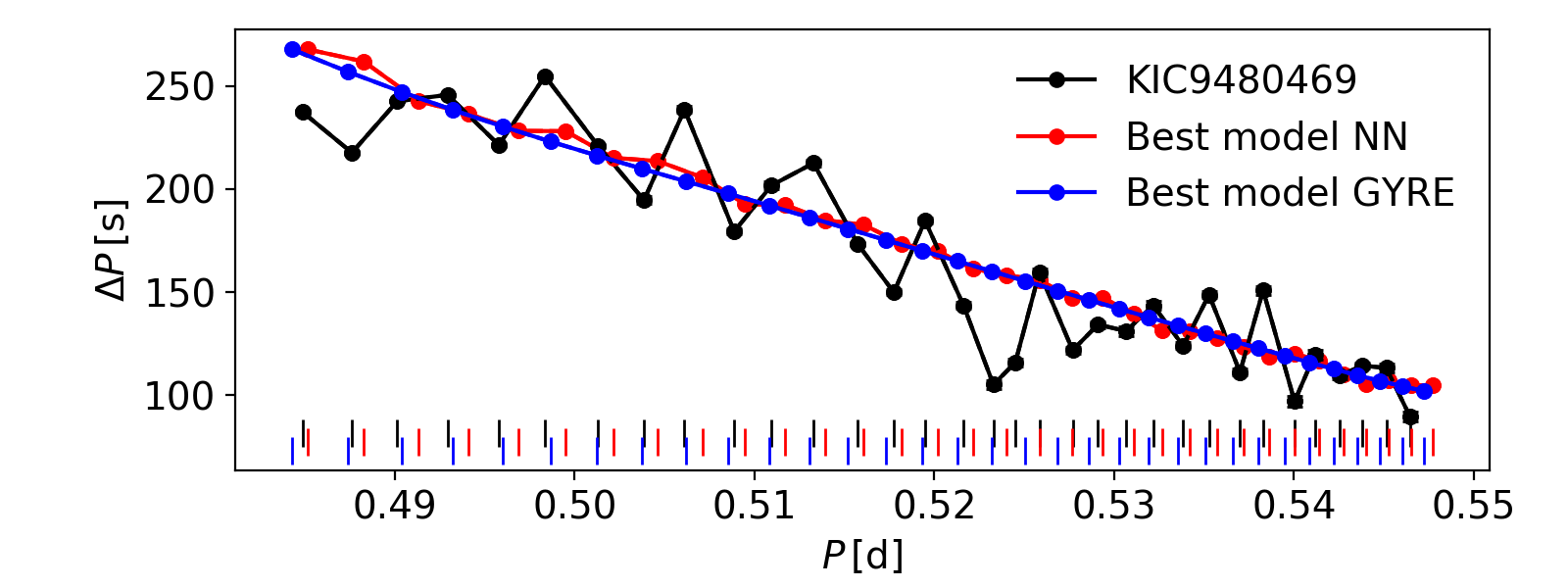}
    \caption{Best-fitting models from the   neural network and \gyre for KIC\,9480469.}
\end{figure*}

\begin{figure*}
    \centering
    \includegraphics[width = 0.9\textwidth]{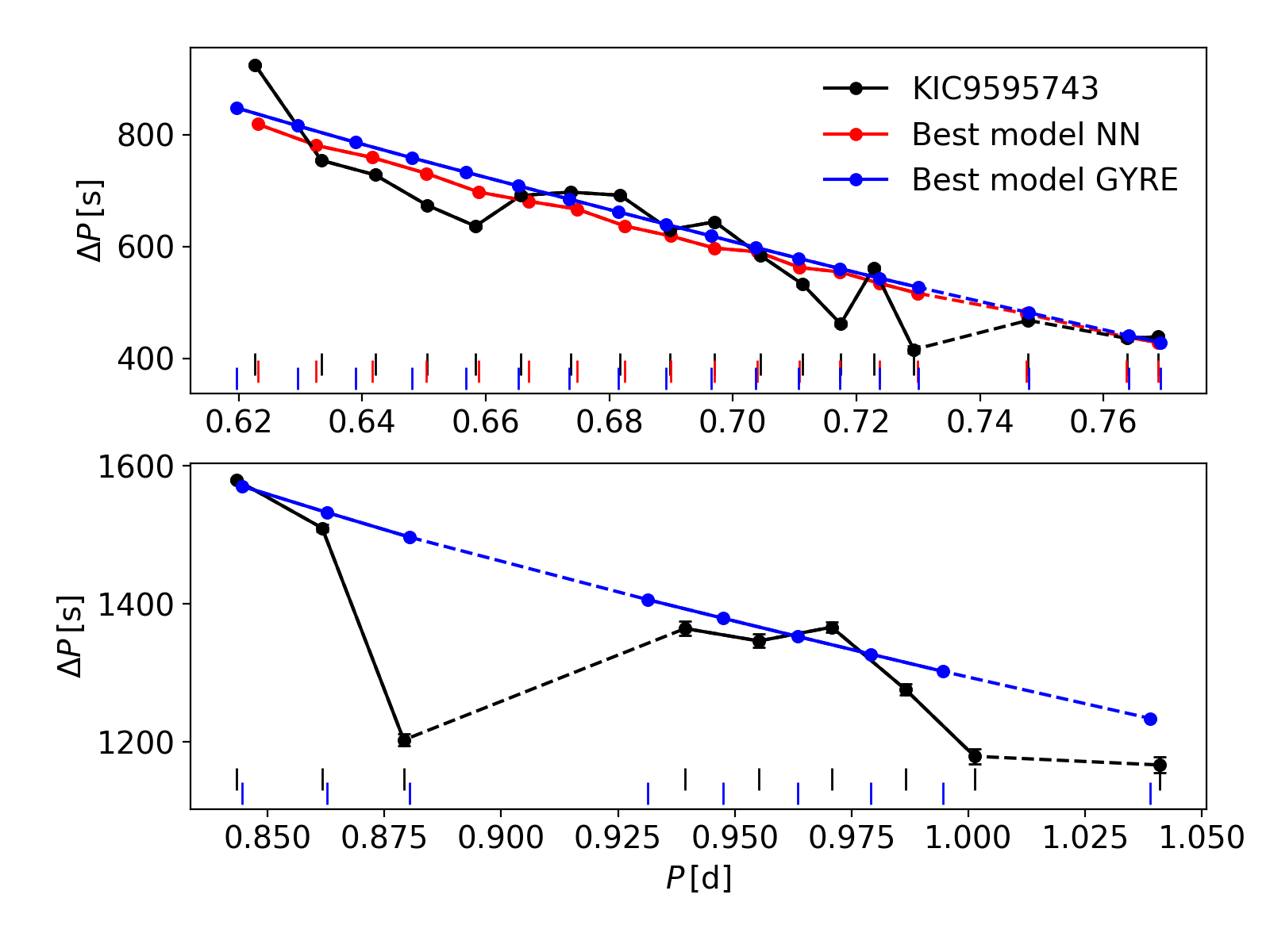}
    \caption{Best-fitting models from the   neural network and \gyre for KIC\,9595743. {Top: $(\ell, m) = (1,1)$, bottom $(\ell, m) = (1,0)$.}}
\end{figure*}

\begin{figure*}
    \centering
    \includegraphics[width = 0.9\textwidth]{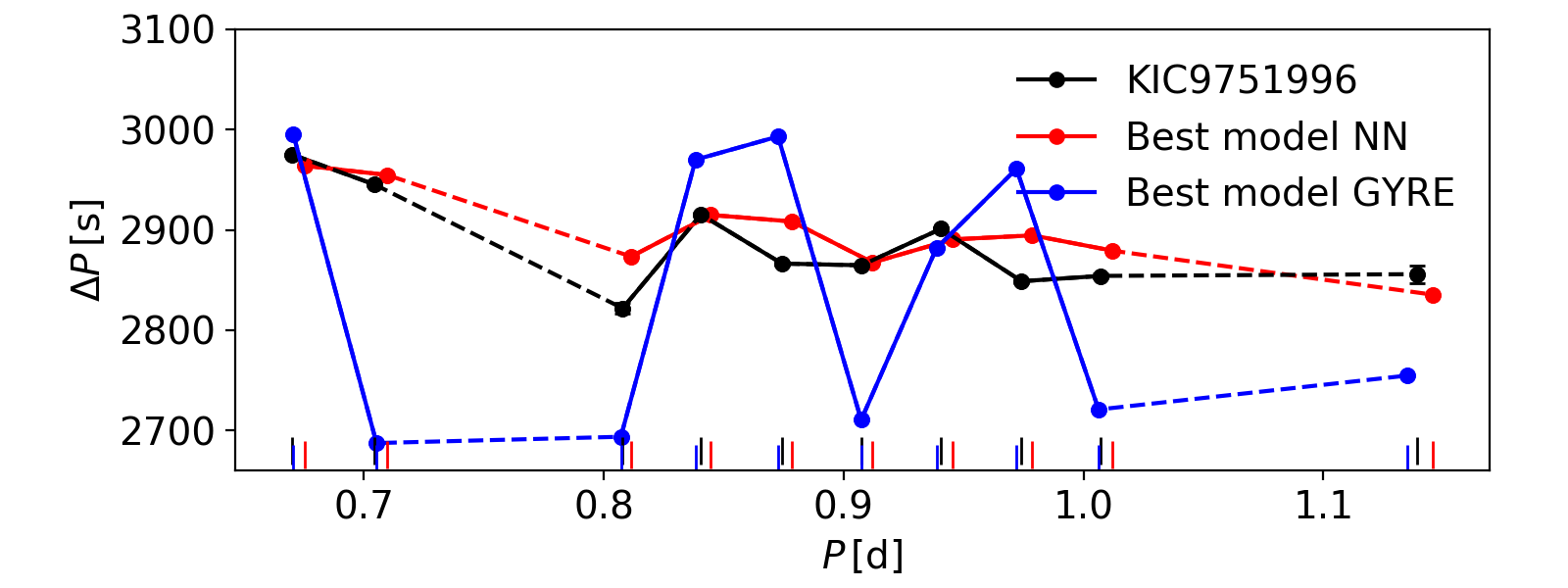}
    \caption{Best-fitting models from the   neural network and \gyre for KIC\,9751996.}
\end{figure*}

\begin{figure*}
    \centering
    \includegraphics[width = 0.9\textwidth]{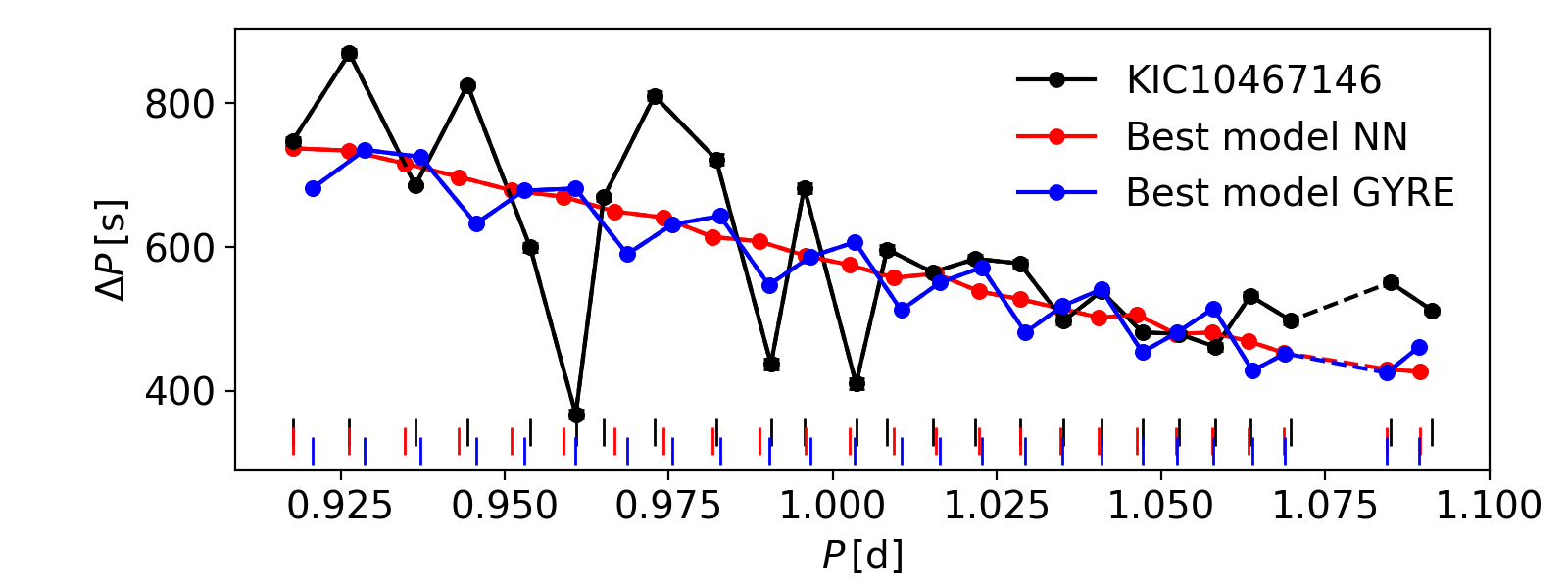}
    \caption{Best-fitting models from the   neural network and \gyre for KIC\,10467146.}
\end{figure*}

\begin{figure*}
    \centering
    \includegraphics[width = 0.9\textwidth]{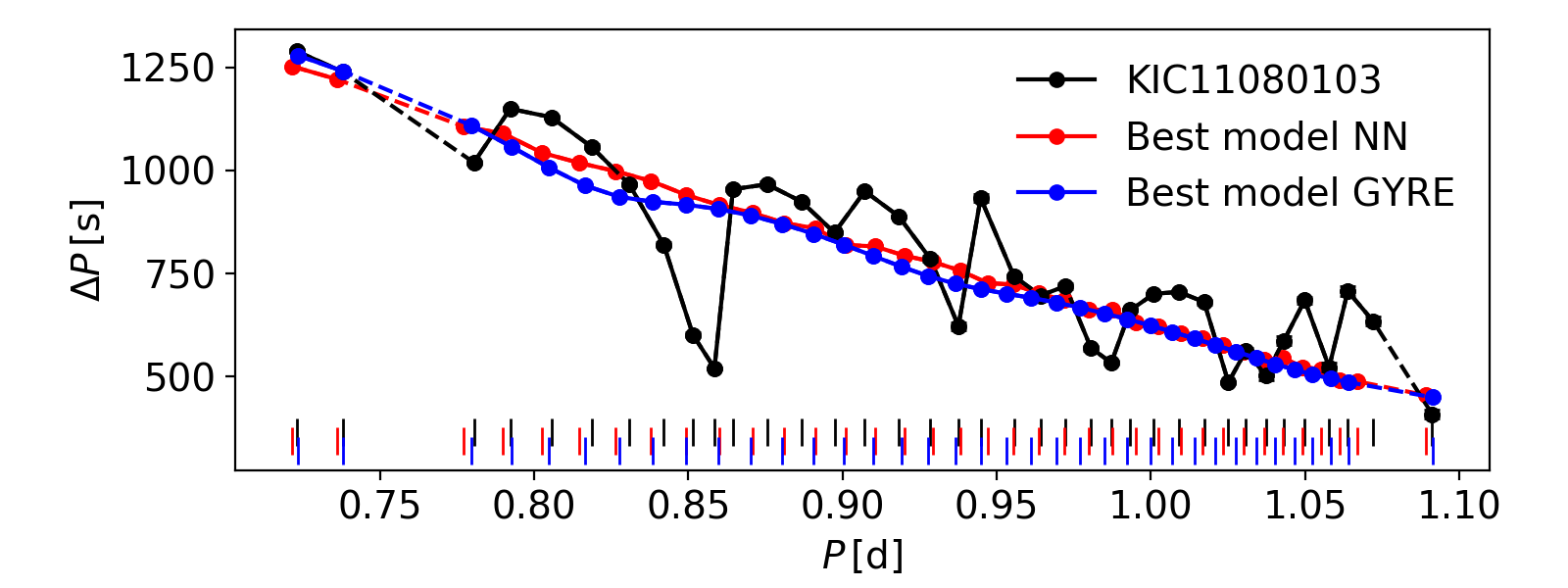}
    \caption{Best-fitting models from the   neural network and \gyre for KIC\,11080103.}
\end{figure*}

\begin{figure*}
    \centering
    \includegraphics[width = 0.9\textwidth]{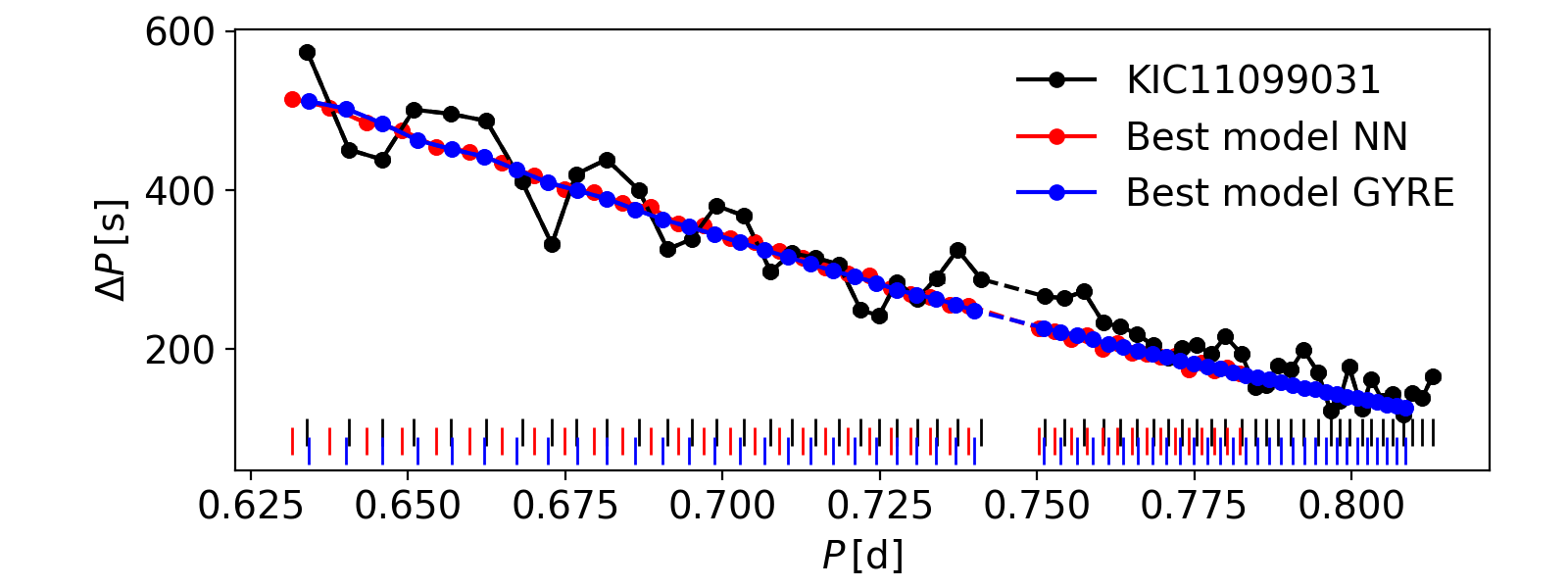}
    \caption{Best-fitting models from the   neural network and \gyre for KIC\,11099031.}
\end{figure*}

\begin{figure*} \label{fig:psp_112_step}
    \centering
    \includegraphics[width = 0.9\textwidth]{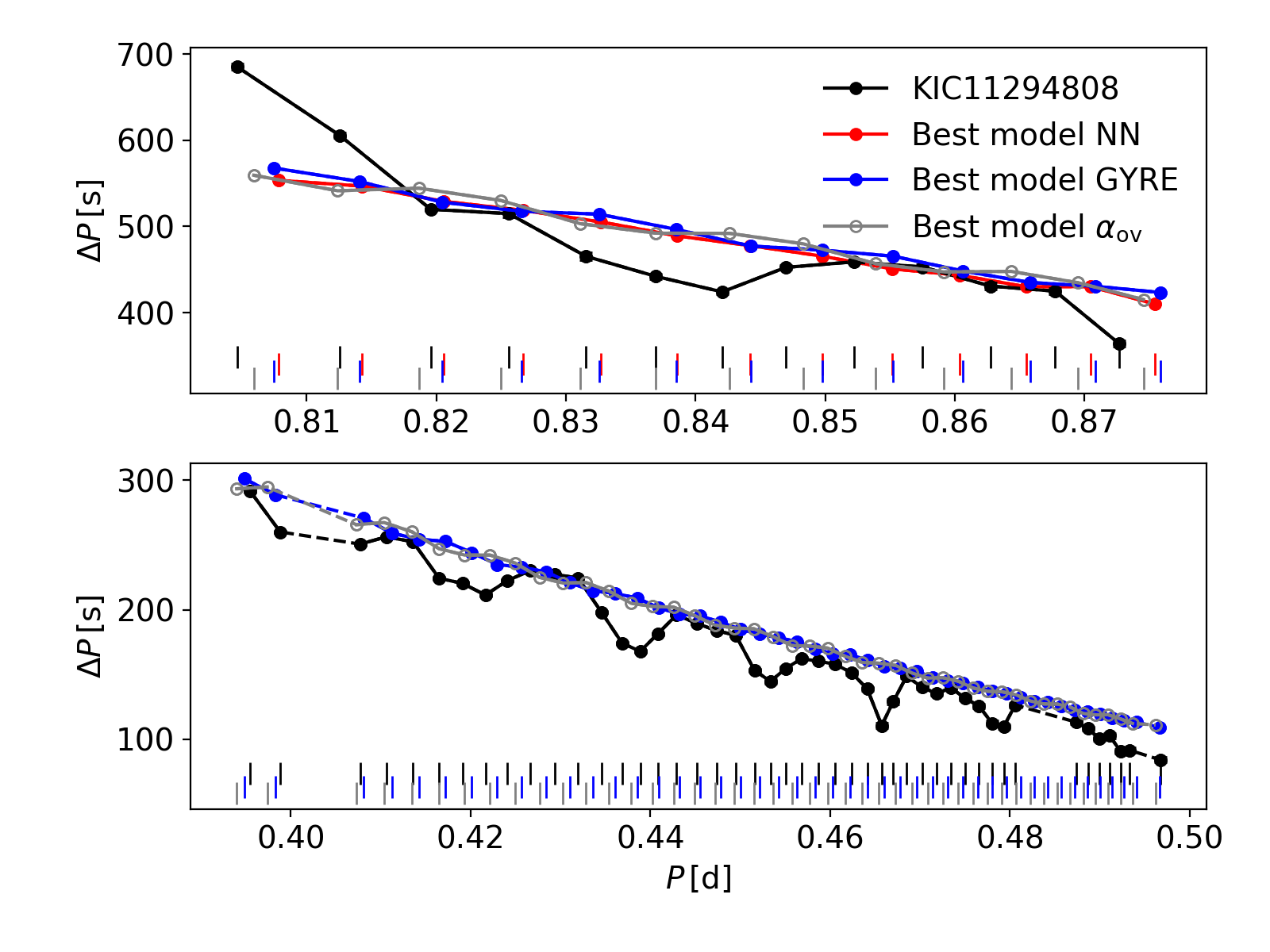}
    \caption{{Best-fitting models from the   neural network and \gyre for KIC\,11294808. Top: $(\ell, m) = (1,1)$, bottom: $(\ell, m) = (2,2)$. Additionally, the best-fitting \gyre model based on equilibrium models with a step-like core overshoot prescription is shown in grey.}}
\end{figure*}

\begin{figure*}
    \centering
    \includegraphics[width = 0.9\textwidth]{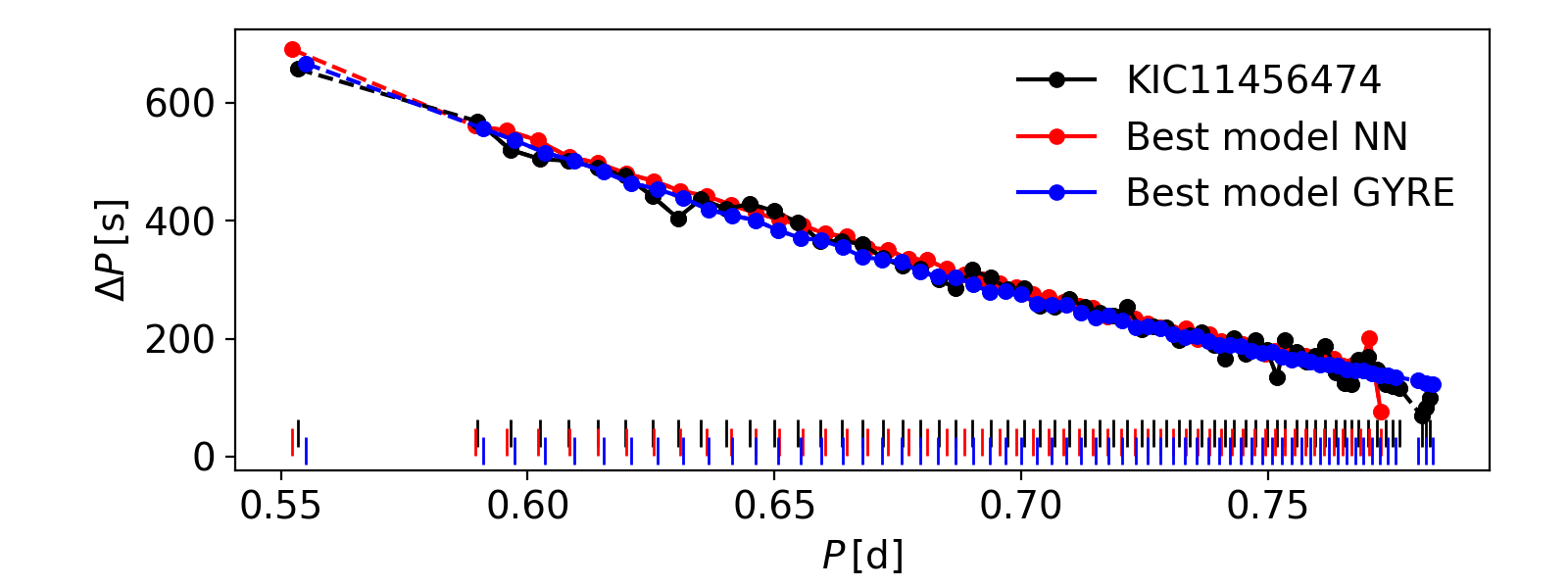}
    \caption{Best-fitting models from the   neural network and \gyre for KIC\,11456474.}
\end{figure*}

\begin{figure*}
    \centering
    \includegraphics[width = 0.9\textwidth]{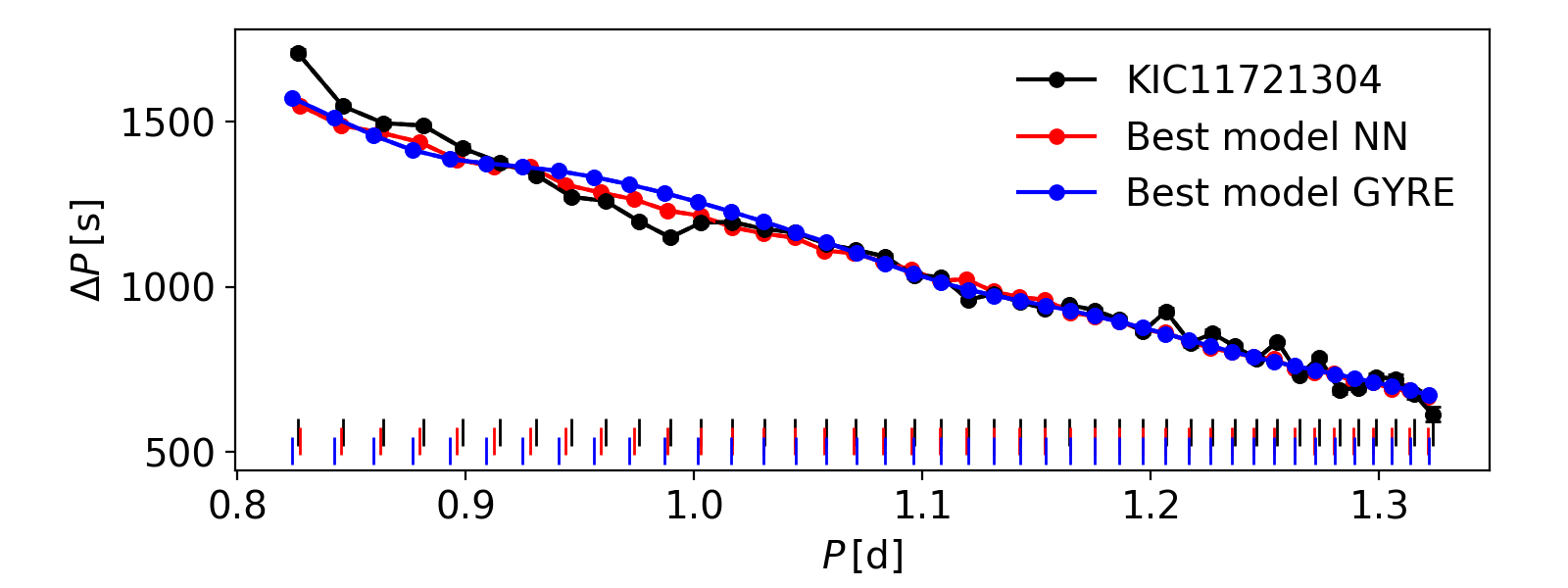}
    \caption{Best-fitting models from the   neural network and \gyre for KIC\,11721304.}
\end{figure*}

\begin{figure*}
    \centering
    \includegraphics[width = 0.9\textwidth]{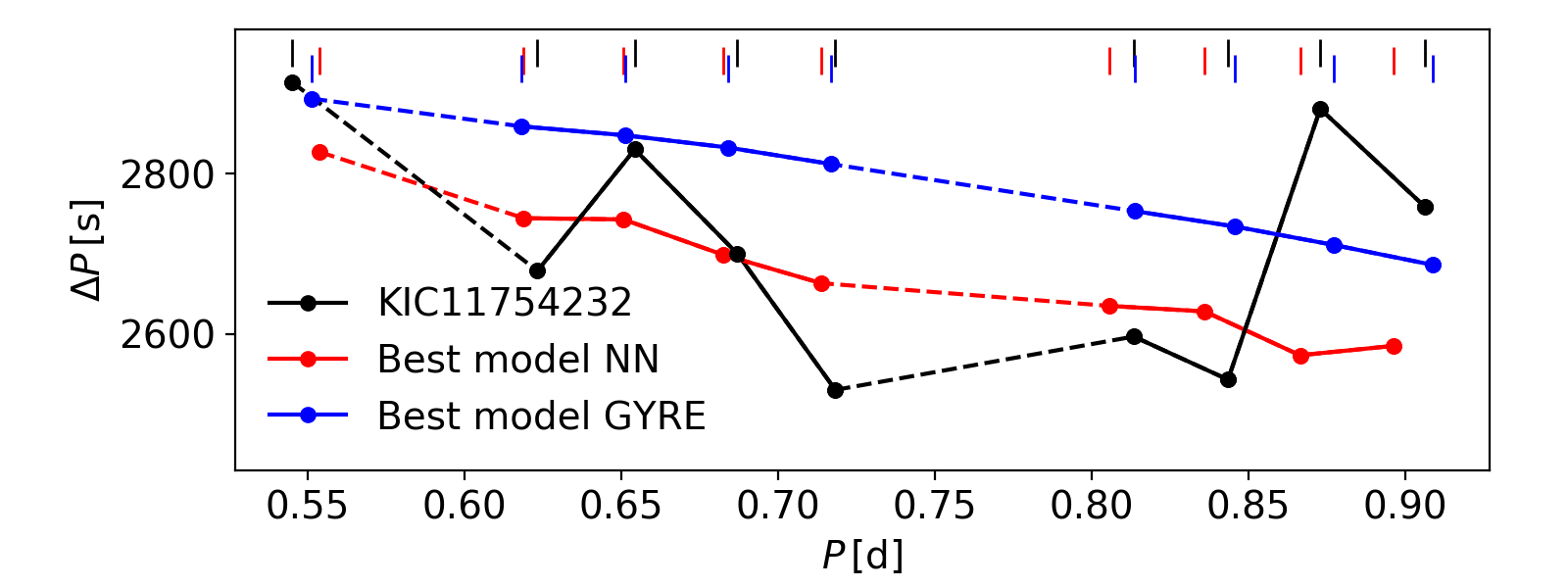}
    \caption{Best-fitting models from the   neural network and \gyre for KIC\,11754232.}
\end{figure*}

\begin{figure*}
    \centering
    \includegraphics[width = 0.9\textwidth]{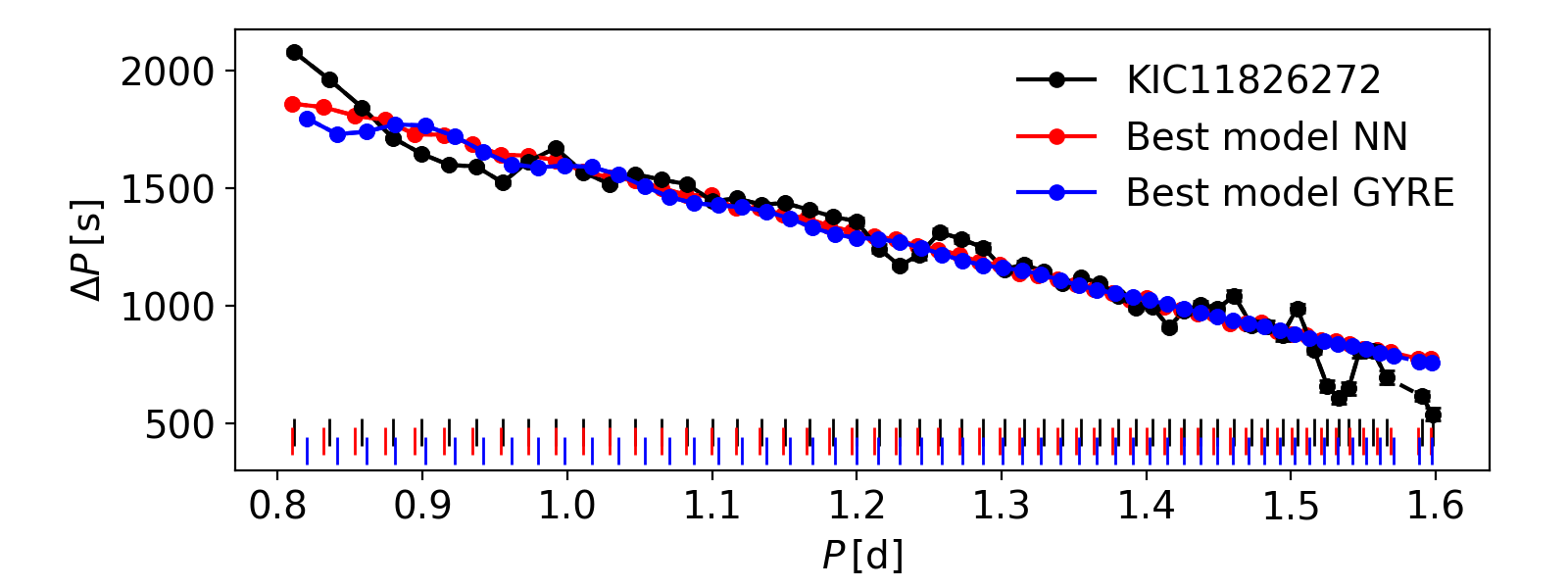}
    \caption{Best-fitting models from the   neural network and \gyre for KIC\,11826272.}
\end{figure*}

\begin{figure*}
    \centering
    \includegraphics[width = 0.9\textwidth]{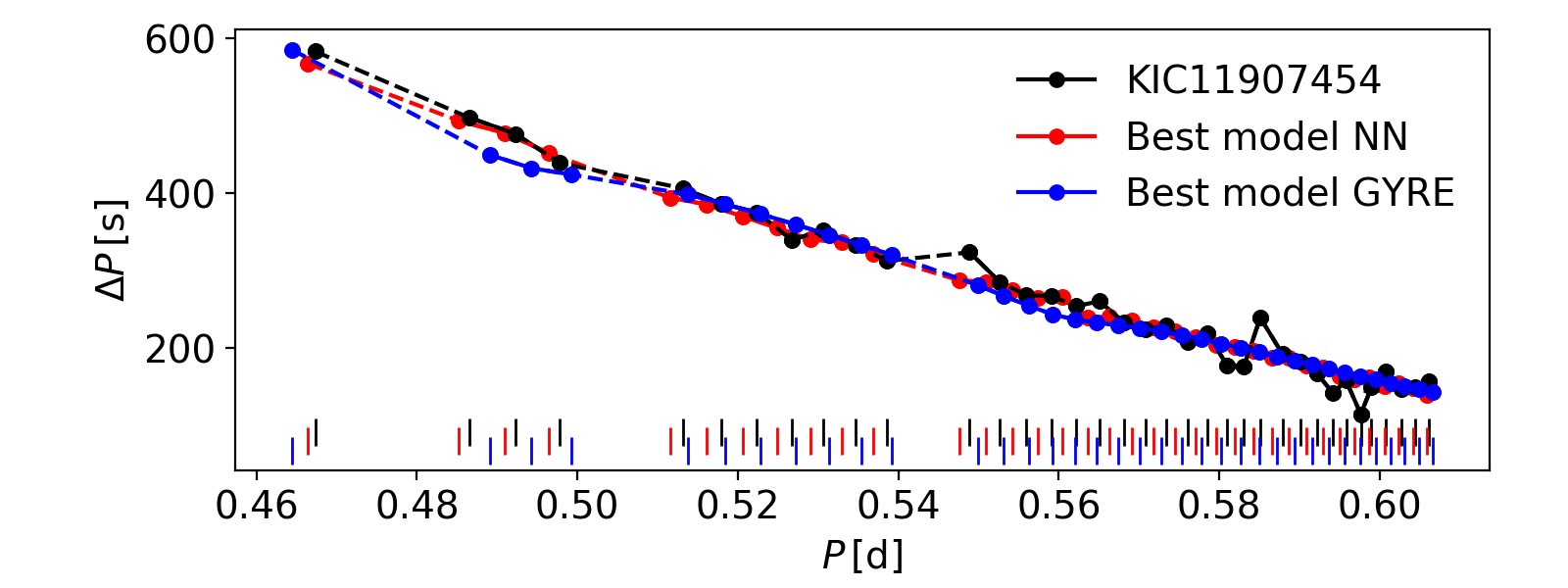}
    \caption{Best-fitting models from the   neural network and \gyre for KIC\,11907454.}
\end{figure*}

\begin{figure*}
    \centering
    \includegraphics[width = 0.9\textwidth]{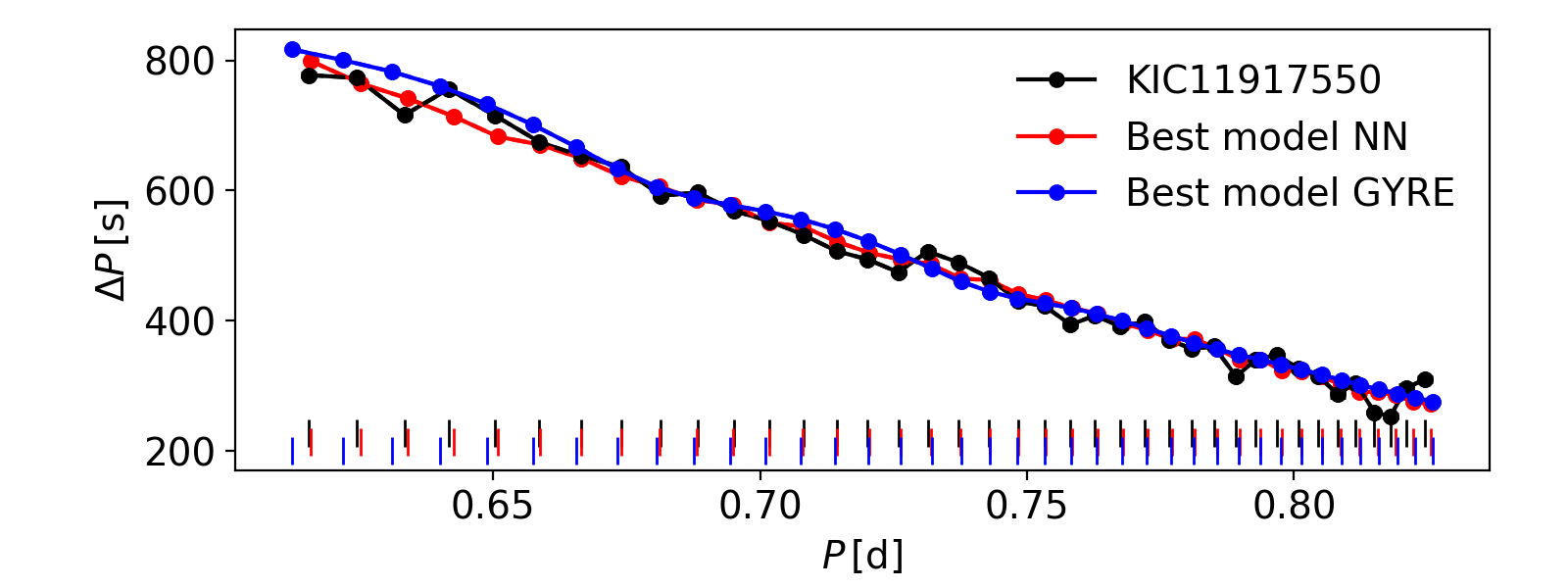}
    \caption{Best-fitting models from the   neural network and \gyre for KIC\,11917550.}
\end{figure*}

\begin{figure*}
    \centering
    \includegraphics[width = 0.9\textwidth]{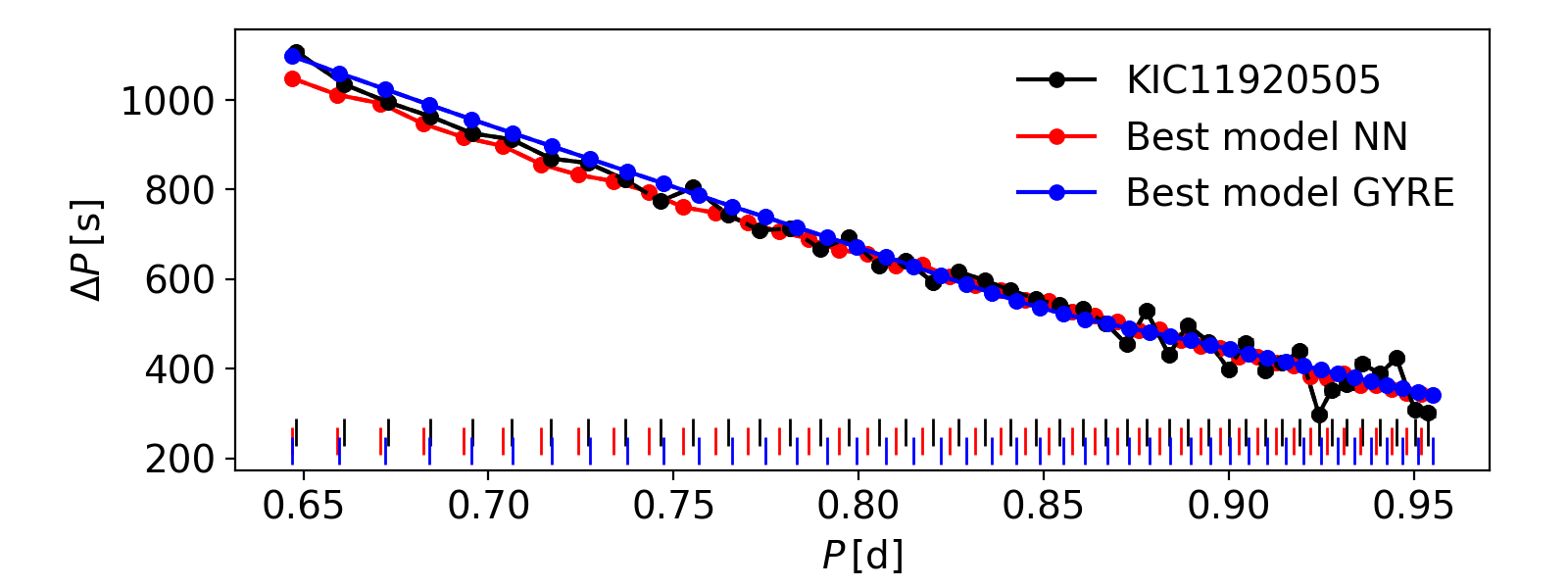}
    \caption{Best-fitting models from the   neural network and \gyre for KIC\,11920505.}
\end{figure*}

\begin{figure*}
    \centering
    \includegraphics[width = 0.9\textwidth]{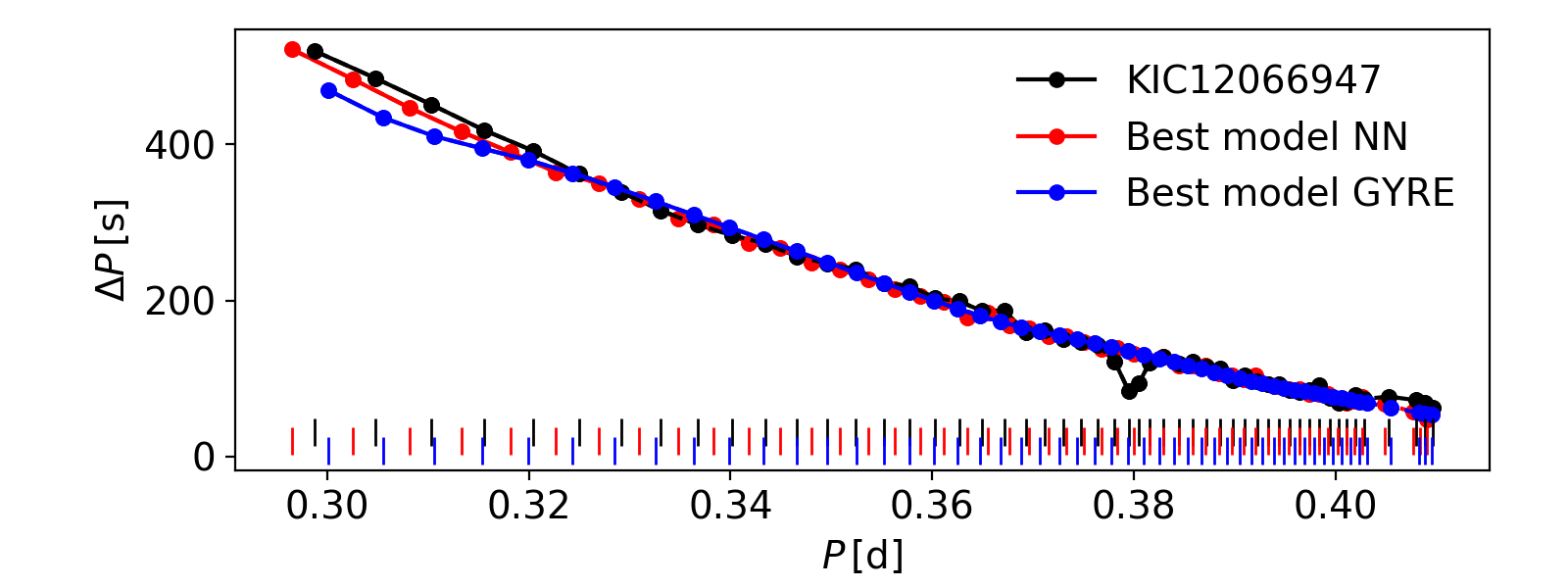}
    \caption{Best-fitting models from the   neural network and \gyre for KIC\,12066947.}
\end{figure*}
\clearpage

\begin{figure*}
    \centering
    \includegraphics[width = \textwidth]{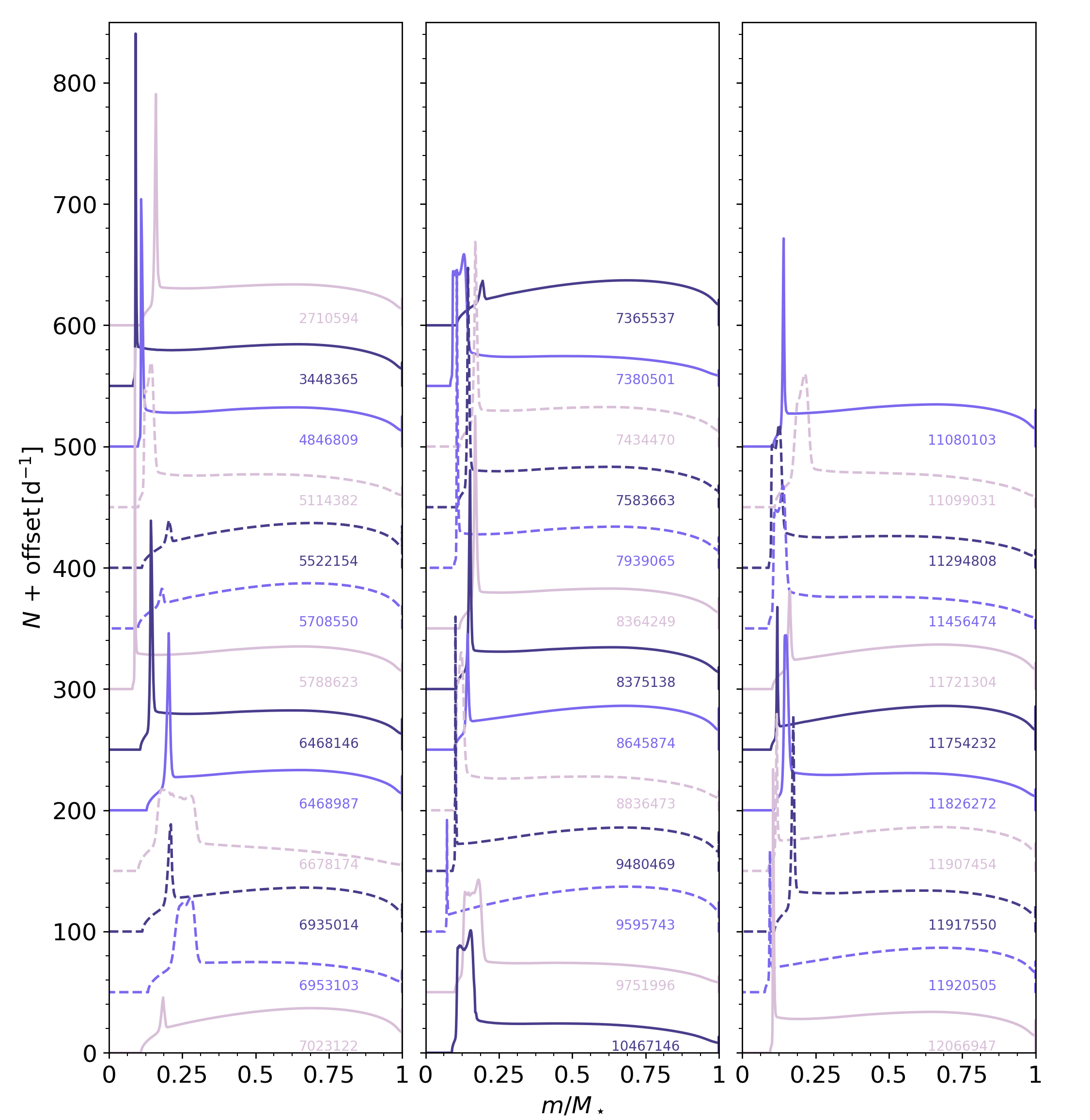}
    \caption{{Brunt-V\"ais\"al\"a frequency profiles of the best model for each star in our sample. For clarity, the profiles are offset by $50\,{\rm d^{-1}}$ from one another.}}
    \label{fig:BV_profiles}
\end{figure*}
\clearpage

\section{Test with $D_0 = 0.05\,{\rm cm^2 s^{-1}}$}
{In the appendix, we show the result of the modelling of KIC\,11294808, where we have fixed $D_0 = 0.05\,{\rm cm^2 s^{-1}}$ instead of $D_0 = 1\,{\rm cm^2 s^{-1}}$ used in the rest of the paper.
\begin{figure*}
    \centering
    \includegraphics[width = 0.9\textwidth]{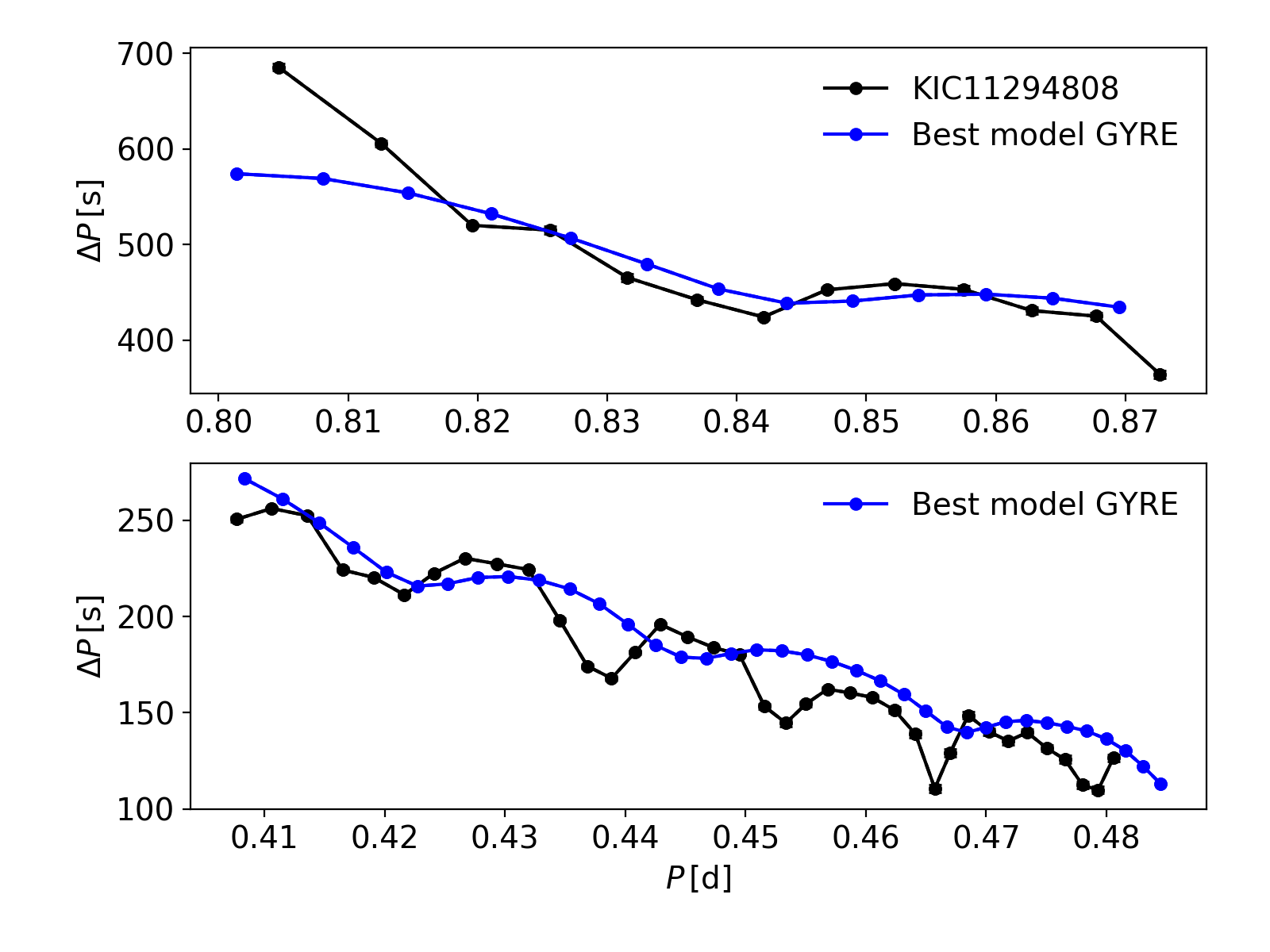}
    \caption{Best-fitting \gyre model for KIC\,11294808, when $D_0 = 0.05\,{\rm cm^2 s^{-1}}$ is assumed. Top: $(\ell, m) = (1,1)$, bottom: $(\ell, m) = (2,2)$.}
    \label{fig:psp_112_lowD}
\end{figure*}
}

\section{The effect of including Gaia luminosities on the derived mass}
\begin{figure*}
    \centering
    \includegraphics[width = 0.9\textwidth]{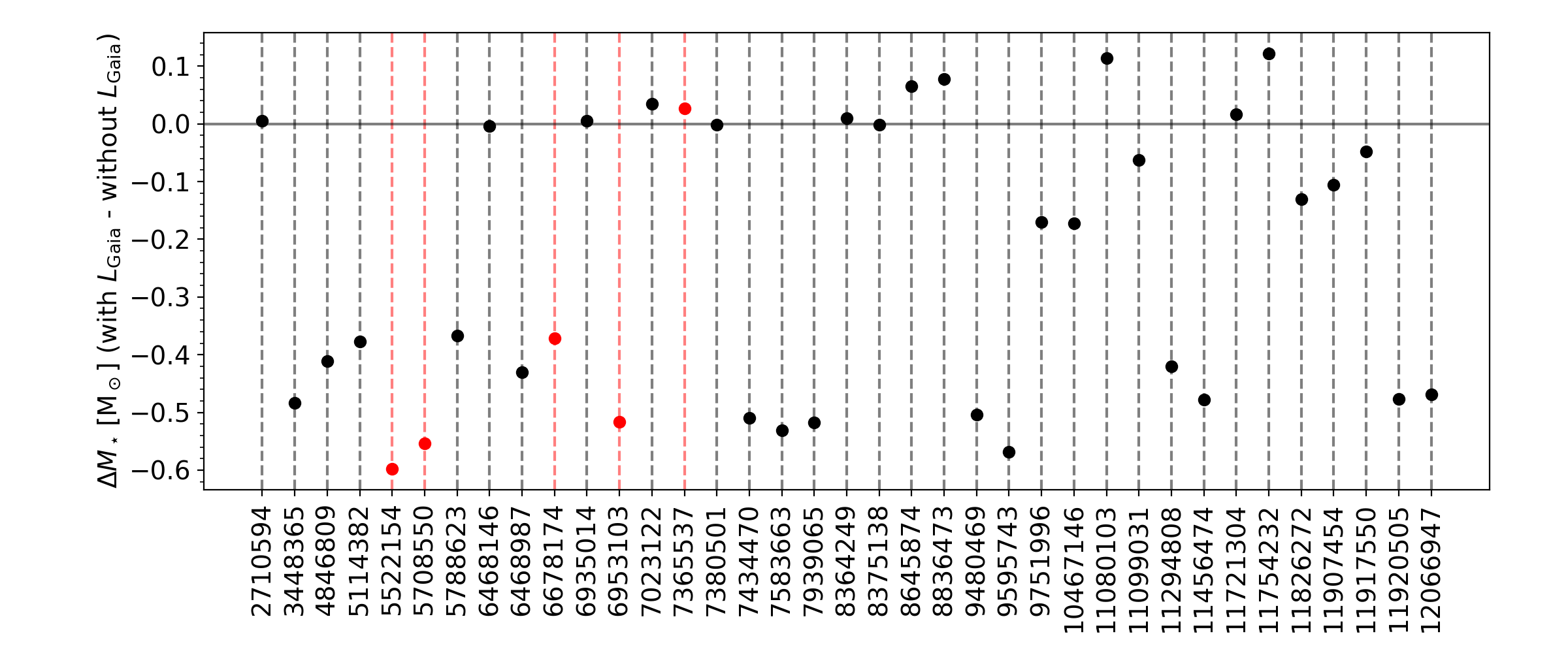}
    \caption{{Differences in inferred mass per star (KIC ID on the abscissa) when the luminosities of the models are not required to comply with the observed luminosity deduced from Gaia DR2 data. Constraints on \teff and \logg are in both cases enforced. The stars for which we have not used the Gaia luminosity in the modelling are indicated in red. For these stars, we have put no constraints on the spectroscopic parameters either in our final MLE (hence this solution is not indicated here).} }
    \label{fig:M_GaiaL}
\end{figure*}

{In Figure\,\ref{fig:M_GaiaL}, we show the difference in inferred stellar mass when we do not include the luminosity from Gaia DR2 in our modelling procedure.}

\end{document}